%% file: main.tex
\documentclass[11pt]{article}
\usepackage[running,letterpaper]{cleanpaper}
 
\title{Graph Similarity Meets Optimal Transport}
\author{%
  \cpauthor{Jens Breitung}{1}%
  \cpauthor{Timo Gervens}{1}%
  \cpauthor{Martin Grohe}{1}}
\affiliation{%
  \cpaffil{1}{RWTH Aachen University, Germany\\
    \hspace*{0.6em}\texttt{\{lastname\}@informatik.rwth-aachen.de}}
    }
\keywords{graph similarity, optimal transport, edit distance, cut distance, spectral distance}
\funding{All three authors are funded by the European Union (ERC, SymSim, 101054974).}
\acknowledgements{We thank Tom\'{a}\v{s} Novotn\'{y} for fruitful discussions.}
 
\runningtitle{Graph Similarity Meets Optimal Transport}
\runningauthors{J.~Breitung, T.~Gervens and M.~Grohe}
 
\input{init/packages}
\input{init/definitions}

\begin{document}
\maketitle
\nolinenumbers
\thispagestyle{empty}\setcounter{page}{0}

\setstretch{1.1}

\begin{abstract}
    A natural way to compare two graphs is to align their vertex sets by a bijection and to measure the difference of the aligned adjacency matrices in a matrix norm.
    Depending on the norm, this yields, for example, the edit distance, the cut distance, the spectral distance, and the local edit distance.
    To make these distances applicable to graphs of different order, one can consider blowups of the graphs or, equivalently, relax bijections to fractional alignments, that is, optimal transport plans between the uniform distributions on the vertex sets, in the spirit of Gromov--Wasserstein distances.

    We investigate how much the optimal transport versions can differ from their bijective counterparts, that is, we quantify their integrality gaps.
    For the edit distance, we improve the best known upper bound on the integrality gap from $3$ to $2$ and the lower bound from $6/5$ to $4/3$.
    For the cut distance,
    we approach Lov\'asz's conjecture that $\nd{\cut} \leq 2\ndOT{\cut}$ from both sides. We improve the previously best known upper bound $\nd{\cut} \leq 32 \cdot (\ndOT{\cut})^{1/67}$ (Borgs, Chayes, Lov\'asz, S\'os and Vesztergombi, Adv.~Math.~2008) to $\nd{\cut} \leq C_\eps \cdot (\ndOT{\cut})^{1-\eps}$ for every $\eps > 0$, and we construct graphs $G,H$ with $\nd{\cut}(G,H) = 2\ndOT{\cut}(G,H)$, establishing a factor-$2$ lower bound that matches Lov\'asz's conjecture.
    For the local edit and spectral distances, we obtain upper bounds of order $\sqrt{t\ln(1/t)}$ and $\sqrt{t}$, respectively, where $t$ is the corresponding optimal transport distance, and we show that, unlike for the edit and cut distances, no linear upper bounds are possible.

    Finally, we show that computing the optimal transport distances is NP-hard, and we give hardness of approximation results for the cut, spectral, and local edit distances.
\end{abstract}
\setstretch{1.0}

\newpage
\input{sections/01_introduction}

\input{sections/02_preliminaries}
\input{sections/03_integrality_gaps}
\input{sections/04_hardness}

\input{sections/05_conclusion}

\newpage
\appendix
\input{appendices/01_extended_preliminaries}
\input{appendices/02_ot_formulation_p_to_q}
\input{appendices/03_edit_distance_bounds}
\input{appendices/04_local_edit_distance_bounds}
\input{appendices/05_cut_distance_bounds}
\input{appendices/06_spectral_distance_bounds}
\input{appendices/07_hardness}

\nolinenumbers
\bibliography{references}

\end{document}

%% file: init/packages.tex
\usepackage[T1]{fontenc}
\usepackage[utf8]{inputenc} 

\usepackage{fontawesome5} 

\usepackage[textsize=tiny,disable]{todonotes}
\usepackage{csquotes}

\usepackage{bm}
\usepackage{bbm}
\usepackage{complexity}
\usepackage{mathtools}
\usepackage{nccmath}
\usepackage{wasysym}
\usepackage{xfrac}

\usepackage[capitalise]{cleveref}
\usepackage{refcount} 
\hypersetup{bookmarksnumbered=true} 

\usepackage{adjustbox}

\usepackage[section]{algorithm}
\usepackage{algorithmicx}
\usepackage{algpseudocode}

\usetikzlibrary{
    patterns,
    patterns.meta,
    calc,
    math,
    arrows,
    arrows.meta,
    automata,
    positioning,
    shapes.symbols,
    shapes.geometric,
    decorations.pathmorphing,
    decorations.pathreplacing,
    shapes,
    nfold,
    fit
}

\AddToHook{cmd/appendix/before}{%
  \crefalias{section}{appendix}%
  \crefalias{subsection}{appendix}%
  \crefalias{subsubsection}{appendix}%
}

%% file: init/definitions.tex
\newcommand{\martin}[1]{\todo[color = red!50, author = \textbf{Martin},size=tiny]{#1}}

\crefname{claim}{Claim}{Claims}
\crefname{figure}{Figure}{Figures}

\newcommand{\headingref}[2]{\texorpdfstring{\cref{#2}}{#1 \getrefnumber{#2}}}
\Crefname{claim}{Claim}{Claims}

\crefname{observation}{Observation}{Observations}
\Crefname{observation}{Observation}{Observations}

\newcommand{\wrt}{\text{with respect to}}
\newcommand{\WLOG}{\text{w.l.o.g.\@}}
\newcommand{\ie}{\text{i.e.\@}}
\newcommand{\eg}{\text{e.g.\@}}

\newcommand{\lovasz}{\text{Lovász}}

\newcommand{\ceq}{\ensuremath{\coloneqq}}
\newcommand{\eps}{\ensuremath{\varepsilon}}
\newcommand{\cst}[2][]{\ensuremath{\zeta_{#2}\if\relax\detokenize{#1}\relax\else^{#1}\fi}}
\newcommand{\scst}[2][]{\ensuremath{\gamma_{#2}\if\relax\detokenize{#1}\relax\else^{#1}\fi}}

\newcommand{\indicator}{\ensuremath{\mathbbm{1}}}
\newcommand{\symdiff}{\ensuremath{\,\triangle\,}}
\newcommand{\supp}{\ensuremath{\textup{supp}}}
\DeclareMathOperator{\diag}{diag}

\DeclarePairedDelimiter{\paren}{(}{)}
\newcommand{\norm}[1]{\ensuremath{\left \lVert #1 \right \rVert}}
\newcommand{\ip}[2]{\langle #1, #2\rangle}
\newcommand{\pos}[1]{\bigl(#1\bigr)_{+}}
\renewcommand{\vec}[1]{\boldsymbol{#1}}

\let\originalexp\exp
\renewcommand\exp{\originalexp\paren}

\newcommand{\N}{\ensuremath{\mathbb{N}}}
\newcommand{\Npos}{\ensuremath{\mathbb{N}_{>0}}}
\newcommand{\Z}{\ensuremath{\mathbb{Z}}}
\newcommand{\Q}{\ensuremath{\mathbb{Q}}}
\renewcommand{\R}{\ensuremath{\mathbb{R}}}

\newcommand{\0}{\ensuremath{\bm{\mathrm{0}}}}
\newcommand{\one}{\ensuremath{\bm{\mathrm{1}}}}
\newcommand{\q}{\bm{q}}
\newcommand{\w}{\ensuremath{\bm{w}}}
\newcommand{\x}{\ensuremath{\bm{x}}}
\newcommand{\y}{\ensuremath{\bm{y}}}
\newcommand{\z}{\ensuremath{\bm{z}}}

\renewcommand{\L}{\ensuremath{\mathcal L}}

\newcommand{\Hs}{\mathcal{H}}
\newcommand{\Ltwo}{\ensuremath{\mathcal{L}^2(n)}}
\newcommand{\ball}{B_{\Hs}}

\newcommand{\id}{\ensuremath{\textup{\textsf{id}}}}
\newcommand{\Sym}{\ensuremath{\textup{Sym}}}
\newcommand{\Bij}{\ensuremath{\textup{Bij}}}
\newcommand{\Perm}{\ensuremath{\textup{Perm}}}
\newcommand{\DS}{\ensuremath{\textup{DS}}}
\newcommand{\DSS}{\ensuremath{\textup{DS}^\ast}}

\renewcommand{\deg}{\ensuremath{\textup{deg}}}

\newcommand{\maxcut}{\ensuremath{\textup{\textsf{maxcut}}}}

\let\d\relax
\NewDocumentCommand{\d}{ g }{%
    \IfNoValueTF{#1}{\ensuremath{\textup{\textsf{d}}}}{\ensuremath{\textup{\textsf{d}}_{#1}}}%
}
\let\nd\relax
\NewDocumentCommand{\nd}{ g }{%
    \IfNoValueTF{#1}{\ensuremath{\delta}}{\ensuremath{\delta_{#1}}}%
}

\newcommand{\OT}{\ensuremath{\textup{\textsf{OT}}}}
\newcommand{\blowup}{\ensuremath{\text{\raisebox{0.1ex}{\scalebox{0.65}{\faBomb}}}}}
\NewDocumentCommand{\dOT}{ g }{%
  \ensuremath{\d^{\OT}\IfValueT{#1}{_{#1}}}%
}
\NewDocumentCommand{\dBU}{ g }{%
  \ensuremath{\d^{\blowup}\IfValueT{#1}{_{#1}}}%
}
\NewDocumentCommand{\ndOT}{ g }{%
  \ensuremath{\nd^{\OT}\IfValueT{#1}{_{#1}}}%
}
\NewDocumentCommand{\ndBU}{ g }{%
  \ensuremath{\nd^{\blowup}\IfValueT{#1}{_{#1}}}%
}

\NewDocumentCommand{\Cost}{ g }{%
  \ensuremath{\mathcal{C}\IfValueT{#1}{_{#1}}}%
}
\newcommand{\LocalCost}[1]{\ensuremath{\mathcal{L}_{#1}}}
\newcommand{\Disc}{D}

\newcommand{\normsubscript}{\ensuremath{\lozenge}} 
\newcommand{\edit}{\mathsf{ed}}
\newcommand{\cut}{\ensuremath{\square}}

\newcommand{\GSim}[1]{%
    \textup{\textsc{GSim}$\big(#1\big)$}%
}

\renewcommand{\O}{\ensuremath{\textup{O}}}
\newcommand{\threepartition}{\textsc{Three-Partition}}

\DeclareMathOperator*{\Expected}{\mathbb{E}}
\RenewDocumentCommand{\E}{o m}{%
  \IfNoValueTF{#1}
  {\ensuremath{\Expected\!\left[#2\right]}}
  {\ensuremath{\Expected{#1}\left[#2\right]}}%
}

\DeclareMathOperator*{\Prob}{Pr}
\RenewDocumentCommand{\Pr}{o m}{%
  \IfNoValueTF{#1}
  {\ensuremath{\Prob\left(#2\right)}}
  {\ensuremath{\Prob_{#1}\left(#2\right)}}%
}

\newcommand{\Unif}{\mathrm{Unif}}
\newcommand{\Ber}{\mathrm{Ber}}

%% file: sections/01_introduction.tex
\section{Introduction}

Metrics on graphs are of fundamental importance for many applications, for example in machine learning \cite{boker2023finegrained,wittig2026which}, and also for an analytical theory such as the theory of graph limits \cite{lovasz_large_networks}. 
Two widely used graph metrics are \emph{graph edit distance} \cite{gao2010survey} and \emph{cut distance} \cite{lovasz_large_networks}. 

A natural way to measure the distance between two graphs $G, H$ with the same number $n$ of vertices is to align their vertex sets via a bijection $\pi$ and to quantify how much the adjacency matrices $A_G,A_H$ differ under this alignment. For a matrix norm $\norm{\cdot}_{\normsubscript}$, we define
\[
    \d{\normsubscript}(G, H) \ceq \min_{\pi} \norm{A_G^\pi - A_H}_{\normsubscript},
\]
minimising over all bijections $\pi:V(G)\to V(H)$. Here, $A_G^\pi$ denotes the adjacency matrix of $G$ with rows and columns permuted according to $\pi$. To obtain an isomorphism-invariant metric, we need to assume that the matrix norm is invariant under simultaneous permutations of rows and columns, which most matrix norms are (including all norms we consider here).
Different matrix norms yield a variety of interesting graph metrics. First of all, the entrywise $1$-norm $\|\cdot\|_{\edit}$ (that is, the $\ell_1$-norm of the vector of length $n^2$ obtained by flattening the matrix) yields edit distance,\footnote{We ignore the factor $2$ the matrix norm induces by counting each undirected edge twice.} and the Frobenius norm yields the square root of graph edit distance. The cut norm $\|\cdot\|_\square$ \cite{frieze1999quick} yields cut distance, which is the metric underlying the theory of graph limits. It has been observed in \cite{gervens2022graphsimilaritybasedmatrix} that the operator $1$-norm $\|\cdot\|_1$ yields \emph{local edit distance}, which measures the maximum number of edge edits necessary at a single vertex to obtain isomorphic graphs. More generally, operator norms are quite interesting for defining graph metrics. For example, based on a result from \cite{nikiforov2009cutnormsspectramatrices}, we observe that the spectral norm (\ie{}, the $2$-operator norm) yields a graph metric that is equivalent to cut distance.

The matrix norms and their graph metrics scale differently in terms of the order $n$ of the matrices or graphs. To allow for comparisons between different metrics, and also to be able to compare graphs of different orders (discussed below), we need to normalise the metrics. The most natural scaling factor might be $1/s(n)$, where $s(n)$ is the maximum distance between graphs of order $n$, often the distance between the empty graph and the complete graph. However, more pragmatically, we just scale by the asymptotic growth rate of $s(n)$, that is, by $1/n^2$ for edit distance and cut distance and by $1/n$ for the metrics based on $p$-operator norms. So for every matrix norm $\|\cdot\|_{\normsubscript}$ we obtain an unnormalised graph metric $\d{\normsubscript}$, which we normalise to the graph metric $\nd{\normsubscript}$.

So far, all these graph metrics have a serious limitation: they only allow for comparing graphs of the same order $n$. This renders them unsuitable for many theoretical considerations, where we may want to approximate large graphs by small ones using tools like the regularity lemma or consider limit objects for sequences of graphs of increasing order, and also for many applications. Lovász~\cite[Chapter~8]{lovasz_large_networks} proposed two different, but actually equivalent, approaches to generalise cut distance to graphs of different orders. Both can be generalised to arbitrary graph metrics based on matrix norms \cite{grohe2025thoughtsgraphsimilarity}, and it turns out they remain equivalent for the metrics considered here. 

\subparagraph*{The Blowup Construction.}
The \emph{$k$-blowup} $G^{\blowup k}$ of a graph $G$ is obtained by replacing every vertex $v$ by a \emph{cloud} of $k$ independent copies of $v$ and each edge by a complete bipartite graph. Now if $G,H$ are graphs of order $m,n$, respectively, the idea is to blow them up to the same order $N$ and then compare them using a normalised metric $\nd{\normsubscript}$. For example, we can take $N$ to be the least common multiple of $m,n$, or we can take $N=mn$. To make the resulting metric independent of any particular choice, we define the \emph{blowup metric} associated with $\nd{\normsubscript}$ by 
\[
    \ndBU{\normsubscript}(G, H) \ceq \lim_{\ell \to \infty} \nd{\normsubscript}\big(G^{\blowup n\ell}, H^{\blowup m\ell}\big).
\]
It is important here that we work with the \emph{normalised} metric, as otherwise the limit would not exist. Even for the normalised metrics it is not clear a priori that the limit exists, but for all matrix norms considered here it does (see \cite{grohe2025thoughtsgraphsimilarity}). Note, however, that for graphs $G,H$ of the same order, it may happen that $\ndBU{\normsubscript}(G, H)<\nd{\normsubscript}(G, H)$, because there are alignments between the blowups $G^{\blowup k}$ and $H^{\blowup k}$ that do not project to alignments between $G$ and $H$. This leads to the idea of considering fractional alignments.

\subparagraph*{Fractional Alignments and Optimal Transport.}
Rather than blowing the graphs up to a common size and then aligning them by a bijection, we can use \emph{fractional alignments}. Assume that the vertex sets of our graphs $G,H$ are $V,W$, and let $m\coloneqq|V|$ and $n\coloneqq|W|$. A fractional alignment between $G$ and $H$ is a matrix $Q^\ast\in\R^{V\times W}$ with nonnegative entries and row sums $\sum_{w\in W}{Q^\ast}(v,w)=\frac{1}{m}$ for all $v\in V$ and column sums $\sum_{v\in V}{Q^\ast}(v,w)=\frac{1}{n}$ for all $w\in W$. Observe that such fractional alignments are precisely couplings between the uniform distributions on $V,W$.
Moreover, note that if $m=n$, then for every permutation matrix $P\in\R^{V\times W}$ describing an alignment between $G$ and $H$ the matrix $\frac{1}{n}P$ is a fractional alignment. In this way, (integral) alignments, that is, bijections between the vertex sets, can also be viewed as special fractional alignments.

Let us now define the optimal transport version of graph edit distance. Recall that the matrix norm $\|\cdot\|_{\edit}$ underlying the definition of $\nd{\edit}$ is the entrywise $1$-norm, that is, the sum of the absolute values of all entries of a matrix, and the normalisation factor is $\frac{1}{n^2}$. Thus for graphs $G$, $H$ of the same order $m=n$ we have
\begin{align*}
    \nd{\edit}(G, H) = \frac{1}{n^2} \min_{P} \sum_{v, v' \in V} \sum_{w, w' \in W} \left| \big( A_G(v, v') - A_H(w, w')\big)  P(v, w) P(v', w')\right|,
\end{align*}
where $P$ ranges over all permutation matrices $P\in\R^{V\times W}$. To understand this expression, note that $P(v, w) P(v', w')=1$ if and only if the permutation underlying $P$ maps $v$ to $w$ and $v'$ to $w'$.
Now for arbitrary $G,H$, we define the optimal transport version of edit distance by 
\begin{equation}\label{eq:edOT}
    \ndOT{\edit}(G, H) = \min_{{Q^\ast}} \sum_{v, v' \in V} \sum_{w, w' \in W} \left| \big( A_G(v, v') - A_H(w, w')\big)  {Q^\ast}(v, w) {Q^\ast}(v', w')\right|,
\end{equation}
where ${Q^\ast}$ ranges over (the compact set of) all fractional alignments between $G$ and $H$. Letting $B(G,H,{Q^\ast})$ be the $mn\times mn$ matrix with row and column indices from $V\times W$ and entries $B(G,H,{Q^\ast})((v,w),(v',w'))=\big( A_G(v, v') - A_H(w, w')\big)  {Q^\ast}(v, w) {Q^\ast}(v', w')$, we can write $\ndOT{\edit}(G, H) = \min_{{Q^\ast}} \|B(G,H,{Q^\ast})\|_{\edit}$. Observe that for graphs of the same order $m=n$, we have $\nd{\edit}(G,H)=\min_{{P^\ast}}\|B(G,H,{P^\ast})\|_{\edit}$, where ${P^\ast}$ ranges over all matrices $\frac{1}{n}P$ for a permutation matrix $P$. This shows that $\ndOT{\edit}(G,H)\le\nd{\edit}(G,H)$. 

A moment of reflection reveals that the definitions of $\ndBU{\edit}$ and $\ndOT{\edit}$ are very similar. Unsurprisingly, it holds that $\ndBU{\edit}=\ndOT{\edit}$ \cite{grohe2025thoughtsgraphsimilarity}.

An optimal-transport version of other graph metrics can be defined similarly. For cut distance, we use exactly the same matrix $B(G,H,{Q^\ast})$ and define
\[\ndOT{\cut}(G,H)=\min_{{Q^\ast}}\|B(G,H,{Q^\ast})\|_{\cut}.\]
Again, we have $\ndBU{\cut}=\ndOT{\cut}$ \cite{lovasz_large_networks}.

It is less obvious how to define an optimal transport version of graph metrics based on operator norms. Recall that, for $1\le p,q\le\infty$, the $p\to q$ operator norm of a matrix $A\in\R^{n\times n}$ is defined by $\|A\|_{p\to q}\coloneqq\max_{\vec x\in\R^n\setminus\{\vec 0\}}\|A\vec x\|_q/\|\vec x\|_p$. In particular, the $1$-operator norm is $\|\cdot\|_1\coloneqq\|\cdot\|_{1\to1}$ and the spectral norm is $\|\cdot\|_2\coloneqq\|\cdot\|_{2\to2}$.
We propose the following definition of an optimal transport version of the graph metric $\nd{p\to q}$ based on $\|\cdot\|_{p\to q}$:
for every fractional alignment $Q^\ast$ between $G$ and $H$, with the convention $0^0 = 0$, we let
\[
    B_{p \to q}(G, H, Q^\ast)\big((v, w), (v', w')\big) \ceq \big( A_G(v, v') - A_H(w, w') \big) Q^\ast(v, w)^{1/q} Q^\ast(v', w')^{1 - 1/p}
\]
(and also write $B_{\cut}(G, H, Q^\ast)\ceq B_{\edit}(G, H, Q^\ast)\ceq B(G, H, Q^\ast) = B_{\infty \to 1}(G, H, Q^\ast)$) and define
\begin{equation}
    \label{equation:p_to_q_OT}
    \ndOT{p \to q}(G, H) \ceq \min_{Q^\ast} \norm{B_{p \to q}(G, H, Q^\ast)}_{p \to q},
\end{equation}
where, as usual, the minimum ranges over all fractional alignments $Q^\ast$.
Note that if the graphs have the same order and we restrict $Q^\ast$ to be $\frac{1}{n} P$ for a permutation matrix $P$, this yields the graph metric $\nd{p\to q}$, where $s(n) = n^{1/q + 1 - 1/p}$.
We show that with this definition, blowup distance and optimal transport distance coincide for all operator norms.

\begin{theorem}
    \label{theorem:blowup_ot_identity}
    For all $1 \leq p, q \leq \infty$, it holds that $\ndBU{p \to q} = \ndOT{p \to q}$.
\end{theorem}

\noindent To prove \cref{theorem:blowup_ot_identity}, we make the correspondence between alignments of blowups and fractional alignments precise.
The main technical work consists in relating optimal test vectors for the operator norms of the aligned blowups to those of $B_{p \to q}(G, H, Q^\ast)$, and in approximating an optimal fractional alignment by one with rational entries.
The proof is given in \cref{appendix:ot_formulation_p_to_q}.

\subparagraph*{Integrality Gaps.}
We have already remarked that, since every bijection is also a fractional alignment, we always have $\ndOT{\normsubscript} \leq \nd{\normsubscript}$ for all pairs of graphs of the same order. It was known that this inequality can be strict for $\normsubscript \in \{\edit, \cut\}$~\cite[Example~3.6]{grohe2025thoughtsgraphsimilarity}, \cite[Section~5.1]{borgs2008convergent}.
We study how large the resulting \emph{integrality gap} of relaxing permutation matrices to doubly stochastic matrices can be, \ie{} whether $\nd{\normsubscript}$ is bounded by a function of $\ndOT{\normsubscript}$ independent of the input graphs and their order.
This question appears in several places in the literature.
    In the theory of graph limits, the fractional distance is the one that extends to graphons, while the integral one is the natural distance between finite graphs; bounding the gap shows that both define the same notion of convergence~\cite{borgs2008convergent,lovasz_large_networks}. It allows transferring statements about limits back to finite graphs, as in Pikhurko's analytic approach to stability~\cite{pikhurko2010analytic}.
    In property testing, Goldreich, Krivelevich, Newman and Rozenberg~\cite{goldreich2008hierarchy} asked how much blowing up two graphs can decrease their edit distance.
    In graph matching, the relaxation of edit distance is known as the \emph{indefinite relaxation}.\martin{Was dann kam war vie zu technisch, aber eine Referenz und vielleicht ein nichttechnischer Satz dazu wäre gut. Oder einfach streichen.}

Some bounds were previously known. For edit distance, Pikhurko~\cite[Lemma~14]{pikhurko2010analytic} proved that for all graphs $G,H$ of the same order it holds that $\nd{\edit}(G,H) \leq 3 \cdot \ndOT{\edit}(G,H)$. Combined with an example of Matsliah~\cite[Appendix~B]{goldreich2008hierarchy}, the optimal factor was known to lie in $[\frac65, 3]$.
For cut distance, Borgs, Chayes, \lovasz{}, Sós and Vesztergombi~\cite[Theorem~2.3]{borgs2008convergent} proved that for all graphs $G,H$ of the same order it holds that $\nd{\cut}(G,H)\leq 32 \cdot \big(\ndOT{\cut}(G,H)\big)^{1/67}$.
\lovasz{} conjectured that the bound can be made linear, more precisely, that $\nd{\cut}(G,H) \leq 2 \cdot \ndOT{\cut}(G,H)$ for all graphs $G,H$ of the same order~\cite[Conjecture~9.28]{lovasz_large_networks}.

We substantially improve both bounds. Furthermore, we obtain the first upper bounds for
local edit and spectral distance.
In the following we abbreviate $\nd_{p\to p}$ and $\ndOT{p \to p}$ to $\nd{p}$ and $\ndOT{p}$, respectively.

\begin{theorem}[Upper Bounds]
    \label{theorem:intro_gaps}
    For all pairs of graphs of the same order and all $\eps > 0$, we have
    \[
        \nd{\edit} \leq 2 \cdot \ndOT{\edit}, \qquad
        \nd{1} \leq 15 \sqrt{\ndOT{1} \ln(4/\ndOT{1})}, \qquad
        \nd{\cut} \leq C_\eps \cdot \big(\ndOT{\cut}\big)^{1-\eps}, \qquad
        \nd{2} \leq \big(1 + 2\sqrt{2}\big)\sqrt{\ndOT{2}},
    \]
    where $C_\eps > 0$ is a constant depending only on $\eps$. Here we just write $\ndOT{\normsubscript}$ instead of $\ndOT{\normsubscript}(G,H)$.
\end{theorem}

\noindent \cref{theorem:intro_gaps} follows from \cref{theorem:edit_factor_two,theorem:local_nonlinear_bound,theorem:cut_almost_linear,theorem:spectral_upper_bound_hard}.

We complement the upper bounds by the following lower bounds. 

\begin{theorem}[Lower Bounds]
    \label{theorem:intro_gaps_lower}
    \leavevmode
    \begin{enumerate}
        \item For every $\eps > 0$, there are graphs $G, H$ such that $\nd{\edit}(G, H) \geq \big(\frac{4}{3} - \eps\big) \cdot \ndOT{\edit}(G, H)$.
        \item For every even $r \geq 2$, there are graphs $G_r, H_r$ such that $\nd{1}(G_r, H_r) = r \cdot \ndOT{1}(G_r, H_r)$.
        \item There exist graphs $G, H$ such that $\nd{\cut}(G, H) = 2 \cdot \ndOT{\cut}(G, H)$.
        \item For every $r \geq 1$, there are graphs $G_r, H_r$ such that $\nd{2}(G_r, H_r) \geq \sqrt{r} \cdot \ndOT{2}(G_r, H_r)$.
    \end{enumerate}
\end{theorem}

\noindent \cref{theorem:intro_gaps_lower} follows from \cref{theorem:edit_gap_four_thirds,theorem:local_unbounded_gap,theorem:cut_factor_two_gap,theorem:spectral_lower_bound}.

\subparagraph*{Computational Complexity.}
In addition to these bounds on the integrality gaps, we prove several hardness results for computing the optimal transport distances.
As nonconvex quadratic programs over the transportation polytope, they are neither covered by the hardness proofs for $\d{\normsubscript}$~\cite{grohe_rattan_woeginger_edit_distance,gervens2022graphsimilaritybasedmatrix} nor by algorithms for linear optimal transport, and their membership in \NP{} is unclear.

\begin{theorem}[\NP-hardness and Hardness of Approximation]
    \label{theorem:intro_hardness}
    \leavevmode
    \begin{enumerate}
        \item Computing $\dOT{\edit}$ is \NP-hard, even when restricted to collections of paths.
        \item For every $1 \leq p \leq \infty$, computing $\dOT{p}$ is \NP-hard. Moreover, unless $\P = \NP$, there is no polynomial-time algorithm approximating $\dOT{p}$ within a multiplicative factor of $2^{\max\{1/p, 1-1/p\}}-\eps$ for any $\eps > 0$. For $p = 1$, this holds even when restricted to graphs of maximum degree two.
        \item Computing $\dOT{\cut}$ is \NP-hard. For every $\eps > 0$, unless $\P = \NP$, there is no polynomial-time algorithm approximating $\dOT{\cut}$ within a multiplicative factor of $\frac{17}{16}-\eps$.
    \end{enumerate}
\end{theorem}
\noindent \cref{theorem:intro_hardness} follows from \cref{theorem:ot_edit_dist_hardness,theorem:ot_local_edit_dist_hardness,theorem:ot_op_p_general_hardness,theorem:ot_cut_hardness}.
For the first two assertions, we encode the partitioning constraints of \threepartition{} in the optimisation over fractional alignments,
while the cut distance hardness is based on approximately fixing the identity bijection between two suitably defined graphs and using that the cut norm cannot be approximated in polynomial time within a multiplicative factor of $\frac{16}{17} + \eps$ \cite{hastad_inapproximability_results} unless $\P = \NP$.

We supply Lean certificates for all our proofs, available at
\url{https://github.com/jensbreitung/graph-similarity-meets-optimal-transport}.

\subsection{Related Work}
Graph metrics are of great practical importance. Comparing graphs by aligning
their vertex sets, known as \emph{graph matching}, has a long history in pattern
recognition and computer vision~\cite{conte2004thirty,foggia2014graph}, and it is
used, for example, to align biological networks and to compare chemical
structures~\cite{emmertstreib2016fifty}; graph edit distance in particular has
become a standard tool~\cite{gao2010survey}. In machine learning, graph
similarity underlies graph kernels~\cite{kriege2020survey} and informs the
analysis of graph neural networks~\cite{boker2023finegrained,wittig2026which}.
In contrast, graph metrics have received limited attention in theory. The
computational complexity of edit distance has been studied under the names
approximate and robust graph
isomorphism~\cite{arvind_approx_graph_isomorphism,odonnell14_edit_distance,grohe_rattan_woeginger_edit_distance},
and Gervens and Grohe~\cite{gervens2022graphsimilaritybasedmatrix} extended this
study to metrics based on other matrix norms. Lyzinski et
al.~\cite{lyzinski2016relax} analysed convex relaxations of graph matching. On the analytic side,
cut distance is the foundation of the theory of graph
limits~\cite{borgs2008convergent,lovasz_large_networks}.

An important inspiration for our ``optimal transport'' metrics is the Gromov--Wasserstein distance on finite metric spaces~\cite{memoli2011gromov}. Of course we can view graphs as metric spaces on their vertex sets with the shortest path distance. We always assume that the probability distribution on both spaces (\ie{} vertex sets) is uniform. In this setting, the Gromov--Wasserstein distance with respect to the $\ell_r$ norm, where $1\le r\le\infty$, can be expressed as 
\[
    \delta_{\mathsf{GW}}(G, H) \ceq \mfrac{1}{2}\min_{Q^\ast} \Big( \sum_{v,v' \in V} \sum_{w, w' \in W} \big| D_G(v, v') - D_H(w, w') \big|^r Q^\ast(v, w) Q^\ast(v', w') \Big)^{1/r},
\]
where $D_G,D_H$ are the distance matrices of our graphs and $Q^\ast$ ranges over all fractional alignments. Compare this with Eq.~\eqref{eq:edOT}, in particular for $r=1$: all we do is replace the adjacency matrices by the distance matrices.
Note that, in contrast to classical optimal transport, where one minimises a cost that is linear in $Q^\ast$~\cite{villani2009optimal, peyre2019computational}, the objective of our distances is nonlinear in $Q^\ast$: whether it is beneficial to transport mass from $v$ to $w$ depends on where the neighbours of $v$ are transported.
The complexity of $\delta_{\mathsf{GW}}$ and other metrics based on the distance matrices is investigated in \cite{breitung_gervens_poerzgen_gromov_wasserstein}.

%% file: sections/02_preliminaries.tex
\section{Preliminaries}
The set of real numbers is denoted $\R$. We write $\log$ and  $\ln$ for the binary and natural logarithm, respectively.
For a positive integer $n$, we let $[n] \ceq \{1, \ldots, n\}$.
We denote vectors by bold symbols and write $x_1, \ldots, x_n$ for the entries of $\x \in \R^n$.

For finite sets $V, W$ of the same size, we write $\Bij(V, W)$ and $\Perm(V, W)$ for the bijections between $V$ and $W$ and their permutation matrices, where $P_\pi(v, w) = 1$ if and only if $\pi(v) = w$. $\DS(V, W)$ denotes the set of doubly stochastic matrices. Hence, if $Q \in \DS(V, W)$, then $\frac{1}{n}Q \in \DSS(V, W)$. The \emph{support} of $Q \in \R^{V\times W}$ is $\supp(Q) \ceq \{ (v, w) \mid Q(v, w) > 0 \}$.

The vertex set of a graph $G$ is denoted $V(G)$, its edge set $E(G)$.
Unless explicitly stated otherwise, graphs are finite, simple and undirected, so $E(G) \subseteq \binom{V(G)}{2}$. 
We denote edges by $vw$ instead of $\{v, w\}$. 
The \emph{order of $G$} is $|V(G)|$.
The \emph{adjacency matrix} $A_G \in \{0, 1\}^{V(G) \times V(G)}$ is defined in the usual way.
For a graph $G$ and a vertex $v \in V(G)$, we denote the \emph{open neighbourhood} of $v$ in $G$ by $N_G(v) \ceq \{w \in V(G) \mid vw \in E(G)\}$.
Unless stated otherwise, graphs $G$ and $H$ have vertex sets $V$ and $W$, respectively.
If the graphs are of the same order, for $Q \in \DS(V, W)$, we define the \emph{(unnormalised) cost} $\Cost{\normsubscript}(Q) \ceq \norm{B_{\normsubscript}(G, H, Q)}_{\normsubscript}$, corresponding to $\d{\normsubscript}$ instead of $\nd{\normsubscript}$.
For $\pi \in \Bij(V, W)$, we let $\Cost_{\normsubscript}(\pi) \ceq \Cost_{\normsubscript}(P_\pi) = \norm{A_G^\pi - A_H}_{\normsubscript}$.
Thus $\d{\normsubscript}(G, H)$ and $\dOT{\normsubscript}(G, H)$ are the minima of $\Cost{\normsubscript}$ over $\Bij(V, W)$ and $\DS(V, W)$, respectively.
The \emph{local cost} of a pair $(v, w) \in V\times W$ is
\[
    \LocalCost{Q}(v, w) \ceq \hspace*{-0.2cm}\sum_{(v', w') \in V \times W} \hspace*{-0.2cm}\big|A_G(v, v') - A_H(w,w')\big| Q(v', w')
    \qquad\text{and}\qquad
    \Cost{1}(Q) = \hspace*{-0.1cm}\max_{(v, w) \in \supp(Q)} \LocalCost{Q}(v, w),
\]
the latter identity is also shown in \cref{equ:local_transport_identity} (compare \cref{section:ot_formulation_p_to_q}).
For $\pi \in \Bij(V, W)$, we let $\LocalCost{\pi} \ceq \LocalCost{P_\pi}$ and note that $\LocalCost{\pi}(v, \pi(v)) = |\pi(N_G(v)) \symdiff N_H(\pi(v))|$ counts the pairs $\{v, v'\}$ \emph{mismatched} by $\pi$, \ie{} with exactly one of $vv' \in E(G)$ and $\pi(v)\pi(v') \in E(H)$, so $\Cost{\edit}(\pi)$ is twice the number of mismatched pairs.

By the Birkhoff--von Neumann theorem~\cite{schrijver_linear_programming}, every $Q \in \DS(V, W)$ can be written as a convex combination of permutation matrices, that is, there are $\pi_1, \ldots, \pi_m \in \Bij(V, W)$ such that
\begin{equation}
    \label{equation:birkhoff_decomposition}
    Q = \sum_{i = 1}^m \lambda_i P_{\pi_i} \qquad \text{with $\lambda_i > 0$ for $i \in [m]$ and $\sum_{i=1}^m \lambda_i = 1$.}
\end{equation}
We call $\{\pi_1, \ldots, \pi_m\}$ the \emph{support} of the decomposition.

%% file: sections/03_integrality_gaps.tex
\section{The Integrality Gaps of Optimal Transport Formulations}
\label{section:integrality_gaps}
In this section, we analyse how far $\nd{\normsubscript}$ and $\ndOT{\normsubscript}$ may deviate from each other.
Clearly, $\ndOT{\normsubscript} \leq \nd{\normsubscript}$, 
so we are interested in the converse direction, that is, in functions $f$ such that $\nd{\normsubscript} \leq f(\ndOT{\normsubscript})$.

For the upper bounds (\cref{theorem:intro_gaps}), we start with an optimal fractional alignment $Q \in \DS(V, W)$ which is subsequently \enquote{rounded} to a bijection whose cost remains bounded in terms of $\Cost{\normsubscript}(Q)$.
Since $\Cost{\normsubscript}$ is not linear in $Q$, a random bijection $\boldsymbol\pi$ from a Birkhoff decomposition~\cref{equation:birkhoff_decomposition}, generally, need not satisfy $\E{\Cost{\normsubscript}(\boldsymbol\pi)} = \Cost{\normsubscript}(Q)$.
We use three increasingly sophisticated rounding schemes to deal with this.
For edit distance, even though the cost is not linear in $Q$, there always exists a bijection within the Birkhoff decomposition with at most twice the cost.
For the local edit distance, we use the dependent rounding scheme of Chekuri, Vondrák and Zenklusen \cite{chekuri2011multibudgeted} to obtain a large matching $M \subseteq V \times W$. We complete this into a bijection by breaking up long \enquote{paths} into short cycles.
For the cut distance, we combine the thin-perfect-matching theorem of Haqi and Oveis Gharan \cite{haqi2026thin} with an iterated rounding scheme that merges two weighted bijections into a single new one in each step. The new bijection is obtained by breaking long permutation cycles into short ones and filling the remaining gaps with an arbitrary bijection.

For the lower bounds (\cref{theorem:intro_gaps_lower}), we construct
graph pairs witnessing gaps between $\nd{\normsubscript}$ and
$\ndOT{\normsubscript}$. Their vertices comprise \emph{core vertices}
and \emph{markers}, which encode structural restrictions on the core vertices.
With markers fixed, suitable fractional alignments of the core
outperform every bijection. To force low-cost bijections to fix
the markers, we attach them to successively longer prefixes of
a large clique, separating their degrees from one another and
from those of core and clique vertices.
The formal statement (\cref{lemma:locking_markers}) and its proof are in \cref{appendix:edit}.

\subsection{Edit Distance}
\label{section:gaps_edit}

Our first result improves on the previously known lower bound of $\frac{6}{5}$ on the integrality gap of Matsliah \cite[Appendix B]{goldreich2008hierarchy} to $\frac{4}{3}$.

\begin{proposition}
    \label{theorem:edit_gap_four_thirds}
    For every odd $N\geq3$, there are graphs $G$ and $H$ with
    $\d{\edit}(G,H)\geq4N-2$ and $\dOT{\edit}(G,H)\leq3N+1$.
    Thus, for every $\eps>0$, there is a sufficiently large $N$ such that
    \[
        \nd{\edit}(G,H)\geq
        \left(\mfrac{4}{3}-\eps\right)\ndOT{\edit}(G,H).
    \]
\end{proposition}

\input{figures/edit_lower_bound_43_cores_all_y0}

\begin{proof}[Proof Idea]
Fix an odd $N\geq3$. Both graphs $G$ and $H$ consist of $N$ core vertices and $N$ markers $\{z_0,\dots,z_{N-1}\}$, which are fixed by \cref{lemma:locking_markers}.
We assign each core vertex a \emph{position} on the cycle $\Z_{2N}$ so that $G$ occupies all even positions and $H$ the odd ones.
In both graphs, for each $r\in\{0,\dots,N-1\}$, the marker $z_r$ is adjacent to all core vertices positioned on the semicircle $\{r, \ldots, r + N - 1\}$ (see \cref{figure:edit_lower_cores}).
Hence, mapping a core vertex at position $i$ in $G$ to a vertex at position $j$ in $H$ incurs a change of adjacency to $\Delta$ markers, where $\Delta$ equals the cyclic distance between $i$ and $j$ in $\Z_{2N}$ (\cref{lemma:edit_lower_anchor_cost}).
Since even and odd positions have distance at least one, the core-marker pairs contribute at least $2N$ to the cost of any alignment (considering both pair orientations).

This lower bound is attained for example by mapping each core vertex one position \enquote{forward}, from $i$ to $i+1$, or \enquote{backward}, from $i$ to $i-1$ in $\Z_{2N}$.
However, we place $(N-1)/2$ edges between core vertices in each graph in a way so that none of them is preserved by either of these mappings, yielding a total cost of $4N-2$ for them.
Furthermore, preserving edges requires any bijection $\pi$ to move core vertices so far that the core-marker pairs keep $\Cost{\edit}(\pi)\geq4N-2$ (\cref{lemma:edit_lower_bijection_cost}).

A fractional alignment $Q$ can map core vertices one position forward with weight $\sfrac{1}{2}$ and one position backward with weight $\sfrac{1}{2}$, which preserves the core edges (except for one) in the two out of four cases in which one endpoint is moved forward while the other is moved backward (see \cref{figure:four_thirds_edge}), yielding $\Cost{\edit}(Q)\leq3N+1$ (\cref{lemma:edit_lower_matching_witness}).
The formal proof is given in \cref{appendix:gaps_edit_lower_four_thirds}.
\end{proof}

For the upper bound, we show that the relaxation of the edit distance can only differ from the bijective variant by a factor of at most two.
Remarkably, this does not require a clever rounding argument: every bijection in the support of a decomposition is already good \emph{on average}.
This improves on the factor 3 of Pikhurko \cite[Lemma 14]{pikhurko2010analytic}.

\begin{theorem}
    \label{theorem:edit_factor_two}
    Let $Q \in \DS(V, W)$ with a decomposition as in \cref{equation:birkhoff_decomposition}. Then
    \[
        \sum_{i = 1}^m \lambda_i \Cost{\edit}(\pi_i) \leq 2 \cdot \Cost{\edit}(Q).
    \]
    In particular, $\d{\edit}(G, H) \leq 2 \cdot \dOT{\edit}(G, H)$ for all graphs $G, H$ of the same order.
\end{theorem}

\noindent The following example shows that any further improvement must either exploit additional properties of $Q$,
such as optimality, or consider bijections outside the support of its decomposition.

\begin{example}
    Let $V=W=\Z_6$, let $G$ be the cycle with edges
    $\{v,v\pm1\}$, and let $H$ consist of two triangles
    with edges $\{v,v\pm2\}$. Set $\pi(v)\ceq v+3$ and
    $Q\ceq\frac12(P_\id+P_\pi)$.
    Since $G$ and $H$ are edge-disjoint and $\pi$ is an
    automorphism of $H$, we have
    $\Cost{\edit}(\id)=\Cost{\edit}(\pi)=24$.
    For each pair of vertices $(v, v')$ in $G$, the fractional
    cost compares their adjacency with that of four equally
    weighted pairs $(v,v')$, $(v+3,v')$, $(v,v'+3)$, and $(v+3,v'+3)$ in $H$.
    The first and last comparisons give the costs of
    $\id$ and $\pi$, respectively. The other two produce
    no disagreement, since shifting exactly one of $v,v'$
    by $3$ turns adjacency in $G$ into adjacency in $H$.
    Thus $\Cost{\edit}(Q)=\frac14(24+0+0+24)=12$,
    whereas both bijections in the prescribed decomposition
    have cost $24$. However, this does not show tightness of the integrality gap since $\d{\edit}(G,H)=8$, attained by mapping two
    consecutive triples of cycle vertices to the two
    triangles.
\end{example}

\subsection{Local Edit Distance}
\label{section:gaps_local}
In contrast to the edit distance, we cannot bound the gap for the local edit distance by any constant.
\begin{proposition}
    \label{theorem:local_unbounded_gap}
    For every even $r \geq 2$, there are graphs $G_r, H_r$ on $\Theta(r^3)$ vertices such that
    \[
        \d{1}(G_r, H_r) = r \qquad\text{and}\qquad \dOT{1}(G_r, H_r) = 1.
    \]
\end{proposition}

\begin{proof}[Proof Idea]
    The graphs $G_r$ and $H_r$ are defined on a common vertex set $V$ and share a set of \emph{core vertices}, consisting of a special vertex $u$ and $r$ \emph{branches} $v_{i, 1}, \ldots, v_{i, \ell - 1}$ for $i \in [r]$ and $\ell \ceq r^2 + 2r$, and a set of \emph{markers} (\cref{figure:local_unbounded_gap}).
    Every edge joins a core vertex to a marker.
    This is done in such a way that every core vertex is described by its \emph{pattern} (a binary vector) indicating its adjacencies to the markers.
    There are three types of markers: $r(r - 1)$ \emph{clock} markers, $r$ \emph{tags} for each branch, and one \emph{bit} for each branch.
    In $G_r$, the vertex $u$ is isolated, while in $H_r$ it is adjacent to all clock markers.
    The construction is finished by fixing the markers. For this, we apply \cref{lemma:locking_markers} with stepsize $r+1$.
    The number of vertices is $\Theta(r^3)$.

    In each \enquote{step} $v_{i,k} \to v_{i, k+1}$ along the branch $i$, only one entry is changed in either graph: 
    Along a fixed branch $i$, 
    first the tags of branch $i$ are switched on one by one, then the clock markers are switched on one by one in $r-1$ blocks of size $r$ where the entry for $b_i$ is toggled on and off between these blocks. Finally, the tags are switched off again one by one.
    Hence, patterns corresponding to consecutive vertices along a branch have Hamming distance one (compare \cref{figure:local_unbounded_gap_pattern}).

    For the upper bound on $\dOT{1}$, for $i \in [r]$, let $\sigma_i$ be the bijection from $G_r$ to $H_r$ that fixes marker vertices and shifts the core vertices of the $i$-th branch along one step, that is $\sigma_i(u) = v_{i,1}$, $\sigma_i(v_{i,k}) = v_{i,k+1}$ for $k \in [\ell-2]$ and $\sigma_i(v_{i,\ell-1}) = u$.
    Let $Q \ceq \frac1r \sum_{i = 1}^r P_{\sigma_i}$.
    Every pair of core vertices in $\supp(Q)$ differs in adjacency to at most one marker, and every marker is changed with total mass at most one: each clock marker changes once on each of the $r$ branches, which receive mass $\frac1r$ each, and the bit of branch $i$ flips $r$ times, but again only with mass $\frac1r$.
    Hence $\Cost{1}(Q) \leq 1$ with equality holding by \cref{claim:local_cost_degree_bound} since $u$ is isolated in $G_r$ but $H_r$ has no isolated vertex.

    \input{figures/fractional_local_example}

    For the lower bound on $\d{1}$, the degrees force every bijection $\pi$ with $\Cost{1}(\pi) < r$ to fix all markers.
    Start at $u$ in $G_r$, map it along $\pi$ to $H_r$, continue from the same vertex in $G_r$, and repeat until reaching $u$ in $H_r$. The patterns along this sequence lead from the $G_r$-pattern to the $H_r$-pattern of $u$, and consecutive patterns differ in fewer than $r$ markers. Thus $\pi(u) \neq u$ and this sequence is forced to traverse a single branch $i$ due to the $r$ tags that branch vertices are connected to.
    Then, the clock markers enforce that an assignment $x \mapsto \pi(x)$ only \enquote{jumps} between consecutive blocks within the branch $i$ because non-adjacent blocks differ in connectivity to at least $r$ clock vertices. However, each assignment $x \mapsto \pi(x)$ that moves one block forward incurs an error of one relative to $b_i$ because $A_{G_r}(x, b_i) \neq A_{H_r}(\pi(x), b_i)$. By construction, this must occur at least $r$ times, so $\Cost{1}(\pi) \geq r$.
    Intuitively, averaging over $r$ branches can spread the cost by a factor of $r$ across the bits of all branches, while a bijection aggregates all errors at the bit of a single branch.
\end{proof}

While \cref{theorem:local_unbounded_gap} rules out any bound of the form $\nd{1} \leq c \cdot \ndOT{1}$, the distances are still related by a uniform, albeit non-linear, bound.
In particular, $\ndOT{1}(G_k, H_k) \to 0$ implies $\nd{1}(G_k, H_k) \to 0$.
\begin{theorem}
    \label{theorem:local_nonlinear_bound}
    For all graphs $G, H$ of the same order, we have
    \[
        \nd{1}(G, H) \leq 15 \sqrt{\ndOT{1}(G, H) \cdot \ln\mfrac{4}{\ndOT{1}(G, H)}},
    \]
    where the right-hand side is defined as $0$ if $\ndOT{1}(G, H) = 0$.
\end{theorem}

\begin{proof}[Proof Idea]
    Let $L \ceq \dOT{1}(G,H)$, fix an optimal $Q$, and choose
    a reference bijection $\sigma \in \Bij(V,W)$ contained in
    $\supp(Q)$ by \cref{equation:birkhoff_decomposition}.
    By \cref{lemma:local_structure}, 
    whenever $(v, w) \in \supp(Q)$, the neighbourhood of $w$ must be similar to $\sigma(v)$,
    that is $|N_H(w) \symdiff N_H(\sigma(v))| \leq 2L$.
    The same lemma shows that $L<1$ implies that $\sigma$ is
    an isomorphism, so we may assume $L\geq1$.
    Viewing $Q$ as a fractional perfect matching in the graph $\supp(Q)$, we apply the
    dependent rounding scheme of Chekuri, Vondrák and Zenklusen \cite{chekuri2011multibudgeted} with
    $\gamma \in (0,\frac12]$.
    The resulting random matching has edge marginals
    $(1-\gamma)Q$, and the local error
    $\Phi_v(M) \ceq \sum_{(v',w') \in M}
    |A_G(v,v')-A_H(\sigma(v),w')|$ of each $v$ (relative to $\sigma$ and $M$)
    has expectation at most
    $\LocalCost{Q}(v,\sigma(v)) \le L$.
    Concentration and a union bound then yield a matching that
    leaves at most $2\gamma n$ vertices unmatched on each side
    and satisfies $\Phi_v(M) \le 2L+\O(\ln(n)/\gamma)$
    simultaneously for all $v$ \cite[Theorem 1.1]{chekuri2011multibudgeted}.
    To complete $M$ to a bijection, we identify $W$ with $V$ via $\sigma$, turning each $(v, w) \in M$ into a directed edge $v \to \sigma^{-1}(w)$.
    The resulting directed graph consists of cycles and
    $\O(\gamma n)$ paths. We keep all cycles, and split the paths
    into blocks of at most $\ell$ vertices (closing each block into a cycle) to obtain a bijection $\pi$.
    Only $\O(\gamma n+n/\ell)$ closing assignments are needed,
    each adding at most one to any local error relative to
    $\sigma$. The neighbourhood bound above, applied along
    each block, also gives
    $|N_H(\sigma(v)) \symdiff N_H(\pi(v))| \leq 2\ell L$.
    Together, these bounds control the actual local costs
    $\LocalCost{\pi}$.
    Balancing the block length $\ell$ and rounding error based on $\gamma$ gives
    $\d{1}(G,H)=\O(L+\sqrt{nL}+\sqrt{n\ln(4n)})$.
    Finally, we use the fact that $L \geq 1$ to bound the remaining logarithmic term by a function of $\ndOT{1}(G,H)$.
\end{proof}

For the graphs $G_r, H_r$ of \cref{theorem:local_unbounded_gap}, we have $\ndOT{1}(G_r, H_r) = \Theta(r^{-3})$ and $\nd{1}(G_r, H_r) = \Theta(r^{-2}) = \Theta\big(\ndOT{1}(G_r, H_r)^{2/3}\big)$.
Hence there is no uniform bound of the form $\nd{1} \leq C \cdot (\ndOT{1})^\alpha$ with $\alpha > \frac{2}{3}$, while \cref{theorem:local_nonlinear_bound} yields such a bound for every $\alpha < \frac{1}{2}$.

\subsection{Cut Distance}
\label{section:gaps_cut}
We first show that if \lovasz{}'s conjecture $\nd{\cut} \leq 2 \cdot \ndOT{\cut}$ \cite[Conjecture~9.28]{lovasz_large_networks} holds, then the factor $2$ is tight.
\begin{proposition}
    \label{theorem:cut_factor_two_gap}
    There are graphs $G, H$ on $145$ vertices with
    $\d{\cut}(G, H) = 8$ and $\dOT{\cut}(G, H) \leq 4$.
\end{proposition}
The graphs are shown in \cref{figure:cut_factor_two_graphs}.
Markers ensure that bijections may only swap vertices within the same pair.
As adjacency between pairs of the pentagons is exactly flipped between $G$ and $H$ and all pairs are connected to each other, any bijection yields a cut norm of $8$. 
A fractional alignment can reduce the value to $4$ by spreading mass uniformly across each pair. This essentially decomposes the twisted-pair pentagons into two individual copies of weight $\frac{1}{2}$ with cut norms scaled by $\frac{1}{4}$ each.
Note that one of our Lean certificates\footnote{\url{https://github.com/jensbreitung/graph-similarity-meets-optimal-transport/blob/main/GsimOT/Proposition10_CutLowerBound/OptimalAlignment.lean}}
verifies that the $\dOT{\cut}$-bound is tight, so $\dOT{\cut}(G,H) = 4$.
\input{figures/cut_factor_two_graphs}

While we were unable to confirm or refute \lovasz{}'s conjecture, we improve the known upper bound significantly and show that arbitrarily good uniform non-linear bounds are possible. 
\begin{theorem}
    \label{theorem:cut_almost_linear}
    There is a universal constant $C > 0$ such that for all graphs $G, H$ of the same order, $\nd{\cut} \ceq \nd{\cut}(G,H)$, and $\ndOT{\cut} \ceq \ndOT{\cut}(G, H)$ we have
    \[
        \ndOT{\cut} \leq \nd{\cut} \leq C \ndOT{\cut} \ln\frac{e}{\ndOT{\cut}} \ln\left(e \ln\frac{e}{\ndOT{\cut}}\right),
    \]
    where $e\approx2.718$ is the Euler number and the right-hand side is defined as $0$ if $\ndOT{\cut} = 0$.
    Consequently, for every $\eps > 0$ we have $\nd{\cut} = \O\big( (\ndOT{\cut})^{1-\eps}\big)$.
\end{theorem}

\paragraph*{Proof Overview.}
As usual, the lower bound is trivial, and we only need to prove the upper bound.
Let $Q\in\DS(V,W)$ be an optimal fractional alignment of normalised cost $\cst{Q}\ceq n^{-2}\Cost{\cut}(Q)=\ndOT{\cut}$. It is our goal to ``round'' $Q$ to a permutation matrix $P$ whose cost $\cst{P}$ is sufficiently close to $\cst{Q}$. We follow two completely different strategies. The first yields a multiplicative upper bound on the cost of an optimal permutation matrix in terms of $\cst{Q}$, and the second an additive upper bound. 
The upper bound of the theorem is then obtained by applying the two estimates in complementary regimes, using the multiplicative bound for small $\cst{Q}$ and the additive bound for larger $\cst{Q}$.

\subparagraph*{The Multiplicative Bound.} 
We prove that we can find a permutation matrix $P$ with $\cst{P}=\O(\cst{Q}\log n\log\log n)$ (\cref{theorem:cutmid_v2_thin}). To obtain this upper bound, we use a recent result due to Haqi and Oveis Gharan~\cite{haqi2026thin} stating that for each fractional perfect matching of a bipartite graph there is a ``thin matching''. In our context, for the fractional matching $Q\in\DS(V,W)$ this yields a perfect matching, described by a permutation matrix $P\in\Perm(V,W)$, such that for every $U\subseteq V\cup W$, the number of matching edges from $U$ to its complement is at most $\alpha$ times the total weight the fractional matching $Q$ assigns to the edges from $U$ to its complement, for some $\alpha\in \O(\log n\log\log n)$. Such a matching $P$ is called \emph{$\alpha$-thin} with respect to $Q$.

We then prove that if $P$ is $\alpha$-thin with respect to $Q$, then $\cst{P}\le (1+4\alpha)\cst{Q}$. The proof turns thinness into a transport inequality $\sum_v|x_v-y_{\pi(v)}|\le \alpha\sum_{v,w}Q_{vw}|x_v-y_w|$ via a level-set decomposition (\cref{lemma:cutmid_v2_transport}), and combines it with the observation that the fractional alignment $Q$ only matches vertices with similar neighbourhoods (\cref{lemma:cutmid_v2_local}).

\subparagraph*{The Additive Bound.} 
We prove that we can find a permutation matrix $P$ with $\cst{P}\le\cst{Q}+\O(n^{-1/6}\sqrt{\log n})$ (\cref{theorem:cut_additive_rounding}). Technically, this part of the proof is significantly harder. The high-level argument goes as follows. We describe $\cst{Q}$ in terms of the $n^2\times n^2$ \emph{discrepancy matrix} $D$ with entries $D_{vw,v'w'}\coloneqq A_{vv'}-B_{ww'}$, where $A$ and $B$ are the adjacency matrices of the graphs $G$ and $H$. For every $R\in\DS(V,W)$ we can rewrite the normalised cost $n^{-2}\Cost{\cut}(R)$ as 
\begin{equation}\label{eq:epsR}\cst{R}=\max_{S,T\subseteq V\times W}\frac {1}{n^2}\bigg|\sum_{(v,w)\in S,(v',w')\in T}R_{vw}R_{v'w'}D_{vw,v'w'}\bigg|.
\end{equation}
The first step uses standard arguments to construct a spectral approximation of $D$. We obtain a family of $n^2$ vectors $\vec z_{vw}\in\R^{O(n)}$ and a matrix $T$ of bounded spectral norm such that if we replace $D$ in \cref{eq:epsR} by the matrix $D'$ with entries $\ip{\vec z_{vw}}{T\vec z_{v'w'}}$ then the resulting value $\cst[\prime]{R}$ approximates $\cst{R}$ sufficiently closely.

Next, for each $R\in\DS(V,W)$ we consider the zonotope $Z_R$ generated by the line segments $[0,\frac{1}{n}R_{vw}\vec z_{vw}]$. The Hausdorff distance between these zonotopes yields a pseudometric $\rho$ on $\DS(V,W)$. We show that matrices $R$ that are $\rho$-close have approximately the same $\cst[\prime]{R}$.
This motivates the main result (\cref{lemma:cutmid_merge}): if $R\in\DS(V,W)$ is a convex combination $R=(1-\theta)P_1+\theta P_2$ of permutation matrices $P_1,P_2$, then there is a permutation matrix $P$ with $\rho(P,R)=\O(n^{-1/3})$.
Viewing $P_1,P_2$ as perfect matchings in $K_{V,W}$, the complete bipartite graph between $V$ and $W$, their union decomposes into alternating cycles, which we cut into blocks of roughly $ n^{1/3}$ consecutive vertices.
Independently for each block, all its vertices follow $P_2$ with probability $\theta$ and $P_1$ otherwise. The resulting random map has expected support function $h_R$, deviates from it by $\O(n^{-1/3})$ in expectation uniformly over all directions (by symmetrisation and Rademacher contraction), and fails to be a bijection only at block boundaries, so it can be repaired to a permutation matrix $P$ at an additional cost of $\O(n^{-1/3})$.

Since every doubly stochastic matrix is a convex combination of permutation matrices, we can use this to prove that every $R\in\DS(V,W)$ is $\rho$-close to a permutation matrix.

\subsection{Spectral Distance}
The spectral distance $\nd{2}$ is closely related to the cut distance. The following proposition shows that the two bound each other uniformly, so in particular $\nd{2}(G_k, H_k) \to 0$ if and only if $\nd{\cut}(G_k, H_k) \to 0$.
We note that a similar relation holds for all $\nd{p}$ with $1 < p < \infty$, but not for $\nd{1}$ or $\nd{\edit}$.

\begin{proposition}
    \label{proposition:cut_spectral_equivalence}
    For all graphs $G, H$ of the same order, it holds that
    \[
        \nd{\cut}(G, H) \leq \nd{2}(G, H) \leq 2\sqrt{\nd{\cut}(G, H)}.
    \]
\end{proposition}
The second inequality uses a result of Nikiforov \cite{nikiforov2009cutnormsspectramatrices}.
Despite their close relation, our following results illustrate a fundamentally different behaviour between their optimal transport formulations.

\begin{proposition}
    \label{theorem:spectral_lower_bound}
    For every $r \geq 1$, there are graphs $G_r, H_r$ on $n =\Theta(r^2)$ vertices such that
    \[
        \d{2}(G_r, H_r) = \sqrt{r} \qquad\text{and}\qquad \mfrac{1}{2\sqrt2} \leq \dOT{2}(G_r, H_r) \leq 1.
    \]
\end{proposition}
Since $\nd{2}(G_r, H_r) = \Theta(n^{-3/4})$ and $\ndOT{2}(G_r, H_r) = \Theta(n^{-1})$, it follows that there is no uniform bound of the form $\nd{2} \leq C_\alpha(\ndOT{2})^{\alpha}$ for any $\alpha > \frac{3}{4}$, whereas for the cut distance, bounds arbitrarily close to linear are possible.

For the upper bound, we closely follow the proof of the multiplicative estimate for the cut distance. However, here a transport inequality similar to \cref{lemma:cutmid_v2_transport} does not hold.
We replace this step with a probabilistic argument combined with spectral truncation as for the additive cut estimate.
\begin{theorem}
    \label{theorem:spectral_upper_bound_hard}
    For all graphs $G, H$  of the same order, it holds that
    \[
        \nd{2}(G, H) \leq \ndOT{2}(G, H) + 2\sqrt{2\, \ndOT{2}(G, H)}.
    \]
\end{theorem}

\noindent As $\ndOT{2} \leq 1$, this implies $\nd{2} \leq \big(1 + 2\sqrt2\big) \sqrt{\ndOT{2}}$ of \cref{theorem:intro_gaps}.
Together with the previous observation, the best exponent $\alpha$ in a uniform bound of the form $\nd{2} \leq C_\alpha \big(\ndOT{2}\big)^\alpha$ therefore lies in $\big[\frac12, \frac34\big]$.

%% file: figures/edit_lower_bound_43_cores_all_y0.tex
\input{figures/edit_lower_bound_43_style}
\begingroup

\def\FigScale{0.75}           
\def\CycleScale{1.00}         
\def\VertexScale{1.00}        
\def\MarkerLabelScale{1.00}   
\def\PositionLabelScale{0.85} 

\def\MarkerX{4.20}           
\def\MarkerSep{0.55}         
\def\PositionLabelGap{0.48}  

\pgfmathsetmacro{\MarkerLabelComp}{\MarkerLabelScale/\FigScale}
\pgfmathsetmacro{\PositionLabelComp}{\PositionLabelScale/\FigScale}
\tikzset{
    lb label/.append style={scale=\MarkerLabelComp},
    lb pos/.append style={scale=\PositionLabelComp},
    lb four thirds vertex/.append style={scale=\VertexScale}
}
\let\lbAngle\lbFourThirdsAngle

\renewcommand{\lbFourThirdsCore}[1]{%
    \draw[lb ring] (0,0) circle (\lbR);
    \foreach \lbu in {0,...,6} {%
        \ifx G#1
            \pgfmathtruncatemacro{\lbx}{2*\lbu}%
            \pgfmathtruncatemacro{\lbo}{2*\lbu+1}%
        \else
            \pgfmathtruncatemacro{\lbx}{2*\lbu+1}%
            \pgfmathtruncatemacro{\lbo}{2*\lbu}%
        \fi
        \node[lb slot] at (p\lbo) {};
        \node[lb vertex,lb #1,lb four thirds vertex] at (p\lbx)
            {\lbFourThirdsVertexLabel{#1}{\lbx}};
    }%
    \foreach \lbx in {0,...,13} {%
        \node[lb pos]
            at ({\lbAngle{\lbx}}:{\lbR+\PositionLabelGap}) {\lbx};
    }%
}

\renewcommand{\lbFourThirdsMarkerColumnPanel}[2]{%
    \pgfmathsetmacro{\FigCycleRadius}{1.75*\CycleScale}%
    \lbSetup{\FigCycleRadius}%
    \begin{scope}[lb band/.append style={opacity=0.15}]
        \lbSemicircle{#2}%
    \end{scope}
    \foreach \lbr in {0,...,6} {%
        \pgfmathsetmacro{\lbmy}{(3-\lbr)*\MarkerSep}%
        \ifnum\lbr=#2
            \lbFourThirdsForFan{#1}{\lbr}{%
                \lbFourThirdsMarkerEdge{\MarkerX}{\lbmy}{\lbx}{lb#1}{\lbfan}}%
        \else
            \lbFourThirdsForFan{#1}{\lbr}{%
                \lbMarkerStub{\MarkerX}{\lbmy}{\lbfan}{lb#1}}%
        \fi
    }%
    \lbFourThirdsCoreEdges{#1}%
    \pgfmathsetmacro{\FigLeft}{-\lbR-0.55}%
    \pgfmathsetmacro{\FigHalfHeight}{max(\lbR+0.55,3*\MarkerSep+0.25)}%
    \pgfresetboundingbox
    \path (\FigLeft,-\FigHalfHeight)
        rectangle (\MarkerX,\FigHalfHeight);
    \lbFourThirdsCore{#1}%
    \foreach \lbr in {0,...,6} {%
        \node[lb marker,minimum size=3.4mm] (z\lbr)
            at (\MarkerX,{(3-\lbr)*\MarkerSep}) {};
        \node[lb label,anchor=west] at (z\lbr.east) {$z_{\lbr}$};
    }%
}

\newcommand{\ScaledCorePanel}[1]{%
    \scalebox{\FigScale}{%
        \begin{tikzpicture}
            \lbFourThirdsMarkerColumnPanel{#1}{0}
        \end{tikzpicture}%
    }%
}

\begin{figure}[tbp]
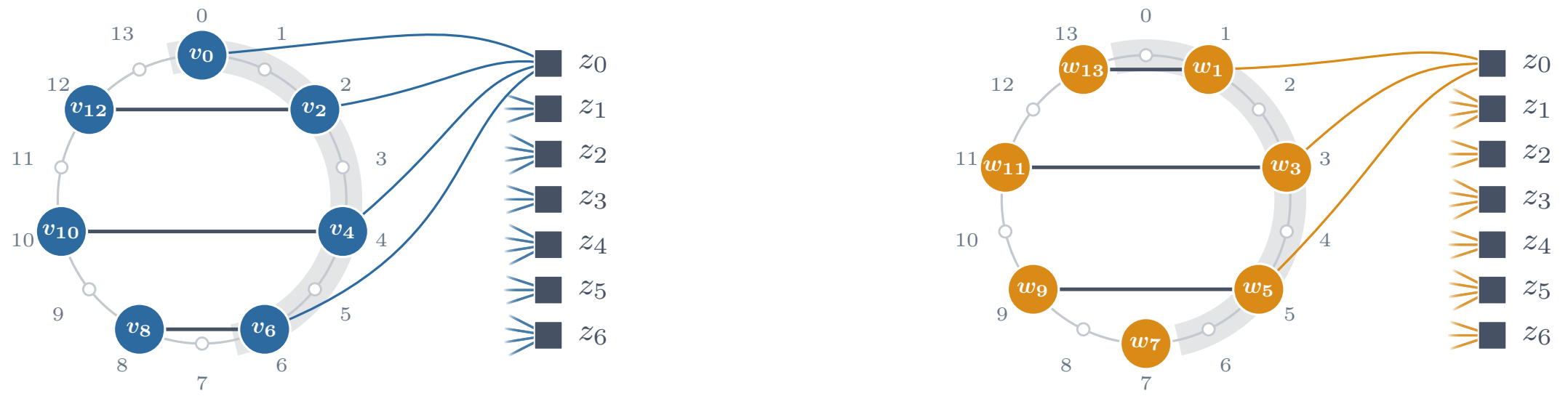

    \centering
    \begin{subfigure}[b]{0.48\textwidth}
        \centering
        \ScaledCorePanel{G}
        \caption{The graph $G$ with core vertices $\{v_0,v_2,\dots,v_{12}\}$.}
        \label{figure:edit_lower_cores_G}
    \end{subfigure}
    \hfill
    \begin{subfigure}[b]{0.48\textwidth}
        \centering
        \ScaledCorePanel{H}
        \caption{The graph $H$ with core vertices $\{w_1,w_3,\dots,w_{13}\}$.}
        \label{figure:edit_lower_cores_H}
    \end{subfigure}
    \caption{The graphs $G$ and $H$ of \cref{theorem:edit_gap_four_thirds} for
    $N=7$, without the clique added by \cref{lemma:locking_markers}.
    According to their positions in $\Z_{14}$, the core vertices are drawn on top of the cycle graph $C_{14}$, which is drawn in light grey (although it is neither part of $G$ nor $H$).
    In both graphs, distinct core vertices are adjacent if their positions sum to $0$ modulo $2N$.
    The core-marker edges are drawn in full only for $z_0$; for the other markers they are indicated by stubs and can be seen in \cref{figure:four_thirds_table}.
    The semicircle corresponding to $z_0$ is drawn shaded.}
    \label{figure:edit_lower_cores}
\end{figure}
\endgroup

%% file: figures/edit_lower_bound_43_style.tex
\input{figures/edit_lower_bound_style}
\ifdefined\lbFourThirdsStyleLoaded \fi
\def\lbFourThirdsStyleLoaded{}

\newcommand{\lbFourThirdsAngle}[1]{90 - (#1) * 360 / 14}

\newcommand{\lbFourThirdsVertexLabel}[2]{{\boldmath$\ifx G#1 v\else w\fi_{#2}$}}
\tikzset{lb four thirds vertex/.style={minimum size=6.2mm}}

\newcommand{\lbFourThirdsCore}[1]{%
    \draw[lb ring] (0,0) circle (\lbR);
    \foreach \lbu in {0,...,6} {%
        \ifx G#1
            \pgfmathtruncatemacro{\lbx}{2*\lbu}%
            \pgfmathtruncatemacro{\lbo}{2*\lbu+1}%
        \else
            \pgfmathtruncatemacro{\lbx}{2*\lbu+1}%
            \pgfmathtruncatemacro{\lbo}{2*\lbu}%
        \fi
        \node[lb slot] at (p\lbo) {};
        \node[lb vertex,lb #1,lb four thirds vertex] at (p\lbx) {\lbFourThirdsVertexLabel{#1}{\lbx}};
    }%
    \foreach \lbx in {0,...,13} {%
        \node[lb pos] at ({\lbAngle{\lbx}}:\lbR+0.48) {\lbx};
    }%
}

\newcommand{\lbFourThirdsCoreEdges}[1]{%
    \ifx G#1
        \foreach \lba/\lbb in {2/12,4/10,6/8} {%
            \draw[lb edge,draw=lbInk] (p\lba)--(p\lbb);%
        }%
    \else
        \foreach \lba/\lbb in {1/13,3/11,5/9} {%
            \draw[lb edge,draw=lbInk] (p\lba)--(p\lbb);%
        }%
    \fi
}

\newcommand{\lbFourThirdsForNeighbours}[3]{%
    \pgfmathtruncatemacro{\lbend}{#2+6}%
    \foreach \lbx in {#2,...,\lbend} {%
        \ifx G#1 \ifodd\lbx\relax\else #3\fi
        \else \ifodd\lbx\relax #3\fi \fi
    }%
}

\newcommand{\lbFourThirdsForFan}[3]{%
    \setcounter{lbfansize}{0}%
    \lbFourThirdsForNeighbours{#1}{#2}{\stepcounter{lbfansize}}%
    \setcounter{lbfanindex}{0}%
    \lbFourThirdsForNeighbours{#1}{#2}{%
        \pgfmathsetmacro{\lbfan}{180-18*((\arabic{lbfansize}-1)/2-\arabic{lbfanindex})}%
        #3%
        \stepcounter{lbfanindex}%
    }%
}

\newcommand{\lbFourThirdsMarkerEdge}[5]{%
    \pgfmathsetmacro{\lbth}{\lbAngle{#3}}%
    \pgfmathsetmacro{\lbtx}{\lbR*cos(\lbth)}%
    \pgfmathsetmacro{\lbty}{\lbR*sin(\lbth)}%
    \pgfmathsetmacro{\lbdelta}{mod(atan2(#2-\lbty,#1-\lbtx)-\lbth+540,360)-180}%
    \pgfmathsetmacro{\lbin}{\lbth+(\lbdelta<0 ? -1 : 1)*min(max(abs(\lbdelta),30),85)}%
    \pgfmathsetmacro{\lbdout}{0.3*veclen(\lbtx-#1,\lbty-#2)}%
    \pgfmathsetmacro{\lbdin}{0.6*veclen(\lbtx-#1,\lbty-#2)}%
    \draw[lb marker edge,draw=#4] (#1,#2)
        to[out=#5,in=\lbin,out distance=\lbdout cm,in distance=\lbdin cm] (p#3);
}

\newcommand{\lbFourThirdsMarkerColumnPanel}[2]{%
    \lbSetup{1.75}%
    \begin{scope}[lb band/.append style={opacity=0.15}]
        \lbSemicircle{#2}%
    \end{scope}
    \def\lbMarkerSep{0.55}%
    \foreach \lbr in {0,...,6} {%
        \pgfmathsetmacro{\lbmy}{(3-\lbr)*\lbMarkerSep}%
        \ifnum\lbr=#2
            \lbFourThirdsForFan{#1}{\lbr}{\lbFourThirdsMarkerEdge{4.2}{\lbmy}{\lbx}{lb#1}{\lbfan}}%
        \else
            \lbFourThirdsForFan{#1}{\lbr}{\lbMarkerStub{4.2}{\lbmy}{\lbfan}{lb#1}}%
        \fi
    }%
    \lbFourThirdsCoreEdges{#1}%
    \lbFourThirdsCore{#1}%
    \pgfresetboundingbox
    \path (-2.3,-2.30) rectangle (4.2,2.30);
    \foreach \lbr in {0,...,6} {%
        \node[lb marker,minimum size=3.4mm] (z\lbr) at (4.2,{(3-\lbr)*\lbMarkerSep}) {};
        \node[lb label,anchor=west] at (z\lbr.east) {$z_{\lbr}$};
    }%
}

%% file: figures/edit_lower_bound_style.tex

\ifdefined\lbStyleLoaded \fi
\def\lbStyleLoaded{}

\input{figures/palette}
\colorlet{lbG}{figBlue}               
\colorlet{lbH}{figOrange}             
\colorlet{lbGLight}{figBlueLight}
\colorlet{lbHLight}{figOrangeLight}
\colorlet{lbInk}{figInk}              
\colorlet{lbMuted}{figMuted}          
\colorlet{lbRing}{figLine}            
\colorlet{lbFaint}{figEmpty}          
\colorlet{lbCopyA}{figInk}            
\colorlet{lbCopyB}{figGreen}          
\colorlet{lbCopyC}{figPurple}         
\colorlet{lbMarker}{figInk}           
\colorlet{lbKept}{figGreen}           
\colorlet{lbLost}{figRed}             

\tikzset{
    lb ring/.style={draw=lbRing, line width=0.8pt},
    lb slot/.style={circle, draw=lbRing, fill=white, line width=0.7pt, inner sep=0pt, minimum size=4.2pt},
    lb vertex/.style={circle, draw=white, line width=0.8pt, inner sep=0pt, minimum size=4.7mm,
        font=\sffamily\bfseries\scriptsize, text=white},
    lb small vertex/.style={lb vertex, minimum size=3.5mm, line width=0.6pt, font=\sffamily\bfseries\tiny},
    lb G/.style={fill=lbG},
    lb H/.style={fill=lbH},
    lb edge/.style={line width=1.25pt, line cap=round},
    lb axis/.style={line width=0.6pt, dash pattern=on 2.2pt off 1.8pt},
    lb pos/.style={font=\tiny, text=lbMuted, inner sep=0pt},
    lb label/.style={font=\scriptsize, text=lbInk},
    lb marker/.style={rectangle, fill=lbMarker, draw=white, line width=0.5pt, inner sep=0pt, minimum size=2.6mm},
    lb marker edge/.style={line width=0.8pt},
    lb band/.style={line width=11pt, draw=lbInk, opacity=0.08},
}

\newcommand{\lbAngle}[1]{90 - (#1 - 1) * 360 / 14}

\newcommand{\lbSetup}[1]{%
    \def\lbR{#1}%
    \foreach \lbx in {0, ..., 13} {%
        \coordinate (p\lbx) at ({\lbAngle{\lbx}}:\lbR);
    }%
}

\newcommand{\lbCore}[1]{%
    \draw[lb ring] (0, 0) circle (\lbR);
    \foreach \lbu in {0, ..., 6} {%
        \ifx G#1
            \pgfmathtruncatemacro{\lbx}{2 * \lbu + 1}\pgfmathtruncatemacro{\lbo}{2 * \lbu}%
        \else
            \pgfmathtruncatemacro{\lbx}{2 * \lbu}\pgfmathtruncatemacro{\lbo}{2 * \lbu + 1}%
        \fi
        \node[lb slot] at (p\lbo) {};
        \node[lb vertex, lb #1] at (p\lbx) {\lbu};
    }%
    \foreach \lbx in {0, ..., 13} {%
        \node[lb pos] at ({\lbAngle{\lbx}}:\lbR + 0.48) {\lbx};
    }%
}

\newcommand{\lbEdge}[4]{%
    \ifx G#1
        \pgfmathtruncatemacro{\lbx}{mod(2 * #2 + 1, 14)}\pgfmathtruncatemacro{\lby}{mod(2 * #3 + 1, 14)}%
    \else
        \pgfmathtruncatemacro{\lbx}{mod(2 * #2, 14)}\pgfmathtruncatemacro{\lby}{mod(2 * #3, 14)}%
    \fi
    \draw[lb edge, draw=#4] (p\lbx) -- (p\lby);
}

\newcommand{\lbAxis}[2]{%
    \draw[lb axis, draw=#2] ({\lbAngle{#1}}:\lbR + 0.2) -- ({\lbAngle{#1} + 180}:\lbR + 0.2);
}

\newcommand{\lbSemicircle}[1]{%
    \draw[lb band] ({\lbAngle{#1 - 0.5}}:\lbR) arc[start angle={\lbAngle{#1 - 0.5}}, end angle={\lbAngle{#1 + 6.5}}, radius=\lbR];
}

\newcommand{\lbCoreEdges}[1]{%
    \ifx G#1
        \foreach \lba/\lbb in {1/6, 2/5, 3/4} {\lbEdge{G}{\lba}{\lbb}{lbInk}}%
    \else
        \foreach \lba/\lbb in {0/1, 2/6, 3/5} {\lbEdge{H}{\lba}{\lbb}{lbInk}}%
    \fi
}

\newcommand{\lbForNeighbours}[3]{%
    \pgfmathtruncatemacro{\lbend}{#2 + 6}%
    \foreach \lbx in {#2, ..., \lbend} {%
        \ifx G#1 \ifodd\lbx\relax #3\fi \else \ifodd\lbx\relax\else #3\fi \fi
    }%
}

\newcommand{\lbMarkerAngles}[3]{%
    \pgfmathsetmacro{\lbth}{\lbAngle{#3}}%
    \pgfmathsetmacro{\lbtx}{\lbR * cos(\lbth)}%
    \pgfmathsetmacro{\lbty}{\lbR * sin(\lbth)}%
    \pgfmathsetmacro{\lbdelta}{mod(atan2(#2 - \lbty, #1 - \lbtx) - \lbth + 540, 360) - 180}%
    \pgfmathsetmacro{\lbin}{\lbth + (\lbdelta < 0 ? -1 : 1) * min(max(abs(\lbdelta), 30), 40)}%
}

\newcounter{lbfansize}
\newcounter{lbfanindex}

\newcommand{\lbForFan}[3]{%
    \setcounter{lbfansize}{0}%
    \lbForNeighbours{#1}{#2}{\stepcounter{lbfansize}}%
    \setcounter{lbfanindex}{0}%
    \lbForNeighbours{#1}{#2}{%
        \pgfmathsetmacro{\lbfan}{180 - 18 * ((\arabic{lbfansize} - 1) / 2 - \arabic{lbfanindex})}%
        #3%
        \stepcounter{lbfanindex}%
    }%
}

\newcommand{\lbMarkerFanEdge}[6][]{%
    \lbMarkerAngles{#2}{#3}{#4}%
    \pgfmathsetmacro{\lbdist}{0.5 * veclen(\lbtx - #2, \lbty - #3)}%
    \draw[lb marker edge, draw=#5] (#2, #3)
        to[out=#6, in=\lbin, out distance=\lbdist cm, in distance=\lbdist cm, #1] (p#4);
}

\newcommand{\lbMarkerStub}[4]{%
    \foreach \lbi in {0, ..., 11} {%
        \draw[lb marker edge, draw=#4, draw opacity=0.16, line cap=butt]
            (#1, #2) ++(#3:0.14) -- ++(#3:{0.25 + 0.0135 * \lbi});
    }%
}

\newcommand{\lbMarkerColumnPanel}[2]{%
    \lbSetup{1.75}%
    \lbSemicircle{#2}%
    \def\lbMarkerSep{0.55}
    \foreach \lbr in {0, ..., 6} {%
        \pgfmathsetmacro{\lbmy}{(3 - \lbr) * \lbMarkerSep}%
        \ifnum\lbr=#2
            \lbForFan{#1}{\lbr}{%
                \ifnum\lbx=1
                    \lbMarkerFanEdge[in distance=1.1cm]{4.2}{\lbmy}{\lbx}{lb#1}{\lbfan}%
                \else
                    \lbMarkerFanEdge{4.2}{\lbmy}{\lbx}{lb#1}{\lbfan}%
                \fi
            }%
        \else
            \lbForFan{#1}{\lbr}{\lbMarkerStub{4.2}{\lbmy}{\lbfan}{lb#1}}%
        \fi
    }%
    \lbCoreEdges{#1}%
    \lbCore{#1}%
    \pgfresetboundingbox
    \path (-2.3, -2.30) rectangle (4.2, 2.30);
    \foreach \lbr in {0, ..., 6} {%
        \node[lb marker, minimum size=3.4mm] (y\lbr) at (4.2, {(3 - \lbr) * \lbMarkerSep}) {};
        \node[lb label, anchor=west] at (y\lbr.east) {$y_{\lbr}$};
    }%
}

%% file: figures/palette.tex

\ifdefined\figPaletteLoaded \fi
\def\figPaletteLoaded{}%

\definecolor{figBlue}{HTML}{2D6A9F}%
\definecolor{figOrange}{HTML}{DA8B17}%
\definecolor{figGreen}{HTML}{279C72}%
\definecolor{figPurple}{HTML}{7D5CC6}%
\definecolor{figRed}{HTML}{DC4445}%
\colorlet{figBlueLight}{figBlue!20}%
\colorlet{figOrangeLight}{figOrange!20}%

\definecolor{figInk}{HTML}{465264}
\definecolor{figMuted}{HTML}{687787}
\colorlet{figLine}{figMuted!40}
\colorlet{figPanel}{black!5}
\colorlet{figEmpty}{black!4}

%% file: figures/fractional_local_example.tex
\begin{figure}[t]
    \centering
    \resizebox{0.9\textwidth}{!}{\input{tikz/fractional_local_example}}
    \caption{The graph $H_4$ of \cref{theorem:local_unbounded_gap}, with the clique fixing markers omitted.
    Of the four branches, only the first and last are shown.
    Each branch is arranged horizontally and split into blocks based on connectivity to the tags, clocks and bits.
    Edges from clocks to core vertices in the fourth branch are faded to improve readability.
    The graph $G_4$ is obtained by removing all edges incident to $u$.
    }
    \label{figure:local_unbounded_gap}
\end{figure}
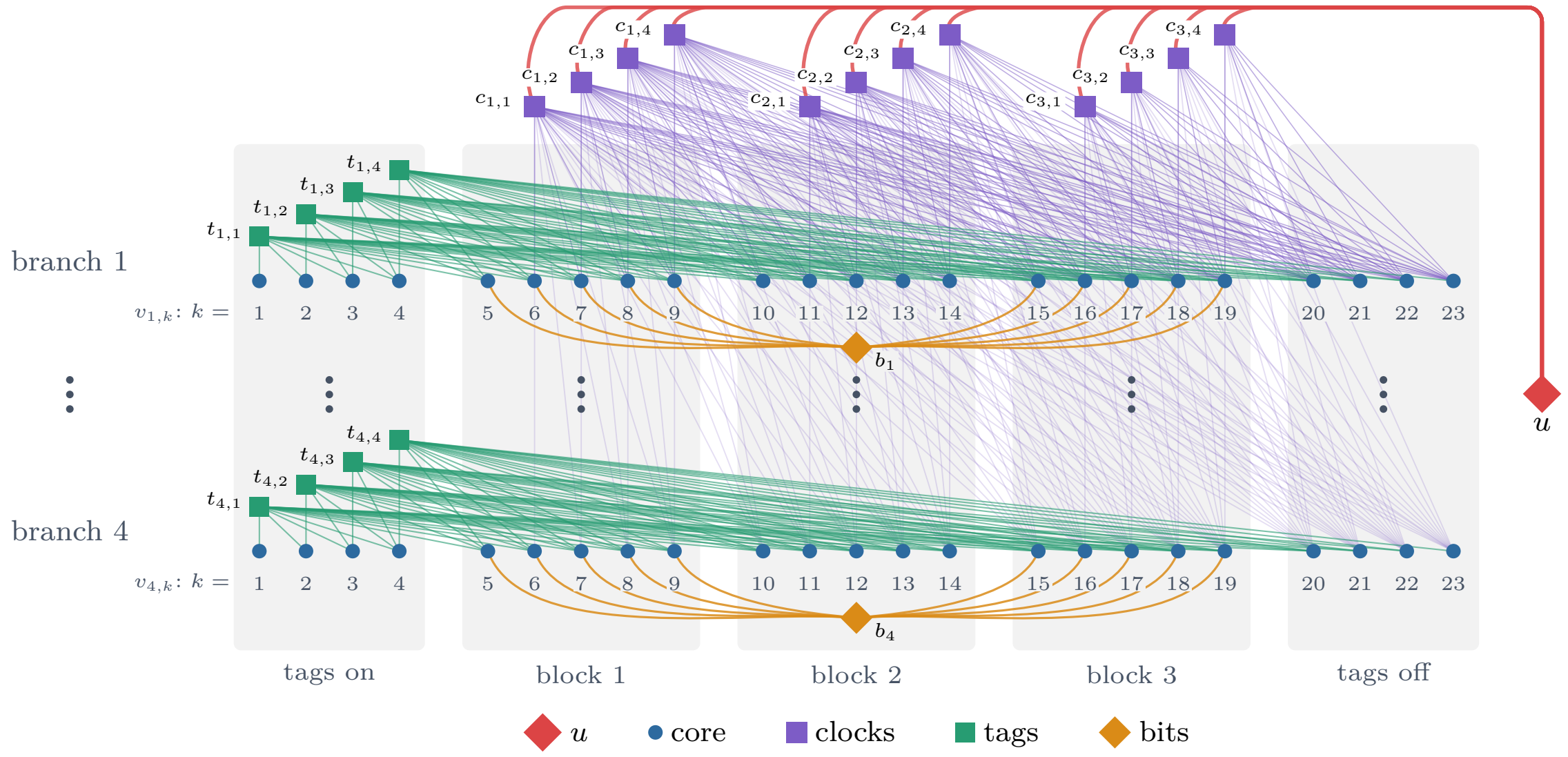

%% file: tikz/fractional_local_example.tex
\begingroup
\providecommand{\hfourClockWidth}{0.3pt}\providecommand{\hfourClockOpacity}{0.5}%
\providecommand{\hfourFadedWidth}{0.35pt}\providecommand{\hfourFadedOpacity}{0.2}%
\providecommand{\hfourTagWidth}{0.4pt}\providecommand{\hfourTagOpacity}{0.6}%
\providecommand{\hfourAddedWidth}{1.0pt}%
\providecommand{\hfourLabels}{2}%
\input{figures/palette}%
\colorlet{HfourData}{figBlue}%
\colorlet{HfourClock}{figPurple}%
\colorlet{HfourTag}{figGreen}%
\colorlet{HfourBit}{figOrange}%
\colorlet{HfourAdded}{figRed}%
\colorlet{HfourText}{figInk}%
\colorlet{HfourMuted}{figMuted}%
\colorlet{HfourPanel}{figPanel}

\begin{tikzpicture}[
  x=0.45cm, y=0.33cm,
  line cap=round, line join=round,
  every node/.style={inner sep=0pt, outer sep=0pt},
  declare function={bx(\k)=\k+0.9*floor(\k/5);},
  h4 data/.style={circle, fill=HfourData, minimum size=1.4mm},
  h4 clock/.style={rectangle, fill=HfourClock, minimum size=2.0mm},
  h4 tag/.style={rectangle, fill=HfourTag, minimum size=1.9mm},
  h4 bit/.style={diamond, aspect=1, fill=HfourBit, minimum size=3.1mm},
  h4 source/.style={diamond, aspect=1, fill=HfourAdded, minimum size=3.7mm},
  h4 marker label/.style={font=\fontsize{5.5}{7}\selectfont, anchor=east},
  h4 clear/.style={fill=#1, inner sep=0.5pt, rounded corners=0.5pt},
  h4 row label/.style={font=\fontsize{8}{10}\selectfont, text=HfourText},
  h4 block label/.style={font=\fontsize{7}{9}\selectfont, text=HfourText},
  h4 legend/.style={font=\scriptsize, anchor=mid west},
  h4 clock edge/.style={draw=HfourClock, line width=\hfourClockWidth, opacity=\hfourClockOpacity},
  h4 tag edge/.style={draw=HfourTag, line width=\hfourTagWidth, opacity=\hfourTagOpacity},
  h4 faded/.style={line width=\hfourFadedWidth, opacity=\hfourFadedOpacity},
  h4 bit edge/.style={draw=HfourBit, line width=0.6pt, opacity=0.85},
  h4 added edge/.style={draw=HfourAdded, line width=\hfourAddedWidth, opacity=0.8, rounded corners=4pt},
  h4 vertex label/.style={font=\fontsize{5.5}{7}\selectfont, text=HfourText, fill=HfourPanel, inner sep=0.35pt}
]
  \foreach \i/\yi in {1/12.1, 4/4.2} {
    \foreach \k in {1,...,23} {
      \coordinate (v-\i-\k) at ({bx(\k)},\yi);
    }
    \foreach \h in {1,...,4} {
      \coordinate (t-\i-\h) at ({bx(\h)},{\yi+1.3+(\h-1)*0.65});
    }
    \coordinate (b-\i) at ({bx(12)},{\yi-1.95});
  }
  \foreach \j in {1,...,3} {
    \foreach \h in {1,...,4} {
      \coordinate (c-\j-\h) at ({bx(5*\j+\h)},{17.2+(\h-1)*0.7});
    }
  }
  \coordinate (u) at ({bx(23)+1.9},8.8);

  \foreach \a/\b/\name in {1/4/tags on, 5/9/block $1$, 10/14/block $2$, 15/19/block $3$, 20/23/tags off} {
    \fill[HfourPanel, rounded corners=2pt] ({bx(\a)-0.55},1.3) rectangle ({bx(\b)+0.55},16.1);
    \node[h4 block label] at ({(bx(\a)+bx(\b))/2},0.6) {\name};
  }

  \foreach \i in {4,1} {
    \foreach \j in {1,...,3} {
      \foreach \h in {1,...,4} {
        \pgfmathtruncatemacro{\hfourStart}{5*\j+\h}
        \foreach \k in {\hfourStart,...,23} {
          \ifnum\i=1
            \draw[h4 clock edge] (c-\j-\h) -- (v-\i-\k);
          \else
            \draw[h4 clock edge, h4 faded] (c-\j-\h) -- (v-\i-\k);
          \fi
        }
      }
    }
  }

  \foreach \i in {1,4} {
    \foreach \h in {1,...,4} {
      \pgfmathtruncatemacro{\hfourEnd}{19+\h}
      \foreach \k in {\h,...,\hfourEnd} {
        \draw[h4 tag edge] (t-\i-\h) -- (v-\i-\k);
      }
    }
  }

  \foreach \i in {1,4} {
    \foreach \start in {5,15} {
      \pgfmathtruncatemacro{\hfourEnd}{\start+4}
      \foreach \k in {\start,...,\hfourEnd} {
        \ifnum\k<12
          \draw[h4 bit edge] (b-\i) to[out=175, in=-70, looseness=0.7] (v-\i-\k);
        \else
          \draw[h4 bit edge] (b-\i) to[out=5, in=-110, looseness=0.7] (v-\i-\k);
        \fi
      }
    }
  }

  \foreach \j in {1,...,3} {
    \foreach \h in {1,...,4} {
      \draw[h4 added edge] (u) |- ($(c-\j-\h |- 0,20.1)+(1,0)$) to[out=180, in=115] (c-\j-\h);
    }
  }

  \foreach \i/\yi in {1/12.1, 4/4.2} {
    \foreach \k in {1,...,23} {
      \node[h4 data] at (v-\i-\k) {};
    }
    \foreach \h in {1,...,4} {
      \node[h4 tag] at (t-\i-\h) {};
      \node[h4 marker label] at ($(t-\i-\h)+(-0.35,0.15)$) {$t_{\i,\h}$};
    }
    \node[h4 bit] at (b-\i) {};
    \node[h4 marker label, anchor=west, h4 clear=HfourPanel] at ($(b-\i)+(0.35,-0.4)$) {$b_{\i}$};
    \node[h4 row label, anchor=east] (bl-\i) at (-1.8,{\yi+0.6}) {branch \i};
    \ifcase\hfourLabels\relax
    \or
      \node[h4 vertex label] at ($(v-\i-1)+(0,-0.95)$) {$v_{\i,1}$};
      \node[h4 vertex label] at ($(v-\i-23)+(0,-0.95)$) {$v_{\i,23}$};
    \or
      \foreach \k in {1,...,23} {
        \node[h4 vertex label] at ($(v-\i-\k)+(0,-0.95)$) {\k};
      }
      \node[h4 vertex label, fill=none, anchor=east] at ($(v-\i-1)+(-0.45,-0.95)$) {$v_{\i,k}$:\ $k={}$};
    \fi
  }
  \coordinate (dots) at (0,9.1);
  \node[h4 row label, font=\Large\bfseries] at (bl-1 |- dots) {\vdots};
  \foreach \a/\b in {1/4, 5/9, 10/14, 15/19, 20/23} {
    \node[text=HfourText, font=\Large\bfseries] at ({(bx(\a)+bx(\b))/2},9.1) {\vdots};
  }

  \foreach \j in {1,...,3} {
    \foreach \h in {1,...,4} {
      \node[h4 clock] at (c-\j-\h) {};
      \node[h4 marker label, h4 clear=white] at ($(c-\j-\h)+(-0.42,0.15)$) {$c_{\j,\h}$};
    }
  }
  \node[h4 source] at (u) {};
  \node[font=\scriptsize] at ($(u)+(0,-0.9)$) {$u$};

  \matrix[ampersand replacement=\&, column sep=0.8mm] at ({(bx(1)+bx(23))/2},-1.1) {
    \node[h4 source] {}; \& \node[h4 legend] {$u$}; \&[5mm]
    \node[h4 data] {}; \& \node[h4 legend] {core}; \&[5mm]
    \node[h4 clock] {}; \& \node[h4 legend] {clocks}; \&[5mm]
    \node[h4 tag] {}; \& \node[h4 legend] {tags}; \&[5mm]
    \node[h4 bit] {}; \& \node[h4 legend] {bits}; \\
  };
\end{tikzpicture}%
\endgroup

%% file: figures/cut_factor_two_graphs.tex
\input{figures/cut_factor_two_style}
\begin{figure}[thbp!]
    \centering
    \renewcommand{\cftPairDistance}{0.5}%
    \begin{subfigure}[b]{0.48\textwidth}
        \centering
        \begin{tikzpicture}
            \cftSetup{1.25}
            \cftFanMarkers{0.68}{e}
            \cftEdges{G}
            \cftVertexNodes
            \cftGapLabels{0.42}
            \cftFanMarkers{0.68}{n}
        \end{tikzpicture}
        \caption{The graph $G$.}
    \end{subfigure}
    \hfill
    \begin{subfigure}[b]{0.48\textwidth}
        \centering
        \begin{tikzpicture}
            \cftSetup{1.25}
            \cftFanMarkers{0.68}{e}
            \cftEdges{H}
            \cftVertexNodes
            \cftGapLabels{0.42}
            \cftFanMarkers{0.68}{n}
        \end{tikzpicture}
        \caption{The graph $H$.}
    \end{subfigure}
    \caption{The graphs $G$ and $H$ of \cref{theorem:cut_factor_two_gap}. The five pairs $\{v_i^+, v_i^-\}$, $i \in \Z_5$, sit at the corners of a pentagon, and each pair has its own set $L_i$ of $9(i + 1)$ markers (grey). Blue core edges join pairs $i$ and $i \pm 1$, orange core edges join pairs $i$ and $i \pm 2$.}
    \label{figure:cut_factor_two_graphs}
\end{figure}
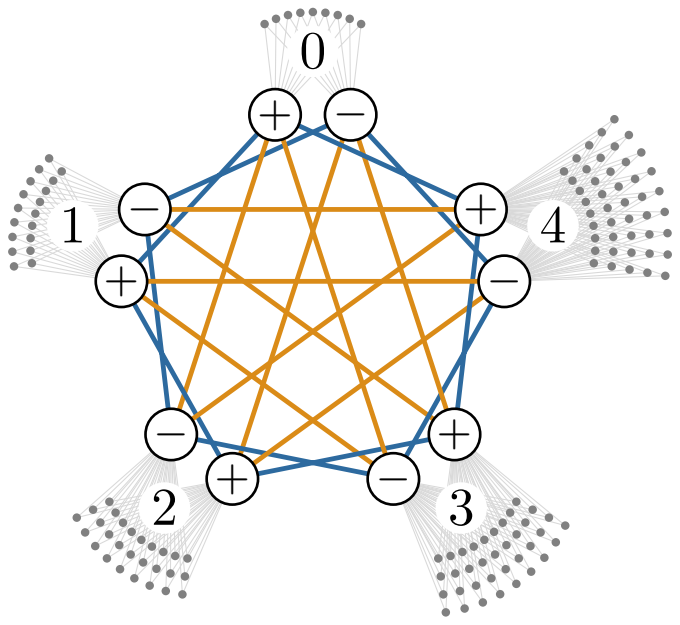
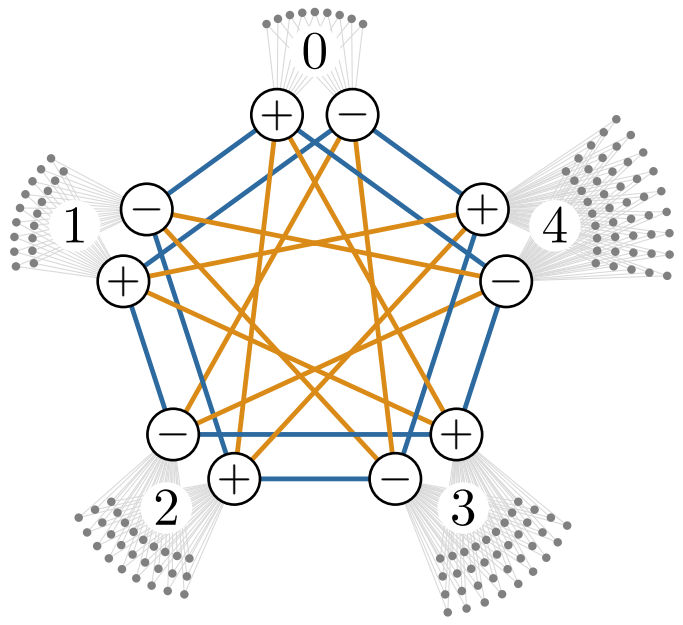

%% file: figures/cut_factor_two_style.tex

\ifdefined\cftSetup \fi

\input{figures/palette}
\tikzset{
    cft vertex/.style={circle, draw=black, fill=white, line width=0.5pt, inner sep=0pt, minimum size=3.4mm, font=\scriptsize},
    cft marker/.style={circle, fill=black!50, inner sep=0pt, minimum size=1.6pt},
    cft marker edge/.style={draw=black!15, line width=0.2pt},
    cft side/.style={draw=figBlue, line width=0.9pt},
    cft diagonal/.style={draw=figOrange, line width=0.9pt},
    cft label/.style={font=\footnotesize, text=black!60},
}

\providecommand{\cftPairDistance}{0.5}

\newcommand{\cftSetup}[1]{%
    \def\cftradius{#1}%
    \foreach \cfti in {0, ..., 4} {%
        \coordinate (c\cfti) at ({90 + 72 * \cfti}:#1);
        \coordinate (p\cfti) at ($(c\cfti) + ({180 + 72 * \cfti}:{\cftPairDistance / 2})$);
        \coordinate (m\cfti) at ($(c\cfti) + ({72 * \cfti}:{\cftPairDistance / 2})$);
    }%
}

\newcommand{\cftMarkers}{%
    \foreach \cfti in {0, ..., 4} {%
        \pgfmathtruncatemacro{\cftrows}{\cfti + 1}%
        \foreach \cftk in {1, ..., \cftrows} {%
            \foreach \cftj in {0, ..., 8} {%
                \coordinate (mk) at ($(c\cfti) + ({90 + 72 * \cfti + (\cftj - 4) * 14}:{0.55 + 0.2 * (\cftk - 1)})$);
                \draw[cft marker edge] (p\cfti) -- (mk) -- (m\cfti);
                \node[cft marker] at (mk) {};
            }%
        }%
    }%
}

\newcommand{\cftEdges}[1]{%
    \foreach \cfti in {0, ..., 4} {%
        \pgfmathtruncatemacro{\cftj}{mod(\cfti + 1, 5)}%
        \pgfmathtruncatemacro{\cftk}{mod(\cfti + 2, 5)}%
        \ifx G#1
            \draw[cft side] (p\cfti) -- (p\cftj) (m\cfti) -- (m\cftj);
            \draw[cft diagonal] (p\cfti) -- (m\cftk) (m\cfti) -- (p\cftk);
        \else
            \draw[cft side] (p\cfti) -- (m\cftj) (m\cfti) -- (p\cftj);
            \draw[cft diagonal] (p\cfti) -- (p\cftk) (m\cfti) -- (m\cftk);
        \fi
    }%
}

\newcommand{\cftVertices}[1]{%
    \foreach \cfti in {0, ..., 4} {%
        \node[cft vertex] at (p\cfti) {$+$};
        \node[cft vertex] at (m\cfti) {$-$};
        \node[cft label] at ({90 + 72 * \cfti}:#1) {$\cfti$};
    }%
}

\newcommand{\cftVertexNodes}{%
    \foreach \cfti in {0, ..., 4} {%
        \node[cft vertex] at (p\cfti) {$+$};
        \node[cft vertex] at (m\cfti) {$-$};
    }%
}

\newcommand{\cftOuterLabels}{%
    \foreach \cfti in {0, ..., 4} {%
        \node[cft label, text=black, font=\small] at ($(c\cfti) + ({90 + 72 * \cfti}:{0.83 + 0.2 * \cfti})$) {$\cfti$};
    }%
}

\newcommand{\cftOverlayLabels}[1]{%
    \foreach \cfti in {0, ..., 4} {%
        \node[cft label, text=black, font=\small, circle, fill=white, inner sep=0.8pt] at ($(c\cfti) + ({90 + 72 * \cfti}:#1)$) {$\cfti$};
    }%
}

\newcommand{\cftLabelledVertexNodes}{%
    \foreach \cfti in {0, ..., 4} {%
        \node[cft vertex, minimum size=4.4mm] at (p\cfti) {$\cfti^{+}$};
        \node[cft vertex, minimum size=4.4mm] at (m\cfti) {$\cfti^{-}$};
    }%
}

\newcommand{\cftFanMarkerCoordinate}[3]{
    \coordinate (mk) at ($(c#1) + ({90 + 72 * #1 + (#3 - 4) * 7}:{\cftfandist + 0.12 * (#2 - 1)})$);%
}
\newcommand{\cftFanMarkers}[2]{
    \def\cftfandist{#1}%
    \foreach \cfti in {0, ..., 4} {%
        \pgfmathtruncatemacro{\cftrows}{\cfti + 1}%
        \foreach \cftk in {1, ..., \cftrows} {%
            \foreach \cftj in {0, ..., 8} {%
                \cftFanMarkerCoordinate{\cfti}{\cftk}{\cftj}%
                \ifx e#2
                    \draw[cft marker edge] (p\cfti) -- (mk) -- (m\cfti);
                \else
                    \node[cft marker] at (mk) {};
                \fi
            }%
        }%
    }%
}

\newcommand{\cftGapLabels}[1]{%
    \foreach \cfti in {0, ..., 4} {%
        \node[circle, fill=white, inner sep=0.6pt, font=\small] at ($(c\cfti) + ({90 + 72 * \cfti}:#1)$) {$\cfti$};
    }%
}

\newcommand{\cftSpreadMarkers}[3]{%
    \foreach \cfti in {0, ..., 4} {%
        \pgfmathtruncatemacro{\cftcount}{9 * (\cfti + 1)}%
        \pgfmathtruncatemacro{\cftrows}{ceil(\cftcount / #2)}%
        \foreach \cftk in {1, ..., \cftrows} {%
            \pgfmathtruncatemacro{\cftinrow}{floor(\cftcount * \cftk / \cftrows) - floor(\cftcount * (\cftk - 1) / \cftrows)}%
            \pgfmathsetmacro{\cftrowradius}{\cftradius + #1 + 0.12 * (\cftk - 1)}%
            \pgfmathsetmacro{\cftstep}{deg(0.1 / \cftrowradius)}%
            \foreach \cftj in {1, ..., \cftinrow} {%
                \coordinate (mk) at ({90 + 72 * \cfti + (\cftj - (\cftinrow + 1) / 2) * \cftstep}:\cftrowradius);%
                \ifx e#3
                    \draw[cft marker edge] (p\cfti) -- (mk) -- (m\cfti);
                \else
                    \node[cft marker] at (mk) {};
                \fi
            }%
        }%
    }%
}

%% file: sections/04_hardness.tex
\section{Computational Hardness of Optimal Transport Formulations}
\label{section:hardness}
In this section, we give an overview of our \NP-hardness results for the optimal transport formulations of the edit distance, the operator $p$-distances, and the cut distance. The full proofs are given in \cref{appendix:hardness}.
Since our reductions always produce instances with pairs of graphs of equal order, we work with $\dOT{\normsubscript}$ rather than $\ndOT{\normsubscript}$ for simplicity.
For $\dOT{\normsubscript}$, we denote by $\GSim{\dOT{\normsubscript}}$ the problem of deciding whether $\dOT{\normsubscript}(G, H) \leq q$ for a given pair of graphs $G, H$ and a rational threshold $q \in \Q$. 

We first establish hardness for $\GSim{\dOT{\edit}}$ using the
reduction of Grohe, Rattan and Woeginger
\cite[Theorem 8]{grohe_rattan_woeginger_edit_distance} from the \NP-hard problem \threepartition{} \cite{garey1979computers} to $\GSim{\d{\edit}}$, with a modified correctness argument.
In \threepartition{}, the input consists of integers $A$ and $a_1, \ldots, a_{3m}$ in unary representation satisfying $\sum_{i=1}^{3m} a_i = mA$ and $A/4 < a_i < A/2$ for $1 \leq i \leq 3m$.
The question then is to decide whether $a_1, \ldots, a_{3m}$ can be partitioned into $m$ triples such that the sum of each triple is exactly $A$.
For such an instance, they construct two graphs $G, H$ where $G$ consists of $3m$ disjoint paths on $a_i$ vertices each and $H$ is the disjoint collection of $m$ paths on $A$ vertices. 
In case of a YES-instance, three paths of $G$ can be mapped into one path of $H$ such that only the two edges joining the three segments in $H$ are not preserved.
For NO-instances, for any bijection $\pi$, at least one path of $G$ must be split across multiple paths in $H$, which incurs additional edge mismatches.

\begin{theorem}
    \label{theorem:ot_edit_dist_hardness}
    \GSim{\dOT{\edit}} is \NP-hard, even when restricted to collections of paths.
\end{theorem}
For the local edit distance, these additional edge mismatches of a NO-instance need not increase the cost, as they do not necessarily increase the maximum number of mismatches occurring at one vertex. A path split across several target paths may have its split points mapped to target endpoints, keeping the local error at one. Replacing the target paths by cycles prevents this: a split then forces at least two mismatches at the same vertex, whereas a valid partition gives at most one per vertex.

\begin{theorem}
    \label{theorem:ot_local_edit_dist_hardness}
    For any $\eps > 0$, unless $\P = \NP$, there is no polynomial-time approximation algorithm
    computing $\dOT{1}$ with multiplicative error guarantee $2-\eps$, even for graphs of maximum degree two.
\end{theorem}
Extending this argument to all operator $p$-norms
requires additional control over fractional alignments.
For $p=1$, the identity in \cref{equ:local_transport_identity}
expresses the objective as a maximum over $\supp(Q)$.
Every supported pair is considered fully, regardless of how little
mass $Q$ assigns to it. For $p>1$, averaging can reduce
the contribution of such pairs. To control this effect,
we augment the construction by attaching sufficiently
many leaves to each vertex.
The resulting approximation threshold depends on $p$. We let
$b(p) \ceq \max\{\frac{1}{p},1-\frac{1}{p}\}$,
where $b(\infty) \ceq 1$.

\begin{theorem}
    \label{theorem:ot_op_p_general_hardness}
    For $1 \leq p \leq \infty$ and any $\eps > 0$, unless $\P = \NP$, there is no polynomial-time approximation algorithm computing $\dOT{p}$ with multiplicative error guarantee $2^{b(p)} - \eps$.
\end{theorem}
Our proof makes use of an auxiliary matrix $A_{G, H}^Q$ to obtain a lower bound on $\dOT{p}$ (compare \cref{lemma:lower_bound_operator_p_blowup}).
Due to its simplified structure, for each NO-instance, we find a column with sufficiently many large entries, separating it from all YES-instances.

While \cref{theorem:ot_op_p_general_hardness} shows that the operator $p$-distances are \NP-hard, it does not settle their membership in \NP, since we do not know whether an optimal fractional alignment can always be encoded with polynomially many bits. For $p=1$ (and by symmetry $p=\infty$) and the edit distance, we can resolve this because of their particularly simple structure. 
In particular, for any two input graphs there is a blowup size, exponential in
their order, at which $\ndBU{1}$ and $\ndBU{\edit}$ are attained. See \cref{proposition:ot_edit_local_containment_in_np}.
Finally, we show hardness for $\dOT{\cut}$. 
\begin{theorem}
    \label{theorem:ot_cut_hardness}
    For every $\eps > 0$, unless $\P = \NP$, there is no
    polynomial-time approximation algorithm computing $\dOT{\cut}$ with multiplicative error guarantee $\frac{17}{16} - \eps$.
\end{theorem}
In the proof, we reduce from \textsc{MaxCut}, adapting the construction of \cite[Theorem~23]{gervens2022graphsimilaritybasedmatrix}, which encodes the maximum cut of an input graph as the cost $\Cost{\cut}(\pi)$ of a canonical bijection $\pi$ between two auxiliary graphs.
Attaching sufficiently many leaves forces even fractional alignments to approximately respect $\pi$: We show that any substantial deviation already incurs a larger cost than $\Cost{\cut}(\pi)$. 
As a result, for a fractional alignment $Q$ that is close to $\pi$, the cut $(S, T)$ witnessing the original distance can be lifted to a cut in the matrix $B_\cut(G, H, Q)$
with their corresponding cut values remaining close. The resulting bounds transfer Håstad’s inapproximability result for \textsc{MaxCut} \cite{hastad_inapproximability_results} to $\GSim{\dOT{\cut}}$.

%% file: sections/05_conclusion.tex
\section{Conclusion}
\label{section:conclusion}

We study optimal transport versions of matrix-based graph metrics and establish a fairly general correspondence between the fractional alignment and the blowup approach.
We prove upper and lower bounds on the integrality gaps of these metrics, significantly improving the known bounds for edit and cut distance and establishing the first bounds for metrics based on operator norms.
The latter separate cut and spectral distance: both induce the same notion of convergence, yet only the former admits almost linear bounds.
Our proof techniques, including spectral methods, are new and fundamentally different from previous approaches.

Determining the exact integrality gaps remains open: the optimal factor for edit distance, the optimal exponents for local edit and spectral distance, and in particular \lovasz{}'s conjectured factor $2$ for cut distance.

\subparagraph*{Statement on AI Use.}
We used ChatGPT 6 Astra to obtain improved bounds relating the integral and fractional versions of the distances studied in this paper. AI-generated results and arguments contributed substantially to the results in \cref{section:integrality_gaps} and their proofs in
\cref{appendix:edit,appendix:local_edit,appendix:cut_factor_two_gap,appendix:spectral}. We subsequently simplified these proofs and improved their presentation significantly.
We take full responsibility for the correctness of all results.

%% file: appendices/01_extended_preliminaries.tex
\section{Extended Preliminaries}
\label{appendix:extended_preliminaries}
We denote the natural numbers by $\N$.
We denote the set of all doubly stochastic $n\times n$ matrices by $\DS(n)\coloneqq\DS([n],[n])$ and the subset of $n\times n$ permutation matrices by $\Perm(n)$.

\subparagraph*{Graphs}
The \emph{degree} of a vertex $v \in V(G)$ is $\deg_G(v) \ceq |N_G(v)|$.
We write $N_G[v] \ceq N_G(v) \cup \{v\}$ for the closed neighbourhood of $v$.

\subparagraph*{Matrix Norms}
For finite sets $V, W$, a matrix $A \in \R^{V \times W}$ and $S \subseteq V$, $T \subseteq W$, we let $A(S, T) \ceq \sum_{s \in S, t \in T} A(s, t)$.
We consider the following norms on $\R^{V \times W}$, see \cite{gervens2022graphsimilaritybasedmatrix} for a systematic treatment.
The \emph{Frobenius norm} is $\norm{A}_F\ceq (\sum_{v \in V, w \in W} |A(v,w)|^2)^{1/2}$, the \emph{entrywise $1$-norm} is $\norm{A}_{\edit} \ceq \norm{A}_{(1)} \ceq \sum_{v \in V, w \in W} |A(v, w)|$ (\ie{} the squared Frobenius norm), 
for $1 \leq p, q \leq \infty$, the \emph{operator $p \to q$-norm} is
\[
    \norm{A}_{p \to q} \ceq \max_{\x \in \R^W,\, \norm{\x}_p = 1} \norm{A\x}_q.
\]
We write $\norm{A}_p \ceq \norm{A}_{p \to p}$. For instance, $\norm{A}_1$ is the maximum absolute column sum of $A$.
The \emph{cut norm} is $\norm{A}_\cut \ceq \max_{S \subseteq V, T \subseteq W} |A(S, T)|$.

For $1 \leq q \leq \infty$, we denote by $q^\ast$ the \emph{Hölder conjugate} of $q$, \ie{} the unique value with $1/q + 1/q^\ast = 1$; thus $1/q^\ast = 1 - 1/q$, $q^\ast = 1$ if $q = \infty$, $q^\ast = \infty$ if $q = 1$, and $(q^\ast)^\ast = q$.
By \emph{Hölder's inequality}, $\sum_{v \in V} |\x(v) \y(v)| \leq \norm{\x}_q \norm{\y}_{q^\ast}$ for every finite set $V$ and all $\x, \y \in \R^V$.
Choosing $\y$ to be the all-ones vector, we obtain
\begin{equation}
    \label{equation:hoelder_special}
    \sum_{v \in V} |\x(v)| \leq \norm{\x}_q \cdot |V|^{1 - 1/q}.
\end{equation}
The $p \to q$-norms have the following well-known characterisation in terms of the Hölder conjugate.
\begin{lemma}[{see, \eg{}, \cite[Section~5.5]{matrix_analysis_second_edition}}]
    \label{lemma:norm_bilinear_identity}
    For $1 \leq p, q \leq \infty$ and all $A \in \R^{V \times W}$ it holds that
    \[
        \norm{A}_{p \to q} = \max_{\norm{\y}_{q^\ast} = 1, \norm{\x}_{p} = 1} \left|\y^\top A \x \right|.
    \]
\end{lemma}
\noindent Indeed, by the duality of the $\ell_q$- and $\ell_{q^\ast}$-norms, $\norm{\z}_q = \max_{\norm{\y}_{q^\ast} = 1} |\y^\top \z|$ for all $\z \in \R^V$, and it suffices to apply this to $\z = A\x$.
As $\y^\top A \x = \x^\top A^\top \y$, it follows that $\norm{A}_{p \to q} = \|A^\top\|_{q^\ast \to p^\ast}$.
In particular, $\norm{A}_p = \|A^\top\|_{p^\ast}$.

\subparagraph*{Optimal Transport Distances}
\label{section:ot_formulation_p_to_q}
For finite sets $V, W$, we denote by $\DSS(V, W)$ the set of nonnegative matrices in $\R^{V\times W}$ with row sums $1/|V|$ and column sums $1/|W|$.
Note that if $|V| = |W| = n$, then $\DS(V, W) = n\DSS(V, W)$.
For matrices $A \in \R^{V \times V}$, $B \in \R^{W \times W}$, the \emph{discrepancy matrix} $\Disc_{A, B} \in \R^{(V \times W) \times (V \times W)}$ has entries
\begin{equation}
    \label{equation:def_disc_matrix}
    \Disc_{A, B}(v, w, v', w') \ceq \Disc_{A, B}\big((v, w), (v', w')\big) \ceq A(v, v') - B(w, w').
\end{equation}
For graphs $G, H$, we let $\Disc_{G, H} \ceq \Disc_{A_G, A_H}$, and we drop the subscripts if $G$ and $H$ are clear from the context.
In this notation, $B_{p \to q}(G, H, Q^\ast)$ has entries $Q^\ast(v, w)^{1/q}\, \Disc_{G, H}(v, w, v', w')\, Q^\ast(v', w')^{1-1/p}$.
For $p = q = 1$, the convention $0^0 = 0$ ensures that column $(v', w')$ of $B_{1 \to 1}(G, H, Q^\ast)$ vanishes whenever $Q^\ast(v', w') = 0$, so
\begin{equation}
    \label{equ:local_transport_identity}
    \ndOT{1}(G, H) = \min_{Q^\ast \in \DSS(V, W)} \max_{(v', w') \in \supp(Q^\ast)} \sum_{(v, w) \in V \times W} \big| \Disc_{G, H}(v, w, v', w') \big| Q^\ast(v, w).
\end{equation}

All minima in the definitions of the graph distances are attained.
For minimisation over $\Bij(V, W)$, this follows from finiteness.
For optimal transport, the coupling polytopes $\DSS(V, W)$ and
$\DS(V, W)$ are compact, and the objectives are continuous for
the entrywise $1$-norm and the cut norm and lower semicontinuous
for all operator $p \to q$-norms.
For the operator norms, let $(Q_k^\ast)$ be a convergent sequence
in $\DSS(V, W)$.
Note that every pair in the support of $\lim_{k \to \infty} Q_k^\ast$ lies
in $\supp(Q_k^\ast)$ for almost all $k$;
restricting the weighted discrepancy matrices to the rows and
columns indexed by this support gives convergent matrices and
cannot increase the operator norm.
The same argument applies to $\DS(V, W)$.

%% file: appendices/02_ot_formulation_p_to_q.tex
\section{Proof of \headingref{Theorem}{theorem:blowup_ot_identity}}
\label{appendix:ot_formulation_p_to_q}
    Let $G, H$ be graphs with vertex sets $V, W$ of orders $m$ and $n$, respectively.
    To simplify the notation below, we let $\kappa \ceq \kappa_{p \to q} = 1/q + 1 - 1/p$ and $\Disc \ceq \Disc_{G, H}$.
    Note that $s(n) = n^{\kappa_{p \to q}}$ for the function $s(n)$ corresponding to the $p \to q$-operator norm.
    Thus, the entries of $B_{p \to q}(G, H, Q^\ast)$ are $Q^\ast(v, w)^{1/q} \, \Disc(v, w, v', w') \, Q^\ast(v', w')^{1-1/p}$.

    We first prove
    \begin{equation}
        \label{equ:forward_dir_v2}
        \ndBU{p \to q}(G, H) \geq \min_{Q^\ast \in \DSS(V, W)} \norm{B_{p \to q}(G, H, Q^\ast)}_{p \to q}.
    \end{equation}
    Let $\ell \in \Npos$ and $G^\blowup \ceq G^{\blowup \ell n}$, $H^\blowup \ceq H^{\blowup \ell m}$. Recall that the vertex sets of $G^\blowup$ and $H^\blowup$ are $V^\blowup \ceq V \times [\ell n]$ and $W^\blowup \ceq W \times [\ell m]$, respectively, and let $N \ceq \ell m n = |V^\blowup| = |W^\blowup|$. We prove that
    \begin{equation}
        \label{equation:forward_2_v2}
        \d{p \to q}(G^\blowup, H^\blowup) \geq N^{\kappa} \cdot \min_{Q^\ast \in \DSS(V, W)} \norm{B_{p \to q}(G, H, Q^\ast)}_{p \to q}
    \end{equation}
    which implies \cref{equ:forward_dir_v2}.

    Choose $\pi \in \Bij(V^\blowup, W^\blowup)$ such that $\d{p \to q}(G^\blowup, H^\blowup) = \norm{M^\pi}_{p \to q}$, where $M^\pi \ceq A_{G^\blowup}^\pi - A_{H^\blowup}$. For all $v \in V$, $w \in W$, let
    \begin{align*}
        I(v, w) &\ceq \left\{ i \in [\ell n] \mid \pi(v, i) \in \{w\} \times [\ell m] \right\}, \\
        J(v, w) &\ceq \big\{ j \in [\ell m] \mid \pi^{-1}(w, j) \in \{v\} \times [\ell n] \big\}.
    \end{align*}
    Then $\pi$ maps $\{v\} \times I(v, w)$ bijectively to $\{w\} \times J(v, w)$, and thus $|I(v, w)| = |J(v,w)|$.
    Note that for $j \in J(v, w)$, $j' \in J(v', w')$ it holds that
    \begin{equation}
        \label{equation:blowup_entries_v2}
        \begin{split}
            M^\pi\big((w, j), (w', j')\big)
            &= A_{G^\blowup}\big(\pi^{-1}(w, j), \pi^{-1}(w', j')\big) - A_{H^\blowup}\big((w, j), (w', j')\big) \\
            &= \Disc(v, w, v', w').
        \end{split}
    \end{equation}
    We define a matrix $Q^\ast \in \R^{V \times W}$ by
    \[
        Q^\ast(v, w) \ceq \frac{|I(v, w)|}{N},
        \quad\text{that is,}\quad
        Q^\ast(v, w) = \frac{|J(v, w)|}{N}.
    \]
    Observe that for all $v \in V$, $w \in W$ we have $\bigcup_{w' \in W} I(v, w') = [\ell n]$ and $\bigcup_{v' \in V} J(v', w) = [\ell m]$. Thus
    \[ 
        \sum_{w' \in W} Q^\ast(v, w') = \frac{\ell n}{N} = \frac{1}{m}
        \qquad\text{and}\qquad
        \sum_{v' \in V} Q^\ast(v', w) = \frac{\ell m}{N} = \frac{1}{n},
    \]
    which implies that $Q^\ast \in \DSS(V, W)$.

    Let $B \ceq B_{p \to q}(G, H, Q^\ast)$. We prove that
    \begin{equation}
        \label{equation:forward_3_v2}
        \norm{M^\pi}_{p \to q} \geq N^{\kappa} \cdot \norm{B}_{p \to q}.
    \end{equation}
    This will imply \cref{equation:forward_2_v2}.
    Using \cref{lemma:norm_bilinear_identity}, let $\x, \y \in \R^{V \times W}$ with $\norm{\y}_{q^\ast} = 1$ and $\norm{\x}_p = 1$ such that $\left| \y^\top B \x \right|$ is maximised.
    Note that we regard $\x$ and $\y$ as vectors of length $mn$ and not as matrices.

    For $w \in W$ and $j \in [\ell m]$ let $v(w, j)$ be the unique $v \in V$ such that $j \in J(v, w)$.
    We define the vectors $\Tilde{\x}, \Tilde{\y} \in \R^{W^\blowup}$ with entries
    \[
        \Tilde{\y}(w, j) \ceq \begin{cases}
            \frac{\y(v(w, j), w)}{|J(v(w, j), w)|^{1-1/q}} & \text{if $|J(v(w, j), w)| > 0$}, \\
            0 & \text{otherwise},
        \end{cases}
    \]
    and
    \[
        \Tilde{\x}(w, j) \ceq \begin{cases}
            \frac{\x(v(w, j), w)}{|J(v(w, j), w)|^{1/p}} & \text{if $|J(v(w, j), w)| > 0$}, \\
            0 & \text{otherwise}.
        \end{cases}
    \]
    In the following, we simply write $v$ instead of $v(w, j)$ whenever the choice is clear from context.
    Let $(\z, r) \in \{(\x, p), (\y, q^\ast)\}$ and note that $1/q^\ast = 1 - 1/q$. If $r < \infty$, it holds that
    \begin{align*}
        \norm{\Tilde{\z}}_r
        &= \left( \sum_{(v, w) \in V \times W} \sum_{j \in J(v, w)} |\Tilde{\z}(w, j)|^r \right)^{1/r}
        \\
        &= \left( \sum_{\substack{(v, w) \in V \times W\\ |J(v, w)| > 0}} \sum_{j \in J(v, w)} |\Tilde{\z}(w, j)|^r \right)^{1/r}
        \\
        &= \left( \sum_{\substack{(v, w) \in V \times W\\ |J(v, w)| > 0}} \sum_{j \in J(v, w)} \left|\frac{\z(v, w)}{|J(v, w)|^{1/r}}\right|^r \right)^{1/r}
        \\
        &= \left( \sum_{\substack{(v, w) \in V \times W\\ |J(v, w)| > 0}} \sum_{j \in J(v, w)} \frac{|\z(v, w)|^r}{|J(v, w)|} \right)^{1/r}
        \\
        &= \left( \sum_{\substack{(v, w) \in V \times W\\ |J(v, w)| > 0}} \frac{1}{|J(v, w)|} \cdot \sum_{j \in J(v, w)} |\z(v, w)|^r \right)^{1/r}
        \\
        &= \left( \sum_{\substack{(v, w) \in V \times W\\ |J(v, w)| > 0}} |\z(v, w)|^r \right)^{1/r}
        \\
        &\leq \left( \sum_{(v, w) \in V \times W} |\z(v, w)|^r \right)^{1/r}
        = \norm{\z}_r
        = 1.
    \end{align*}
    Note that the second equality holds because the inner sum is empty whenever $|J(v, w)| = 0$.
    If $r = \infty$, then $\Tilde{\z}(w, j) = \z(v, w)$ for all $j \in J(v, w)$, and thus $\norm{\Tilde{\z}}_\infty \leq \norm{\z}_\infty = 1$.

    Then we have
    \begin{align*}
        &\hphantom{{}={}} \norm{M^\pi}_{p \to q}
        \\
        &\geq \left| \Tilde{\y}^\top M^\pi \Tilde{\x} \right|
        \\
        &= \left|
            \sum_{(w, j), (w', j') \in W^\blowup}
            \Tilde{\y}(w, j) \,
            M^\pi\big((w, j), (w', j')\big) \,
            \Tilde{\x}(w', j')
        \right|
        \\
        &= \left|
            \sum_{(v, w), (v', w') \in V \times W}
            \sum_{\substack{j \in J(v, w)\\ j' \in J(v', w')}}
            \frac{\y(v, w)}{|J(v, w)|^{1-1/q}} \,
            \Disc(v, w, v', w') \,
            \frac{\x(v', w')}{|J(v', w')|^{1/p}}
        \right|
        \\
        &= \left|
            \sum_{(v, w), (v', w') \in V \times W}
            \Disc(v, w, v', w') \,
            |J(v, w)|^{1/q} \,
            |J(v', w')|^{1-1/p} \,
            \y(v, w) \,
            \x(v', w')
        \right|
        \\
        &= N^{\kappa} \cdot \left|
            \sum_{(v, w), (v', w') \in V \times W}
            \Disc(v, w, v', w') \,
            Q^\ast(v, w)^{1/q} \,
            Q^\ast(v', w')^{1-1/p} \,
            \y(v, w) \,
            \x(v', w')
        \right|
        \\
        &= N^{\kappa} \cdot \left|
            \sum_{(v, w), (v', w') \in V \times W}
            \y(v, w) \,
            \Disc(v, w, v', w') \,
            Q^\ast(v, w)^{1/q} \,
            Q^\ast(v', w')^{1-1/p} \,
            \x(v', w')
        \right|
        \\
        &= N^{\kappa} \cdot \left|\y^\top B \x\right|
        \\
        &= N^{\kappa} \cdot \norm{B}_{p \to q}.
    \end{align*}
    Here, the first inequality uses \cref{lemma:norm_bilinear_identity} and $\norm{\Tilde{\y}}_{q^\ast}, \norm{\Tilde{\x}}_p \leq 1$, the third equality uses \cref{equation:blowup_entries_v2}, and the fifth equality uses that terms with $|J(v, w)| = 0$ or $|J(v', w')| = 0$ vanish (recall the convention $0^0 = 0$).
    The last equality holds because $\x$ and $\y$ were chosen such that they maximise the quantity $\left|\y^\top B \x\right|$, which is $\norm{B}_{p \to q}$.

    For the converse inequality
    \begin{equation}
        \label{equation:backward_v2}
        \ndBU{p \to q}(G, H) \leq \min_{Q^\ast \in \DSS(V, W)} \norm{B_{p \to q}(G, H, Q^\ast)}_{p \to q},
    \end{equation}
    let $\eps > 0$ and $Q^\ast \in \DSS(V, W)$ be a rational matrix such that
    \begin{equation}
        \label{equation:rational_matrix}
        \norm{B_{p \to q}(G, H, Q^\ast)}_{p \to q} \leq \min_{Q_0^\ast \in \DSS(V, W)} \norm{B_{p \to q}(G, H, Q_0^\ast)}_{p \to q} + \eps.
    \end{equation}
    Such a matrix exists, this can be shown similarly to \cite[Corollary~3.12]{grohe2025thoughtsgraphsimilarity}.
    \begin{claim}
        The rational matrix $Q^\ast \in \DSS(V, W)$ fulfilling \cref{equation:rational_matrix} exists.
    \end{claim}
    \begin{claimproof}
        Let $Q_0^\ast \in \DSS(V, W)$
        be a minimiser and put $S \ceq \supp(Q_0^\ast)$.
        The couplings in $\DSS(V, W)$ whose entries outside $S$ are
        zero form a rational polytope.
        Hence there is a sequence of rational matrices
        $Q_k^\ast \in \DSS(V, W)$ with $\supp(Q_k^\ast) = S$ and $Q_k^\ast \to Q_0^\ast$.
        On this fixed support, the entries of
        $B_{p \to q}(G, H, Q_k^\ast)$ converge to those of
        $B_{p \to q}(G, H, Q_0^\ast)$, also when $p = 1$ or $q = \infty$:
        entries outside $S \times S$ remain zero by the convention
        $0^0 = 0$, and the remaining entries depend continuously
        on the positive entries of $Q_k^\ast$.
        By continuity of the norm,
        \[
            \norm{B_{p \to q}(G, H, Q_k^\ast)}_{p \to q}
            \to \norm{B_{p \to q}(G, H, Q_0^\ast)}_{p \to q}.
        \]
        Taking $Q^\ast \ceq Q_k^\ast$ for sufficiently large $k$ therefore
        gives the required inequality.
    \end{claimproof}
    Let $B \ceq B_{p \to q}(G, H, Q^\ast)$.
    We shall prove
    \begin{equation}
        \label{equation:backward_2_v2}
        \ndBU{p \to q}(G, H) \leq \norm{B}_{p \to q}.
    \end{equation}
    As $\eps > 0$ was arbitrary, this will imply \cref{equation:backward_v2} and hence the assertion.
    Let $k$ be the least common multiple of the denominators of all entries $Q^\ast(v, w)$ of $Q^\ast$.
    We shall prove that for every $s \in \Npos$, it holds that
    \begin{equation}
        \label{equation:backward_3_v2}
        \nd{p \to q}\big(G^{\blowup nks}, H^{\blowup mks}\big) \leq \norm{B}_{p \to q}.
    \end{equation}
    Then
    \begin{align*}
        \ndBU{p \to q}(G, H)
        &= \lim_{\ell \to \infty} \nd{p \to q}\big(G^{\blowup n\ell}, H^{\blowup m\ell}\big) \\
        &= \lim_{s \to \infty} \nd{p \to q}\big(G^{\blowup nks}, H^{\blowup mks} \big) \leq \norm{B}_{p \to q},
    \end{align*}
    which proves \cref{equation:backward_2_v2}.
    Thus, it remains to prove \cref{equation:backward_3_v2}.

    Let $s \in \Npos$ and $\ell \ceq ks$, and $G^\blowup \ceq G^{\blowup n\ell}$, $H^{\blowup} \ceq H^{\blowup m\ell}$. Then the vertex sets of $G^\blowup$ and $H^\blowup$ are $V^{\blowup} \ceq V \times [\ell n]$ and $W^\blowup \ceq W \times [\ell m]$, respectively, of size $N\ceq \ell mn = |V^\blowup| = |W^\blowup|$.
    For all $v \in V$, $w \in W$ we let $q(v, w) \ceq N \cdot Q^\ast(v, w)$. As $N$ is a multiple of $k$, we have $q(v, w) \in \N$. For all $v \in V$ we have
    \begin{align*}
        \sum_{w \in W}q(v, w)
        &= N \cdot \sum_{w' \in W} Q^\ast(v, w') = \frac{N}{m} = \ell n.
    \end{align*}
    We choose an arbitrary partition of $[\ell n]$ into sets $I(v, w)$ of size $|I(v, w)| \ceq q(v, w)$ for $w \in W$.
    Similarly, for $w \in W$ we have $\sum_{v \in V} q(v, w) = \ell m$, and we choose an arbitrary partition of $[\ell m]$ into sets $J(v, w)$ of size $|J(v, w)| \ceq q(v, w)$, for $v \in V$.
    For all $v \in V$, $w \in W$ we let $\pi_{vw}$ be an arbitrary bijection from $I(v, w)$ to $J(v, w)$. Then we define a bijection $\pi: V^\blowup \to W^\blowup$ by letting $\pi(v, i) \ceq (w, \pi_{vw}(i))$ for the unique $w$ such that $i \in I(v, w)$.
    Let $M^\pi \ceq A_{G^\blowup}^\pi - A_{H^\blowup}$.
    Observe that \cref{equation:blowup_entries_v2} holds for this $\pi$ as well, that is, for $j \in J(v, w)$, $j' \in J(v', w')$ it holds that
    \[
        M^\pi\big((w, j), (w', j')\big) = \Disc(v, w, v', w').
    \]
    Let $\Tilde{\x}, \Tilde{\y} \in \R^{W^\blowup}$ with $\norm{\Tilde{\y}}_{q^\ast} = 1$ and $\norm{\Tilde{\x}}_p = 1$ such that $\left|\Tilde{\y}^\top M^\pi \Tilde{\x}\right|$ is maximised.
    We define vectors $\x, \y \in \R^{(V \times W)}$ by letting
    \[
        \y(v, w) \ceq \begin{cases}
            \frac{
                \sum_{j \in J(v, w)} \Tilde{\y}(w, j)
            }{|J(v, w)|^{1/q}} & \text{if $|J(v, w)| > 0$}, \\
            0 & \text{otherwise},
        \end{cases}
    \]
    and
    \[
        \x(v, w) \ceq \begin{cases}
            \frac{
                \sum_{j \in J(v, w)} \Tilde{\x}(w, j)
            }{|J(v, w)|^{1-1/p}} & \text{if $|J(v, w)| > 0$}, \\
            0 & \text{otherwise}.
        \end{cases}
    \]
    Let $(\z, r) \in \{(\x, p), (\y, q^\ast)\}$ and note that $1 - 1/q^\ast = 1/q$. If $r < \infty$, it holds that
    \begin{align*}
        \norm{\z}_r &=
        \left(
            \sum_{(v, w) \in V \times W} |\z(v, w)|^r
        \right)^{1/r}
        \\
        &=
        \left(
            \sum_{\substack{(v, w) \in V \times W\\ |J(v, w)| > 0}}
            \left|
                \frac{
                    \sum_{j \in J(v, w)}
                    \Tilde{\z}(w, j)
                }{|J(v, w)|^{1-1/r}}
            \right|^r
        \right)^{1/r}
        \\
        &=
        \left(
            \sum_{\substack{(v, w) \in V \times W\\ |J(v, w)| > 0}}
            \frac{1}{|J(v,w)|^{r-1}} \cdot
            \left|
                \sum_{j \in J(v, w)}
                \Tilde{\z}(w, j)
            \right|^r
        \right)^{1/r}
        \\
        &\stackrel{(1)}{\leq}
        \left(
            \sum_{\substack{(v, w) \in V \times W\\ |J(v, w)| > 0}}
            \frac{1}{|J(v,w)|^{r-1}} \cdot
            \left(
                \sum_{j \in J(v, w)}
                \left| \Tilde{\z}(w, j) \right|
            \right)^r
        \right)^{1/r}
        \\
        &\stackrel{(2)}{\leq}
         \left(
            \sum_{\substack{(v, w) \in V \times W\\ |J(v, w)| > 0}}
            \frac{1}{|J(v,w)|^{r-1}} \cdot
            \left(
                \left(
                    \sum_{j \in J(v, w)}
                    \left| \Tilde{\z}(w, j) \right|^r
                \right)^{1/r}
                \cdot
                |J(v, w)|^{\frac{r-1}{r}}
            \right)^r
        \right)^{1/r}
        \\
        &=
        \left(
            \sum_{\substack{(v, w) \in V \times W\\ |J(v, w)| > 0}}
            \frac{1}{|J(v,w)|^{r-1}} \cdot
            \left(
                \sum_{j \in J(v, w)}
                \left| \Tilde{\z}(w, j) \right|^r
            \right)
            \cdot
            |J(v, w)|^{r-1}
        \right)^{1/r}
        \\
        &=
        \left(
            \sum_{\substack{(v, w) \in V \times W\\ |J(v, w)| > 0}}
                \sum_{j \in J(v, w)}
                \left| \Tilde{\z}(w, j) \right|^r
        \right)^{1/r}
        \\
        &=
        \left(
            \sum_{(v, w) \in V \times W}
                \sum_{j \in J(v, w)}
                \left| \Tilde{\z}(w, j) \right|^r
        \right)^{1/r}
        = \norm{\Tilde{\z}}_r
        = 1,
    \end{align*}
    where $(1)$ holds because taking the absolute value of each individual summand can only increase the value of the sum and $(2)$ is an application of Hölder's inequality in the form of \cref{equation:hoelder_special}.
    If $r = \infty$, then $\z(v, w)$ is the average of the values $\Tilde{\z}(w, j)$, $j \in J(v, w)$, and thus $\norm{\z}_\infty \leq \norm{\Tilde{\z}}_\infty = 1$.

    Putting things together we obtain:
    \begin{align*}
        &\hphantom{{}={}} \d{p \to q}(G^\blowup, H^\blowup)
        \\
        &\leq \norm{M^\pi}_{p \to q}
        \\
        &= \left| \Tilde{\y}^\top M^\pi \Tilde{\x} \right|
        \\
        &=
        \left|
            \sum_{(w, j), (w', j') \in W^\blowup}
            \Tilde{\y}(w, j) \,
            M^\pi\big((w, j), (w', j')\big) \,
            \Tilde{\x}(w', j')
        \right|
        \\
        &=
        \left|
            \sum_{(v, w), (v', w') \in V \times W}
            \sum_{\substack{j \in J(v, w)\\ j' \in J(v', w')}}
            \Tilde{\y}(w, j) \,
            M^\pi\big((w, j), (w', j')\big) \,
            \Tilde{\x}(w', j')
        \right|
        \\
        &=
        \left|
            \sum_{(v, w), (v', w') \in V \times W}
            \Disc(v, w, v', w')
            \cdot
            \left(
                \sum_{\substack{j \in J(v, w)\\ j' \in J(v', w')}}
                \Tilde{\y}(w, j) \,
                \Tilde{\x}(w', j')
            \right)
        \right|
        \\
        &=
        \left|
            \sum_{(v, w), (v', w') \in V \times W}
            \Disc(v, w, v', w')
            \cdot
            \left(
                \sum_{j \in J(v, w)}
                \Tilde{\y}(w, j)
            \right)
            \left(
                \sum_{j' \in J(v', w')}
                \Tilde{\x}(w', j')
            \right)
        \right|
        \\
        &=
        \left|
            \sum_{(v, w), (v', w') \in V \times W}
            \Disc(v, w, v', w')
            \cdot
            |J(v, w)|^{1/q} \, \y(v, w)
            \cdot |J(v', w')|^{1-1/p} \, \x(v', w')
        \right|
        \\
        &=
        N^{\kappa} \cdot
        \left|
            \sum_{(v, w), (v', w') \in V \times W}
            \Disc(v, w, v', w')
            \cdot
            Q^\ast(v,w)^{1/q} \, \y(v, w)
            \cdot Q^\ast(v', w')^{1-1/p} \, \x(v', w')
        \right|
        \\
        &=
        N^{\kappa} \cdot
        \left|
            \sum_{(v, w), (v', w') \in V \times W}
            \y(v, w)
            \underbrace{
                \Disc(v, w, v', w') \,
                Q^\ast(v,w)^{1/q} \,
                Q^\ast(v', w')^{1-1/p}
            }_{
                =\, B((v, w), (v', w'))
            }
            \x(v', w')
        \right|
        \\
        &=
        N^{\kappa} \cdot \left| \y^\top B \x \right|
        \\
        &\leq N^{\kappa} \cdot \max_{\norm{\y'}_{q^\ast}\leq1, \norm{\x'}_p \leq 1} \left|\y'^\top B \x'\right|
        \\
        &= N^{\kappa} \cdot \max_{\norm{\y'}_{q^\ast} = 1, \norm{\x'}_p = 1} \left|\y'^\top B \x'\right|
        \\
        &= N^{\kappa} \cdot \norm{B}_{p \to q}.
    \end{align*}
    Here, the fourth equality uses \cref{equation:blowup_entries_v2}, and the second to last equality holds because the maximum of the bilinear form over the unit balls is attained on the unit spheres.
    Since $\nd{p \to q}(G^\blowup, H^\blowup) = N^{-\kappa} \cdot \d{p \to q}(G^\blowup, H^\blowup)$, this proves \cref{equation:backward_3_v2}.
    This concludes the proof.

%% file: appendices/03_edit_distance_bounds.tex
\section{The Integrality Gap of the Edit Distance}
\label{appendix:edit}
The following two results are used throughout the lower bound constructions for the edit and local edit distances.
We first prove a simple degree bound 
for $\LocalCost{Q}$.
\begin{observation}
    \label{claim:local_cost_degree_bound}
    For all graphs $G, H$ of the same order, all $Q \in \DS(V, W)$ and $(v, w) \in V \times W$, we have $\LocalCost{Q}(v, w) \geq \big| \deg_G(v) - \deg_H(w) \big|$.
\end{observation}
\begin{proof}
    By the triangle inequality and since all row and column sums of $Q$ are $1$,
    \begin{align*}
        \LocalCost{Q}(v, w)
        &\geq \Bigg| \sum_{v' \in V} \sum_{w' \in W} \Disc(v, w, v', w') Q(v', w') \Bigg| \\
        &= \Bigg| \sum_{v' \in V} A_G(v, v') - \sum_{w' \in W} A_H(w, w') \Bigg|
        = \big|\deg_G(v) - \deg_H(w)\big|. \qedhere
    \end{align*}
\end{proof}
Based on this observation, the following lemma allows us to fix the marker vertices in our constructions.
This is achieved by attaching the markers, ordered by degree, to a large clique in a staircase whose stepsize is independent of the marker degrees in the original graph.
\begin{lemma}[Locking Markers]
    \label{lemma:locking_markers}
    Let $G, H$ be graphs on a common vertex set $V$ with maximum degree at most $\Delta$, and let $z_1, \ldots, z_k \in V$ be \emph{markers}, ordered such that $\deg_G(z_1) \leq \cdots \leq \deg_G(z_k)$.
    Let $\tau \geq 1$ and $\delta \geq 0$ be integers with $|\deg_G(z_j) - \deg_H(z_j)| \leq \delta$ for all $j \in [k]$.
    Obtain $G'$ and $H'$ (with vertex set $V'$) from $G$ and $H$, respectively, by adding the same clique $\mathcal{K}$ on $m \ceq 2\Delta + 2\tau + k(\tau + \delta) + 1$ new vertices $y_1, \ldots, y_m$ and joining every marker $z_j$ to $y_1, \ldots, y_{\Delta + \tau + j(\tau + \delta)}$.
    Then the following hold for $\normsubscript \in \{\edit, 1\}$.
    \begin{enumerate}
        \item Every bijection $\pi \in \Bij(V', V')$ with $\Cost{\normsubscript}(\pi) < \tau$ \wrt{} $(G', H')$ fixes every marker and satisfies $\pi(V) = V$.
        \item Let $Q \in \DS(V, V)$ with $Q(z_j, z_j) = 1$ for all $j \in [k]$, and extend $Q$ by the identity on $\mathcal{K}$ to $Q'$. Then $\Cost{\normsubscript}(Q')$ \wrt{} $(G', H')$ equals $\Cost{\normsubscript}(Q)$ \wrt{} $(G, H)$.
    \end{enumerate}
\end{lemma}
\begin{proof}
    Let $Z \ceq \{z_1, \ldots, z_k\}$ and $s \ceq \tau + \delta$.
    Both costs are determined by the local costs, namely
    \begin{equation}
        \label{equation:locking_local_costs}
        \Cost{\edit}(Q) = \sum_{(v, w) \in V \times W} Q(v, w) \LocalCost{Q}(v, w)
        \qquad\text{and}\qquad
        \Cost{1}(Q) = \max_{(v, w) \in \supp(Q)} \LocalCost{Q}(v, w).
    \end{equation}
    So by \cref{claim:local_cost_degree_bound}, a bijection $\pi$ with $\Cost{\normsubscript}(\pi) < \tau$ maps every vertex $v$ to a vertex $\pi(v)$ with $|\deg_{G'}(v) - \deg_{H'}(\pi(v))| < \tau$.

    In $G'$ and $H'$, vertices in $V \setminus Z$ have degree at most $\Delta$, clique vertices have degree at least $m - 1$, and the degree of $z_j$ lies in $[\Delta + \tau, 2\Delta + \tau + ks] = [\Delta + \tau, m - 1 - \tau]$.
    Moreover, for $i < j$, the order of the markers yields
    \[
        \deg_{G'}(z_j) - \deg_{H'}(z_i)
        = (j - i)s + \deg_G(z_j) - \deg_H(z_i)
        \geq s + \deg_G(z_j) - \deg_G(z_i) - \delta
        \geq \tau,
    \]
    and symmetrically $\deg_{H'}(z_j) - \deg_{G'}(z_i) \geq \tau$.
    Hence $\pi$ maps every vertex of $V \setminus Z$ into $V \setminus Z$ and every marker $z_j$ to $z_j$, so $\pi(V) = V$.
    This proves the first item.

    For the second item, $Q(z_j, z_j) = 1$ implies that every $(v, w) \in \supp(Q')$ satisfies $v = w \in Z \cup \mathcal{K}$ or $v, w \in V \setminus Z$.
    In both cases, $v$ in $G'$ and $w$ in $H'$ have the same neighbours in $\mathcal{K}$.
    So for $(v, w), (v', w') \in \supp(Q')$, the entry $\Disc_{G', H'}(v, w, v', w')$ is zero if one of the pairs lies in $\mathcal{K} \times \mathcal{K}$, and $\Disc_{G, H}(v, w, v', w')$ otherwise.
    Thus $\LocalCost{Q'}$ agrees with $\LocalCost{Q}$ on $\supp(Q)$ and is zero on $\supp(Q') \setminus \supp(Q)$. 
    The claim then follows from \cref{equation:locking_local_costs}.
\end{proof}

\subsection{The Lower Bound}
\label{appendix:gaps_edit_lower}
\label{appendix:gaps_edit_lower_four_thirds}
In this section, we prove \cref{theorem:edit_gap_four_thirds}, following the proof overview in \cref{section:gaps_edit}.

Fix an odd $N \geq 3$.
We start with the formal construction of the graphs $G$ and $H$.
To this end, we first define graphs $G'$ and $H'$ which 
are defined on a common set of marker vertices 
$M \ceq \{z_0,\ldots,z_{N-1}\}$ and a set of core vertices $S_G \ceq \{v_0, v_2, \ldots,v_{2N-2}\}$ and $S_H \ceq \{w_1, w_3,\ldots, w_{2N-1}\}$, respectively.
In each graph, distinct core vertices $v_i, v_{i'}$ (respectively $w_i, w_{i'}$) are connected by an edge if $i+i'\equiv 0 \pmod{2N}$ (compare \cref{figure:edit_lower_cores}).
To describe the connections between core and marker vertices,
for every $r \in \{0,\ldots,N-1\}$, we let
\[
    I_r \ceq \{(r+s)\bmod(2N) \mid s \in \{0,\ldots,N-1\}\} \subseteq \Z_{2N}
\]
be the set of positions along a semicircle in the cycle $\Z_{2N}$.
A \emph{core-marker pair} is connected by an edge $v_iz_r$ (or $w_iz_r$) in either graph if the index of the core vertex is contained in the semicircle corresponding to the marker, that is $i \in I_r$.
This is visualised for $z_0$ in \cref{figure:edit_lower_cores} and for all $z\in M$ in \cref{figure:four_thirds_table}.
\input{figures/edit_lower_bound_43_table}

Finally, we obtain $G$ and $H$ by applying \cref{lemma:locking_markers} to $G'$ and $H'$ (where we identify $v_i$ of $G'$ with $w_{i+1}$ of $H'$), with the markers $M$ ordered nondecreasingly by their degree in $G'$ and parameters $\Delta \ceq N+1$, $\tau \ceq 4N-1$ and $\delta \ceq 1$.
These parameters are valid because every vertex has degree at most $N+1$ in $G'$ and $H'$, and the degree of $z_r$ in the two graphs differs by exactly one for each $r$.

All in all, the subgraph induced by the core vertices consists of $(N-1)/2$ disjoint edges and a single isolated vertex in both $G$ and $H$. However, the positions assigned to these vertices (and thus their connections to marker vertices) are shifted by one step along the cycle $\Z_{2N}$.

Write $V \ceq V(G)$, $W \ceq V(H)$ and call a fractional alignment $Q \in \DS(V, W)$ \emph{compliant} if $Q$ acts as the identity on the marker vertices and the clique added by \cref{lemma:locking_markers}.
For $v_i \in S_G$ and $w_j \in S_H$, we define
$\ell(v_i,w_j) \ceq \min_{k\in\mathbb{Z}} \bigl| i-j + 2kN \bigr|$.
Intuitively, this corresponds to the shortest path distance between $v_i$ and $w_j$ when placed at their assigned position in a common cycle graph $C_{2N}$. 

\begin{lemma}
    \label{lemma:edit_lower_anchor_cost}
    For every compliant $Q \in \DS(V, W)$, the core-marker pairs contribute exactly
    \[
        2\sum_{v_i \in S_G,\, w_j \in S_H} Q(v_i,w_j)\,\ell(v_i,w_j)
    \]
    to its edit cost $\Cost{\edit}(Q)$.
\end{lemma}
\begin{proof}
    Mapping $v_i$ to $w_j$ moves its position on the cycle $C_{2N}$ by $\ell(v_i,w_j) \leq N$ (compare \cref{figure:edit_lower_cores}), 
    incurring a change of adjacency to exactly $\ell(v_i,w_j)$ markers (compare \cref{figure:four_thirds_table}).
    Summing these differences weighted by $Q(v_i,w_j)$ and counting both orientations of each core-marker pair gives the formula.
\end{proof}

\begin{lemma}
    \label{lemma:edit_lower_matching_witness}
    $\dOT{\edit}(G,H) \leq 3N+1$.
\end{lemma}
\begin{proof}
    Let $\pi_-, \pi_+ \in \Bij(V, W)$ be the compliant bijections with 
    \begin{align*}
    \pi_-(v_i) &\ceq w_{(i-1)\bmod 2N}, \\
    \pi_+(v_i) &\ceq w_{(i+1)\bmod 2N}
    \end{align*}
    for each $i \in \{0, 2, \ldots, 2N-2\}$ and let
    $Q \ceq \mfrac12\big(P_{\pi_-}+P_{\pi_+}\big)$.
    Hence $Q$ sends half of the mass of $v_i$ \emph{backward} to $\pi_-(v_i)$ and half of it \emph{forward} to $\pi_+(v_i)$.
    By \cref{lemma:locking_markers}, we may compute the cost on $G'$ and $H'$.
    Pairs of markers contribute nothing to $\Cost{\edit}(Q)$, and since $\ell(v_i, \pi_-(v_i)) = \ell(v_i,\pi_+(v_i))=1$, \cref{lemma:edit_lower_anchor_cost} gives a contribution of $2N$ from the core-marker pairs.

    Each edge of $G'$ is preserved under an alignment if one endvertex is moved forward and the other is moved backward.
    As vertices are mapped forward or backward by $Q$ independently with probability $\frac{1}{2}$, 
    this happens in two of the four equally probable cases, so with probability $\frac{1}{2}$.
    Hence, each edge is not preserved with probability $\frac{1}{2}$.
    This is visualised in \cref{figure:four_thirds_edge_typical}.
    \input{figures/edit_lower_bound_43_witness_edge}

    There are a total of $\frac{N-1}{2}$ core edges in $G$ and considering both orientations for each thus yields a total cost of $2 \cdot \frac{N-1}{2} \cdot \frac{1}{2} = \frac{N-1}{2}$.
    Applying the same reasoning for the $\frac{N-1}{2}$ core edges of $H$ (with the inverses of $\pi_-$ and $\pi_+$) yields the same quantity again.
    The only imprecision of this argument occurs when moving forward or backward involves one of the vertices $v_0, w_N$ since these are isolated in the two graphs and hence cannot preserve edges.
    This is visualised in \cref{figure:four_thirds_edge_exception}.
    The cost incurred by this is at most one for each of these two vertices.

    In total, the cost is at most $2N + (N-1) + 2 = 3N+1$, which concludes the proof.
\end{proof}

\begin{lemma}
    \label{lemma:edit_lower_bijection_cost}
    $\d{\edit}(G,H) \geq 4N-2$.
\end{lemma}
\begin{proof}
    Let $\pi \in \Bij(V,W)$ be arbitrary.
    If $\Cost{\edit}(\pi) \geq \tau = 4N-1$, there is nothing to prove.
    Otherwise, \cref{lemma:locking_markers} implies that $\pi$ fixes every marker and maps $S_G$ onto $S_H$.
    We may assume that $\pi$ is compliant, as replacing its action on the added clique by the identity
    cannot increase the cost.
    For every $v_i \in S_G$, let $d_i \in \{-N+2,-N+4,\ldots,N\}$ be the signed \emph{displacement} of $v_i$ under $\pi$, so that
    \[
        \pi(v_i)=w_{(i+d_i)\bmod 2N}
    \]
    for each $v_i \in S_G$. For example, moving $v_i$ to the opposite end of the cycle corresponds to $d_i = N$.
    
    Note that $\ell(v_i,\pi(v_i))=|d_i|$ for each $v_i \in S_G$, so \cref{lemma:edit_lower_anchor_cost} gives a core-marker contribution of $2U$, where
    \[
        U \ceq \sum_{v_i\in S_G}|d_i|.
    \]
    Let $k$ be the number of core edges of $G'$ that $\pi$ maps onto core edges of $H'$.

    \begin{claim}
        \label{claim:edit_lower_displacement}
        We have $U \geq N+2k$.
    \end{claim}
    \begin{claimproof}
        If $d_i=N$ for some $v_i \in S_G$, then $U\geq N+(N-1)=2N-1\geq N+2k$, since every other displacement has absolute value at least $1$ and $\frac{N-1}{2} \geq k$.
        So we can assume that every $d_i$ lies in the interval $(-N, N)$.
        Since $\pi$ maps the even positions bijectively onto the odd ones, we have
        \[
            \sum_{v_i\in S_G}d_i
            \equiv \sum_{r=0}^{N-1}(2r+1)-\sum_{r=0}^{N-1}2r
            =N \pmod{2N},
        \]
        so the absolute value of this sum is at least $N$.
        
        For a core edge $\{v_i, v_{i'}\}$ preserved under $\pi$,  both the source positions and the target positions sum to $0$ modulo $2N$. Since $d_i, d_{i'} \in (-N, N)$, their sum (as integers) is zero as well, while $|d_i| + |d_{i'}| \geq 2$ since all displacements are nonzero.

        Let $S\subseteq S_G$ be the set of endpoints of the $k$ preserved edges, so $|S| = 2k$.
        Then
        \begin{align*}
            U \geq \sum_{v_i \in S} |d_i| + \sum_{v_i \in S_G \setminus S} |d_i| \geq 2k +\bigg|\sum_{v_i\in S_G\setminus S}d_i\bigg|
            = 2k +\bigg|\sum_{v_i\in S_G}d_i -\underbrace{\sum_{v_i\in S}d_i}_{=0}\bigg|
            \geq 2k + N
        \end{align*}
        where the second sum in the last expression is zero by pairing the endvertices of the $k$ preserved edges.
    \end{claimproof}

    Both cores have $(N-1)/2$ edges, of which exactly $k$ are preserved.
    Their ordered pairs therefore contribute $2(N-1-2k)$ to the cost.
    Together with the marker contribution and \cref{claim:edit_lower_displacement}, this gives
    \[
        \Cost{\edit}(\pi)
        \geq 2(N-1-2k)+2U
        \geq 4N-2. \qedhere
    \]
\end{proof}

\begin{proof}[Proof of \cref{theorem:edit_gap_four_thirds}]
    By \cref{lemma:edit_lower_matching_witness,lemma:edit_lower_bijection_cost},
    \begin{align*}
        \d{\edit}(G,H)
        \geq 4N-2
        &=\left(\frac43-\frac{10}{3(3N+1)}\right)(3N+1)\\
        &\geq \left(\frac43-\frac{10}{3(3N+1)}\right)\dOT{\edit}(G,H).
    \end{align*}
    The two distances have the same normalisation factor.
    For any $\eps>0$, choosing an odd $N$ sufficiently large so that $\frac{10}{3(3N+1)} \leq\eps$ proves the claimed bound.
\end{proof}

\subsection{The Upper Bound}
\label{appendix:gaps_edit_upper}
We denote the Frobenius inner product by $\langle \cdot, \cdot \rangle_F$.

\begin{proof}[Proof of \cref{theorem:edit_factor_two}]
    Let $\boldsymbol\pi$ be the random bijection with $\Pr{\boldsymbol\pi = \pi_i} = \lambda_i$ and write $\boldsymbol P \ceq P_{\boldsymbol\pi}$, so that $\E{\boldsymbol P} = Q$ for a Birkhoff decomposition guaranteed by \cref{equation:birkhoff_decomposition}.
    Since all entries of $A_G$ and $A_H$ are in $\{0, 1\}$, we have $|a - b| = a^2 + b^2 - 2ab$ for all entries $a$ of $A_G$ and $b$ of $A_H$.
    Expanding $\Cost{\edit}$ accordingly, the row and column sums of $Q$ reduce the first two terms to $\norm{A_G}_F^2 + \norm{A_H}_F^2$, so that
    \begin{equation}
        \label{equation:edit_cost_quadratic}
        \Cost{\edit}(Q)
        = \norm{A_G}_F^2 + \norm{A_H}_F^2
          - 2\langle A_G Q,Q A_H\rangle_F.
    \end{equation}
    We therefore have
    \begin{align*}
        \E{\Cost{\edit}(\boldsymbol\pi)} 
        ={}&\E{\norm{A_G - \boldsymbol P A_H \boldsymbol P^\top}_F^2} \\
        ={}&\E{
            \norm{A_G \boldsymbol P - \boldsymbol P A_H}_F^2
        } \tag{a}\\
        ={}&\E{
            \norm{
                (A_G \boldsymbol P - Q A_H)
                + (Q A_H - \boldsymbol P A_H)
            }_F^2
        } \\
        \leq{}&2\E{
            \norm{A_G \boldsymbol P - Q A_H}_F^2
            + \norm{Q A_H - \boldsymbol P A_H}_F^2
        } \tag{b}\\
        ={}&2\E{
            \norm{A_G \boldsymbol P - Q A_H}_F^2
            + \norm{Q A_H}_F^2
            + \norm{\boldsymbol P A_H}_F^2
            - 2\langle Q A_H,\boldsymbol P A_H\rangle_F
        } \tag{c}\\
        ={}&2\E{
            \norm{A_G \boldsymbol P - Q A_H}_F^2
            + \norm{Q A_H}_F^2
            + \norm{\boldsymbol P A_H}_F^2
            - 2\langle Q A_H,Q A_H\rangle_F
        } \tag{d}\\
        ={}&2\E{
            \norm{A_G \boldsymbol P - Q A_H}_F^2
            - \norm{Q A_H}_F^2
            + \norm{\boldsymbol P A_H}_F^2
        } \tag{e}\\
        ={}&2\E{
            \norm{A_G \boldsymbol P}_F^2
            + \norm{Q A_H}_F^2
            - 2\langle A_G \boldsymbol P,Q A_H\rangle_F
            - \norm{Q A_H}_F^2
            + \norm{\boldsymbol P A_H}_F^2
        } \tag{f}\\
        ={}&2\E{
            \norm{A_G \boldsymbol P}_F^2
            + \norm{\boldsymbol P A_H}_F^2
            - 2\langle A_G \boldsymbol P,Q A_H\rangle_F
        } \\
        ={}&2\Big(
            \norm{A_G}_F^2 + \norm{A_H}_F^2
            - 2\langle A_G Q,Q A_H\rangle_F
        \Big) \tag{g}\\
        ={}&2\Cost{\edit}(Q).
        \tag*{\cref{equation:edit_cost_quadratic}}
    \end{align*}
    Step~(a) uses permutation invariance
    of the Frobenius norm.
    Step~(b) is the squared triangle inequality
    $\norm{X+Y}_F^2\leq 2\norm{X}_F^2+2\norm{Y}_F^2$.
    Step~(c) expands the second square.
    Step~(d) uses $\E{\boldsymbol P}=Q$, linearity
    of the inner product and linearity of expectation, while step~(e) simplifies
    $\langle Q A_H,Q A_H\rangle_F=\norm{Q A_H}_F^2$.
    Step~(f) expands the first square.
    In step~(g), we use
    $\E{\boldsymbol P}=Q$ and permutation invariance of $\norm{\cdot}_F$.

    For the second assertion, let $Q$ be a minimiser of
    $\Cost{\edit}$ over $\DS(V,W)$ and sample
    $\boldsymbol\pi$ from a Birkhoff
    decomposition~\cref{equation:birkhoff_decomposition} of $Q$.
    Then
    \[
        \d{\edit}(G,H)
        \leq \E{\Cost{\edit}(\boldsymbol\pi)}
        \leq 2\Cost{\edit}(Q)
        = 2\dOT{\edit}(G,H).
        \qedhere
    \]
\end{proof}

Note that \cref{theorem:edit_factor_two} also holds for arbitrary matrices $A, B$ since the argument does not rely on any properties of the input matrices.
Hence, applying it to the threshold matrices $\indicator[A > \tau]$ and $\indicator[B > \tau]$ for a fixed decomposition $Q$ and every $\tau \in \R$ and integrating over $\tau$ yields the claim since $|a - b| = \int_{\R} |\indicator[a > \tau] - \indicator[b > \tau]| \,d\tau$.

%% file: figures/edit_lower_bound_43_table.tex
\input{figures/edit_lower_bound_43_style}
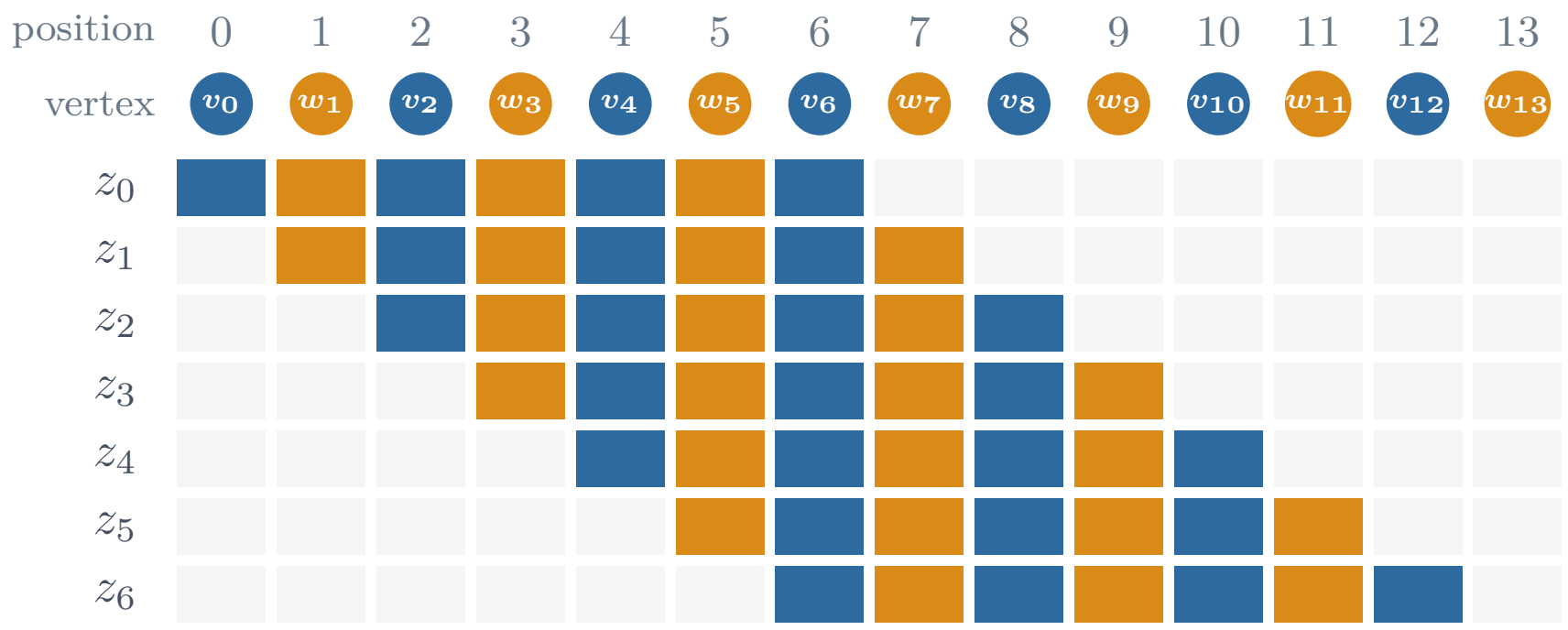
\begin{figure}[tbp]
    \centering
    \begin{tikzpicture}
        \def\lbcp{0.72}\def\lbrp{0.49}\def\lbcw{0.64}\def\lbch{0.41}
        \pgfmathsetmacro{\lbdy}{\lbch/2+0.4}
        \pgfmathsetmacro{\lbpy}{\lbdy+0.52}
        \pgfmathsetmacro{\lblx}{-\lbcw/2-0.15}
        \node[lb pos,anchor=east,font=\scriptsize] at (\lblx,\lbpy) {position};
        \node[lb pos,anchor=east,font=\scriptsize] at (\lblx,\lbdy) {vertex};
        \foreach \r in {0,...,6} {
            \node[lb label,anchor=east,font=\small] at ({\lblx},{-\r*\lbrp}) {$z_{\r}$};
        }
        \foreach \x in {0,...,13} {
            \ifodd\x
                \def\gr{H}\def\fillc{lbH}\def\vertexlabel{w}
            \else
                \def\gr{G}\def\fillc{lbG}\def\vertexlabel{v}
            \fi
            \node[lb pos,font=\footnotesize] at ({\x*\lbcp},{\lbpy}) {\x};
            \node[lb small vertex,lb \gr,minimum size=4.75mm] at ({\x*\lbcp},{\lbdy}) {{\boldmath$\vertexlabel_{\x}$}};
            \foreach \r in {0,...,6} {
                \pgfmathtruncatemacro{\inside}{(\x>=\r)&&(\x<=\r+6)}
                \ifnum\inside=1 \def\lbcell{\fillc}\else\def\lbcell{lbFaint}\fi
                \fill[\lbcell] ({\x*\lbcp-\lbcw/2},{-\r*\lbrp-\lbch/2}) rectangle ++(\lbcw,\lbch);
            }
        }
    \end{tikzpicture}
    \caption{
    Visualisation of the adjacencies between core vertices and marker vertices in $G$ and $H$ of \cref{theorem:edit_gap_four_thirds} for 
    $N=7$.
    A coloured cell indicates adjacency.
    Two columns differ in exactly as many rows as the cyclic distance of their positions, so mapping $v_i$ to $w_j$ changes the adjacency to $\ell(v_i,w_j)$ markers (\cref{lemma:edit_lower_anchor_cost}).}
    \label{figure:four_thirds_table}
\end{figure}

%% file: figures/edit_lower_bound_43_witness_edge.tex
\input{figures/edit_lower_bound_43_style}
\tikzset{
    lb split/.style={-{Stealth[length=4.4pt,width=4.2pt]},line width=1.1pt,draw=lbG},
    lb kept/.style={line width=1.7pt,line cap=round,draw=lbKept},
    lb lost/.style={line width=1.2pt,line cap=round,draw=lbLost,dash pattern=on 2.6pt off 1.8pt},
    lb weight/.style={font=\sffamily\scriptsize,inner sep=1.2pt},
}
\newcommand{\lbFourThirdsSplit}[2]{%
    \pgfmathsetmacro{\lbsa}{\lbAngle{#1}+(#2-#1)*(-5.5)}%
    \pgfmathsetmacro{\lbsb}{\lbAngle{#2}-(#2-#1)*(-5.5)}%
    \draw[lb split] ({\lbsa}:\lbR+0.4) arc[start angle=\lbsa,end angle=\lbsb,radius=\lbR+0.4];
}
\newcommand{\lbFourThirdsWitnessBackground}{%
    \draw[lb ring] (0,0) circle (\lbR);
    \foreach \a/\b in {1/13,3/11,5/9} {\draw[lb edge,draw=lbH!30] (p\a)--(p\b);}%
}
\newcommand{\lbFourThirdsWitnessBox}{\path (-3.3,-3.3) rectangle (3.3,2.5);}
\newcommand{\lbFourThirdsWitnessVertices}{%
    \foreach \u in {0,...,6} {%
        \pgfmathtruncatemacro{\g}{2*\u}%
        \pgfmathtruncatemacro{\h}{2*\u+1}%
        \node[lb slot] at (p\g) {};
        \node[lb vertex,lb H,lb four thirds vertex] at (p\h) {\lbFourThirdsVertexLabel{H}{\h}};
    }%
}
\begin{figure}[tbp]
    \centering
    \let\lbAngle\lbFourThirdsAngle
    \begin{subfigure}[b]{0.48\textwidth}
        \centering
        \begin{tikzpicture}
            \lbSetup{2.4}
            \lbFourThirdsWitnessBox
            \lbFourThirdsWitnessBackground
            \draw[lb edge,draw=lbG,dash pattern=on 3pt off 2pt] (p4)--(p10);
            \draw[lb lost] (p3)--(p9);
            \draw[lb lost] (p5)--(p11);
            \draw[lb kept] (p3)--(p11);
            \draw[lb kept] (p5)--(p9);
            \lbFourThirdsSplit{4}{3} \lbFourThirdsSplit{4}{5}
            \lbFourThirdsSplit{10}{9} \lbFourThirdsSplit{10}{11}
            \lbFourThirdsWitnessVertices
            \node[lb vertex,lb G,lb four thirds vertex] at (p4) {\lbFourThirdsVertexLabel{G}{4}};
            \node[lb vertex,lb G,lb four thirds vertex] at (p10) {\lbFourThirdsVertexLabel{G}{10}};
            \foreach \x in {3.5,4.5,9.5,10.5} {
                \node[lb weight,text=lbG] at ({\lbAngle{\x}}:\lbR+0.72) {$\frac12$};
            }
        \end{tikzpicture}
        \caption{The ordinary edge $\{v_4,v_{10}\}$.}
        \label{figure:four_thirds_edge_typical}
    \end{subfigure}
    \hfill
    \begin{subfigure}[b]{0.48\textwidth}
        \centering
        \begin{tikzpicture}
            \lbSetup{2.4}
            \lbFourThirdsWitnessBox
            \lbFourThirdsWitnessBackground
            \draw[lb edge,draw=lbG,dash pattern=on 3pt off 2pt] (p6)--(p8);
            \draw[lb lost] (p5)--(p7);
            \draw[lb lost] (p9)--(p7);
            \draw[lb kept] (p5)--(p9);
            \draw[lb lost,solid] (p7) .. controls +(-122:1.0) and +(-58:1.0) .. (p7);
            \lbFourThirdsSplit{6}{5} \lbFourThirdsSplit{6}{7}
            \lbFourThirdsSplit{8}{7} \lbFourThirdsSplit{8}{9}
            \lbFourThirdsWitnessVertices
            \node[lb vertex,lb G,lb four thirds vertex] at (p6) {\lbFourThirdsVertexLabel{G}{6}};
            \node[lb vertex,lb G,lb four thirds vertex] at (p8) {\lbFourThirdsVertexLabel{G}{8}};
            \foreach \x in {5.5,6.5,7.5,8.5} {
                \node[lb weight,text=lbG] at ({\lbAngle{\x}}:\lbR+0.72) {$\frac12$};
            }
        \end{tikzpicture}
        \caption{The exceptional edge $\{v_6,v_8\}$.}
        \label{figure:four_thirds_edge_exception}
    \end{subfigure}

    \medskip
    \begin{tikzpicture}[lb label]
        \draw[lb edge,draw=lbG,dash pattern=on 3pt off 2pt] (0,0)--(0.55,0) node[right,text=lbInk] (e) {core edge of $G$};
        \draw[lb kept] ([xshift=0.5cm] e.east)--++(0.55,0) node[right,text=lbInk] (k) {image is a core edge of $H$};
        \draw[lb lost] ([xshift=0.5cm] k.east)--++(0.55,0) node[right,text=lbInk] {image is not a core edge of $H$};
    \end{tikzpicture}
    \caption{The fractional alignment $Q=\frac12(P_{\pi_-}+P_{\pi_+})$ from \cref{lemma:edit_lower_matching_witness} on a core edge of $G$ for $N=7$.\\
    (a) The edge is preserved, with image $\{w_5,w_9\}$ or $\{w_3,w_{11}\}$, in the two cases in which one endpoint is moved forward and the other backward, so with probability $\frac12$; otherwise its image is $\{w_5,w_{11}\}$ or $\{w_3,w_9\}$.\\
    (b) Near the isolated core vertex $w_7$, only moving $v_6$ backward and $v_8$ forward preserves the edge; the other cases give $\{w_7,w_9\}$, $\{w_5,w_7\}$, and the pair $(w_7,w_7)$, drawn as a loop, which is a nonedge of $H$.
    Symmetrically, the isolated core vertex $v_0$ of $G$ sends mass to both endpoints of the core edge $\{w_{13},w_1\}$ of $H$.
    }
    \label{figure:four_thirds_edge}
\end{figure}

%% file: appendices/04_local_edit_distance_bounds.tex
\section{The Integrality Gap of the Local Edit Distance}
\label{appendix:local_edit}
\subsection{The Lower Bound}
\label{appendix:gaps_local_lower}

We start by giving the formal construction of the graphs $G_r$ and $H_r$. (See \cref{figure:local_unbounded_gap} for a visualisation of the case $r=4$.)
Fix an even integer $r \geq 2$ and let
\[
    \ell \ceq r^2 + 2r,
    \qquad
    n_X \ceq 1 + r(\ell - 1),
    \qquad
    n_M \ceq 2r^2.
\]
The $n_X$ \emph{core vertices} are
\[
    X \ceq \{u\} \cup \{ v_{i, k} \mid i \in [r], k \in [\ell - 1] \},
\]
and the $n_M$ \emph{markers} are the disjoint union $M \ceq C \cup T \cup B$ of the \emph{clock markers} $C \ceq \{c_{j, h} \mid j \in [r - 1], h \in [r]\}$, the \emph{tags} $T \ceq \{t_{i, h} \mid i, h \in [r]\}$, and the \emph{bits} $B \ceq \{b_i \mid i \in [r]\}$.
We first define a bipartite graph $G^0_r$ on $X \cup M$ by the following rules:
\begin{center}
    \begin{tabular}{l l l}
        \toprule
        Core vertex & Marker & Condition for adjacency \\
        \midrule
        $v_{i, k}$ & $c_{j, h}$ & $k \geq j(r + 1) + h$ \\
        $v_{i, k}$ & $t_{i', h}$ & $i = i'$ and $h \leq k < \ell - r + h$ \\
        $v_{i, k}$ & $b_{i'}$ & $i = i'$ and $\lfloor k/(r + 1) \rfloor$ is odd \\
        \bottomrule
    \end{tabular}
\end{center}
In particular, $u$ is an isolated vertex of $G^0_r$.
Let $H^0_r$ be obtained from $G^0_r$ by adding all edges between $u$ and the clock markers $C$.
Hence $\deg_{H^0_r}(z) = \deg_{G^0_r}(z) + \indicator[z \in C]$ for every marker $z$.

To force every bijection of small cost to fix markers and preserve core vertices,
let $G_r$ and $H_r$ be obtained from $G^0_r$ and $H^0_r$, respectively, by applying \cref{lemma:locking_markers} with their markers enumerated $z_1, \ldots, z_{n_M}$ (ordered non-decreasingly in order of their degrees in $G^0_r$), $\Delta \ceq n_X$, $\tau \ceq r$ and $\delta \ceq 1$. (Formally, \cref{lemma:locking_markers} requires both graphs to be defined on the same vertex set, which can be achieved by relabelling vertices accordingly.)
Here we use that all vertices of $G^0_r$ and $H^0_r$ have degree at most $n_X$ and that core vertices are only adjacent to markers. Moreover, only the degree of a clock marker differs (by one) between the two graphs because of $u$.

For a core vertex $x \in X$, we let $\chi_{G_r}(x), \chi_{H_r}(x) \in \{0, 1\}^M$ be its \emph{patterns}, \ie{} the indicator vectors of its neighbours in $M$ in $G_r$ and $H_r$, respectively.
By construction, we have $\chi_{G_r}(v_{i, k}) = \chi_{H_r}(v_{i, k})$ for all $i, k$, so in the following we only write $\chi(v_{i, k})$. Moreover, $\chi_{G_r}(u) = \0$ and $\chi_{H_r}(u) = \indicator_C$.
Fix a branch $i \in [r]$ and consider the sequence $\chi_{G_r}(u), \chi(v_{i, 1}), \ldots, \chi(v_{i, \ell - 1}), \chi_{H_r}(u)$ of patterns at \emph{times} $0, 1, \ldots, \ell$.
Consecutive patterns differ in exactly one coordinate:
at time $h \in [r]$, the tag $t_{i, h}$ is switched on;
at time $j(r + 1)$ for $j \in [r]$, the bit $b_i$ is flipped;
at time $j(r + 1) + h$ for $j \in [r - 1]$ and $h \in [r]$, the clock marker $c_{j, h}$ is switched on;
and at time $\ell - r + h$ for $h \in [r]$, the tag $t_{i, h}$ is switched off.
These times partition $[\ell]$, and since $r$ is even, the bit $b_i$ is $0$ at time $\ell$, consistent with $\chi_{H_r}(u) = \indicator_C$. A visualisation of this is given in \cref{figure:local_unbounded_gap_pattern}.

The following two lemmas show the desired gap.

\input{figures/fractional_local_pattern}

\begin{lemma}
    \label{lemma:local_gap_fractional}
    $\dOT{1}(G_r, H_r) = 1$.
\end{lemma}
\begin{proof}
    For $i \in [r]$, let $\sigma_i$ be the bijection from $G_r$ to $H_r$ that acts as the identity on the marker vertices and shifts the core vertices of the $i$-th branch along one step, that is $\sigma_i(u) = v_{i,1}$, $\sigma_i(v_{i,k}) = v_{i,k+1}$ for $k \in [\ell-2]$ and $\sigma_i(v_{i,\ell-1}) = u$.
    Define
    \[
        Q_r \ceq \frac{1}{r} \sum_{i = 1}^r P_{\sigma_i}.
    \]
    Intuitively, $Q_r$ spreads $u$ (in $G_r$) uniformly across the $r$ start vertices of the $r$ branches. Each branch vertex maps a $1/r$ fraction of its mass to the next vertex along the branch; the $r$ endpoints then map mass $1/r$ each into $u$ (in $H_r$).
    
    Since $Q_r$ fixes all markers and all vertices of $\mathcal{K}$, the second item of \cref{lemma:locking_markers} shows that $\Cost{1}(Q_r)$ is the same \wrt{} $(G_r, H_r)$ and \wrt{} $(G^0_r, H^0_r)$, so we may compute it in the latter pair.
    Since core vertices are only adjacent to markers, the local cost of a supported pair $(x, y)$ of core vertices is the Hamming distance of $\chi_{G_r}(x)$ and $\chi_{H_r}(y)$.
    The supported pairs of core vertices are either of the form $(v_{i, k}, v_{i, k})$, of cost $0$, or consecutive vertices along a branch, of cost $1$.
    For a marker $z$, the local cost $\LocalCost{Q_r}(z, z)$ is the total $Q_r$-mass of supported core pairs whose patterns differ at $z$, where every pair of consecutive vertices along a branch has mass $\frac1r$.
    Each clock marker changes once on each of the $r$ branches, each bit changes $r$ times on its own branch, and each tag changes twice on its own branch.
    Hence the local costs at clock markers, tags, and bits are $1$, $\frac2r$, and $1$, respectively.
    Thus $\Cost{1}(Q_r) \leq 1$. Equality follows from \cref{lemma:local_structure} below since $G_r \not\cong H_r$.
\end{proof}

\begin{lemma}
    \label{lemma:local_gap_integral}
    $\d{1}(G_r, H_r) \geq r$.
\end{lemma}
\begin{proof}
    Suppose that $\pi$ is a bijection with $\Cost{1}(\pi) < r$.
    By construction and \cref{lemma:locking_markers} with $\tau \ceq r$, $\pi$ fixes all markers and permutes $X$.
    Hence, for every core vertex $x \in X$, $\LocalCost{\pi}(x, \pi(x))$ is the Hamming distance of the patterns $\chi_{G_r}(x)$ and $\chi_{H_r}(\pi(x))$, which is less than $r$.

    Starting at $u$ in $G_r$, follow $\pi$ to $\pi(u)$ in $H_r$, continue from $\pi(u)$ in $G_r$, and so on, until $\pi^s(u) = u$ is reached in $H_r$. Consider the resulting sequence of patterns $x_0 \ceq \chi_{G_r}(u)$, $x_\iota \ceq \chi(\pi^\iota(u))$ for $0 < \iota < s$, and $x_s \ceq \chi_{H_r}(u)$.
    By assumption, consecutive patterns in this sequence have Hamming distance less than $r$.
    For a pattern $x$, let $c(x)$ be the number of clock markers in $x$, its \emph{clock level}.
    The clock markers of every pattern form a prefix of the order $c_{1, 1}, \ldots, c_{1, r}, c_{2, 1}, \ldots, c_{r - 1, r}$, so consecutive clock levels in the sequence differ by less than $r$, and we have $c(x_0) = 0$ and $c(x_s) = r(r - 1)$.

    Let $\beta$ be the first index with $c(x_\beta) = r(r - 1)$ and let $\alpha < \beta$ be the last index before $\beta$ with $c(x_\alpha) = 0$.
    Every pattern $x_\iota$ with $\alpha < \iota < \beta$ has clock level strictly between $0$ and $r(r - 1)$, so it is the pattern of a vertex $v_{i, k}$ with $r + 2 \leq k \leq r^2 + r - 2$, at which all $r$ tags of branch $i$ are switched on and all other tags are switched off.
    Thus two such patterns on different branches have Hamming distance at least $2r$, and hence all patterns $x_\iota$ with $\alpha < \iota < \beta$ belong to the same branch $i$.

    For $j \in [r - 1]$, let $\iota_j$ be the first index in $(\alpha, \beta]$ with $c(x_{\iota_j}) \geq jr$.
    As consecutive clock levels differ by at most $r - 1$, we have $c(x_{\iota_j - 1}) \in ((j - 1)r, jr)$, so $x_{\iota_j - 1}$ is the pattern of a vertex $v_{i, k}$ with $k = j(r + 1) + h'$ for some $h' \in [r - 1]$, and thus its value at the bit $b_i$ is $j \bmod 2$.
    Since $\iota_1 < \iota_2 < \cdots < \iota_{r - 1}$, and since $x_0$ and $x_s$ both have value $0$ at $b_i$, the sequence of values at $b_i$ contains the subsequence $(0, 1, 0, 1, \ldots, 1, 0)$ of length $r+1$ and $r$ alternations.
    Thus at least $r$ distinct core vertices $x$ satisfy $A_{G_r}(b_i, x) \neq A_{H_r}(b_i, \pi(x))$.
    Hence, with $b_i = \pi(b_i)$, we have $\LocalCost{\pi}(b_i, \pi(b_i)) \geq r$, a contradiction.
\end{proof}

\begin{proof}[Proof of \cref{theorem:local_unbounded_gap}]
    By \cref{lemma:local_gap_fractional}, $\dOT{1}(G_r, H_r) = 1$.
    The bijection $\sigma_1$ fixes all markers and all vertices of $\mathcal{K}$, so by the second item of \cref{lemma:locking_markers} its local costs can be computed in $(G^0_r, H^0_r)$ as in the proof of \cref{lemma:local_gap_fractional} yielding $\d{1}(G_r, H_r) = r$ by \cref{lemma:local_gap_integral} and since $\LocalCost{\sigma_1}(b_1, b_1) = r$.
    Finally, $|G^0_r| = \Theta(r^3)$, and \cref{lemma:locking_markers} only adds $\O(r^3)$ vertices, so $|V(G_r)| = |V(H_r)| = \Theta(r^3)$.
\end{proof}

\subsection{The Upper Bound}
\label{appendix:gaps_local_upper}
Throughout this subsection, let $L \ceq \dOT{1}(G, H)$ and fix an optimal $Q \in \DS(V, W)$, \ie{} $\LocalCost{Q}(v, w) \leq L$ for all $(v, w) \in \supp(Q)$.
The pseudometric $\rho_H(w, w') \ceq |N_H(w) \symdiff N_H(w')|$ on $W$ measures how much the neighbourhood changes if $w$ is replaced by $w'$.
By \cref{equation:birkhoff_decomposition}, $\supp(Q)$ contains the graph of a bijection $\sigma \in \Bij(V, W)$, that is,
\begin{equation}
    \label{equation:local_reference_bijection}
    Q(v, \sigma(v)) > 0 \quad\text{and}\quad \LocalCost{Q}(v, \sigma(v)) \leq L \qquad\text{for all } v \in V.
\end{equation}
We fix $\sigma$ and use it as a reference: $\sigma(v)$ is the vertex of $H$ whose neighbourhood we compare with the neighbourhood of $v$ in $G$.
\begin{lemma}
    \label{lemma:local_structure}
    If $Q(v, w) > 0$, then $\rho_H(\sigma(v), w) \leq 2L$.
    Moreover, if $L < 1$, then $\sigma$ is an isomorphism from $G$ to $H$. In particular, if $G$ and $H$ are not isomorphic, then $L \geq 1$.
\end{lemma}
\begin{proof}
    Let $v \in V$ and $w, w' \in W$.
    Using the column sums of $Q$ and the triangle inequality, we obtain
    \begin{align*}
        \rho_H(w, w')
        &= \sum_{v' \in V} \sum_{w'' \in W} Q(v', w'') \big|A_H(w, w'') - A_H(w', w'')\big| \\
        &\leq \sum_{v' \in V} \sum_{w'' \in W} Q(v', w'') \Big( \big|A_G(v, v') - A_H(w, w'')\big| + \big|A_G(v, v') - A_H(w', w'')\big| \Big) \\
        &= \LocalCost{Q}(v, w) + \LocalCost{Q}(v, w').
    \end{align*}
    For $w \ceq \sigma(v)$ and $(v, w') \in \supp(Q)$, this is at most $2L$ by \cref{equation:local_reference_bijection}.

    Now suppose that $L < 1$.
    By \cref{claim:local_cost_degree_bound}, $\deg_G(v) = \deg_H(w)$ for every $(v, w) \in \supp(Q)$.
    Applying this to $(v, \sigma(v))$ and $(v, w)$ shows that $\deg_H(\sigma(v)) = \deg_H(w)$ for all $(v, w) \in \supp(Q)$, hence $\rho_H(\sigma(v), w)$ is even, and as $\rho_H(\sigma(v), w) \leq 2L < 2$, it is $0$.
    Thus $A_H(w', w) = A_H(w', \sigma(v))$ for all $w' \in W$ and $(v, w) \in \supp(Q)$.
    Hence, for every $v' \in V$, using the row sums of $Q$,
    \[
        \LocalCost{Q}(v', \sigma(v'))
        = \sum_{v \in V} \sum_{w \in W} Q(v, w) \big|A_G(v', v) - A_H(\sigma(v'), w)\big|
        = \sum_{v \in V} \big|A_G(v', v) - A_H(\sigma(v'), \sigma(v))\big|.
    \]
    Since the right-hand side is an integer and strictly less than one, it is $0$.
    It follows that $A_G(v', v) = A_H(\sigma(v'), \sigma(v))$ for all $v, v' \in V$, \ie{} $\sigma$ is an isomorphism from $G$ to $H$.
\end{proof}

We view $\supp(Q) \subseteq V \times W$ as a bipartite graph with sides $V$ and $W$, and $Q$ as a fractional perfect matching in it.
For a matching $M \subseteq \supp(Q)$ and $v \in V$, let
\[
    \Phi_v(M) \ceq \sum_{(v', w') \in M} \big|A_G(v, v') - A_H(\sigma(v), w')\big|
\]
be the local cost of $v$ under $\sigma$ relative to $M$.
Note that $\Phi_v(M)$ is a linear statistic of $M$ with coefficients in $\{0, 1\}$, and that
\[
    \sum_{v' \in V} \sum_{w' \in W} \big|A_G(v, v') - A_H(\sigma(v), w')\big|Q(v', w') = \LocalCost{Q}(v, \sigma(v)) \leq L.
\]
We use the following special case of the dependent rounding scheme of Chekuri, Vondrák and Zenklusen.
\begin{lemma}[{\cite[Theorem~1.1]{chekuri2011multibudgeted}}]
    \label{lemma:cvz_rounding}
    Let $\q$ be a point in the bipartite matching polytope and $\gamma \in (0, \frac12]$.
    Then there is a random matching $\boldsymbol M$ with $\E{\indicator_{\boldsymbol M}} = (1 - \gamma)\q$ such that, for every linear statistic $Z \ceq \sum_e \alpha_e \indicator[e \in \boldsymbol M]$ with $\alpha_e \in [0, 1]$, every $\mu \geq \E{Z}$ with $\mu > 0$, and every $\tau \geq 2$,
    \begin{equation}
        \label{equation:cvz_concentration}
        \Pr{Z \geq \tau\mu} \leq \exp*{-\frac{\mu\gamma(2\tau - 3)}{20}}.
    \end{equation}
\end{lemma}

\begin{lemma}
    \label{lemma:local_rounding}
    Suppose that $L > 0$.
    For every $\gamma \in (0, \frac12]$, there is a matching $M \subseteq \supp(Q)$ such that
    \[
        n - |M| \leq 2\gamma n
        \qquad\text{and}\qquad
        \Phi_v(M) \leq 2L + \frac{10 \ln(4n)}{\gamma} \quad\text{for all } v \in V.
    \]
\end{lemma}
\begin{proof}
    We apply \cref{lemma:cvz_rounding} to $\q \ceq Q$, which is a convex combination of perfect matchings in $\supp(Q)$, and let $\boldsymbol M$ be the resulting random matching.
    For every $v \in V$, we have $\E{\Phi_v(\boldsymbol M)} = (1 - \gamma)\LocalCost{Q}(v, \sigma(v)) \leq L$.
    For $Z \ceq \Phi_v(\boldsymbol M)$, $\mu \ceq L$, and $\tau \ceq 2 + 10\ln(4n)/(\gamma L)$, \cref{equation:cvz_concentration} yields
    \[
        \Pr{\Phi_v(\boldsymbol M) \geq 2L + \frac{10\ln(4n)}{\gamma}} \leq \frac{\exp{-\gamma L/20}}{4n} \leq \frac{1}{4n}.
    \]
    By the union bound, the probability that this condition fails for some $v \in V$ is at most $\frac14$.
    Moreover, $\E{|\boldsymbol M|} = (1 - \gamma)n$, so Markov's inequality applied to $n - |\boldsymbol M|$ yields $\Pr{n - |\boldsymbol M| > 2\gamma n} \leq \frac12$.
    Hence, with positive probability, neither event occurs. This yields the desired matching.
\end{proof}

\begin{proposition}
    \label{proposition:local_completion}
    Suppose that $L > 0$.
    For all $\gamma \in (0, \frac12]$ and $\ell \in \Npos$,
    \[
        \d{1}(G, H) \leq 2L + \frac{10\ln(4n)}{\gamma} + 2\gamma n + \frac{n}{\ell} + 2\ell L.
    \]
\end{proposition}
\begin{proof}
    Let $M$ be the matching of \cref{lemma:local_rounding} and $m \ceq n - |M|$, so that exactly $m$ vertices of $V$ and $m$ vertices of $W$ are unmatched.
    To turn $M$ into a bijection, we use $\sigma$ to go back from $W$ to $V$:
    let $\mu$ be the partial injective map on $V$ given by $\mu(v) \ceq \sigma^{-1}(w)$ for every $(v, w) \in M$.
    In the directed graph on $V$ with the edges $v \to \mu(v)$, every vertex has in- and outdegree at most $1$.
    Thus it decomposes into directed cycles (including loops) and exactly $m$ directed paths (including isolated vertices), since every path ends in exactly one vertex on which $\mu$ is undefined.
    We keep all cycles, cut every path into consecutive blocks of $\ell$ vertices (plus a final shorter block), discard all edges between blocks, and close every block $v_1 \to \cdots \to v_k$ by adding the \emph{closing edge} $v_k \to v_1$, where a single vertex is closed by a loop.
    The resulting directed cycles define a permutation $\tau$ of $V$, and we let $\pi \ceq \sigma \circ \tau \in \Bij(V, W)$.
    Hence $\pi(v) = w$ if $(v, w) \in M$, so all edges of the form $v \to \mu(v)$ are retained, and $\pi(v_k) = \sigma(v_1)$ for a closing edge $v_k \to v_1$.
    The number of closing edges is at most $\sum_j \lceil p_j/\ell \rceil \leq m + n/\ell$, where $p_1, \ldots, p_m$ are the numbers of vertices of the paths.

    We claim that $\rho_H(\sigma(v), \pi(v)) \leq 2\ell L$ for all $v \in V$.
    If the edge $v \to \mu(v)$ is retained, then $(v, \pi(v)) \in M \subseteq \supp(Q)$, and \cref{lemma:local_structure} yields $\rho_H(\sigma(v), \pi(v)) \leq 2L$.
    For a closing edge $v_k \to v_1$ of a block $v_1 \to \cdots \to v_k$, we have $(v_j, \sigma(v_{j + 1})) \in M$ for all $j < k$, so \cref{lemma:local_structure} yields $\rho_H(\sigma(v_j), \sigma(v_{j + 1})) \leq 2L$, and the triangle inequality then gives
    \[
        \rho_H(\sigma(v_k), \pi(v_k)) = \rho_H(\sigma(v_k), \sigma(v_1)) \leq \sum_{j=1}^{k-1} \rho_H(\sigma(v_j), \sigma(v_{j + 1})) \leq 2(k - 1)L.
    \]
    Finally, for every $v \in V$, inserting $A_H(\sigma(v), \pi(v'))$ and using the triangle inequality, we obtain
    \begin{align*}
        \LocalCost{P_\pi}(v, \pi(v))
        &= \sum_{v' \in V} \big|A_G(v, v') - A_H(\pi(v), \pi(v'))\big| \\
        &\leq \sum_{v' \in V} \big|A_G(v, v') - A_H(\sigma(v), \pi(v'))\big| + \sum_{v' \in V} \big|A_H(\sigma(v), \pi(v')) - A_H(\pi(v), \pi(v'))\big|,
    \end{align*}
    where the second sum equals $\rho_H(\sigma(v), \pi(v)) \leq 2\ell L$ since $\pi$ is a bijection.
    In the first sum, the summands for the $v'$ whose edge $v' \to \mu(v')$ is retained satisfy $(v', \pi(v')) \in M$, so their sum is at most $\Phi_v(M)$.
    All other $v'$ are the starting points of closing edges, hence there are at most $m + n/\ell$ such vertices, with their respective summands being at most $1$.
    This yields $\LocalCost{P_\pi}(v, \pi(v)) \leq \Phi_v(M) + m + n/\ell + 2\ell L$ for all $v \in V$, and therefore $\d{1}(G, H) \leq \Cost{1}(\pi) = \max_{v \in V} \LocalCost{P_\pi}(v, \pi(v))$ is bounded by the same quantity.
    \Cref{lemma:local_rounding} completes the proof.
\end{proof}

\begin{proof}[Proof of \cref{theorem:local_nonlinear_bound}]
    Let $f \ceq \ndOT{1}(G, H) = L/n$ and $b \ceq \nd{1}(G, H)$.
    If $L = 0$, then $G$ and $H$ are isomorphic by \cref{lemma:local_structure}, so $b = 0$.
    Hence, suppose that $L > 0$ and let $\ell \ceq \lceil \sqrt{n/(2L)} \rceil$ and $\gamma^\ast \ceq \sqrt{5\ln(4n)/n}$.
    Then $\frac{n}{\ell} + 2\ell L \leq 2\sqrt{2nL} + 2L$.
    If $\gamma^\ast \leq \frac12$, then \cref{proposition:local_completion} with $\gamma \ceq \gamma^\ast$ yields
    \begin{equation}
        \label{equation:local_unnormalised_bound}
        \d{1}(G, H) \leq 4L + 2\sqrt{2nL} + 4\sqrt{5n\ln(4n)}.
    \end{equation}
    If $\gamma^\ast > \frac12$, then $4\sqrt{5n\ln(4n)} = 4n\gamma^\ast > n \geq \d{1}(G, H)$, so \cref{equation:local_unnormalised_bound} holds as well.
    As $L > 0$, the graphs are not isomorphic, so $L \geq 1$ by \cref{lemma:local_structure}, \ie{} $n \geq 1/f \geq 1$.
    Since $x \mapsto \ln(4x)/x$ is decreasing for $x \geq 1$, we obtain $\ln(4n)/n \leq \frac{\ln(4/f)}{1/f} = f\ln(4/f)$.
    Dividing \cref{equation:local_unnormalised_bound} by $n$ yields
    \[
        b \leq 4f + 2\sqrt{2f} + 4\sqrt{5}\sqrt{f \ln(4/f)}.
    \]
    For $0 < f \leq 1$ and $h \ceq \ln(4/f) \geq \ln 4$, we have
    \[
        \frac{b}{\sqrt{fh}} \leq \frac{4f + 2\sqrt{2f} + 4\sqrt{5fh}}{\sqrt{fh}} = 4\sqrt{\frac{f}{h}} + \frac{2\sqrt{2}}{\sqrt{h}} + 4\sqrt{5} \leq \frac{4 + 2\sqrt{2}}{\sqrt{\ln 4}} + 4\sqrt{5} < 15.
    \]
    Multiplying both sides by $\sqrt{fh}$ proves the theorem.
\end{proof}

%% file: figures/fractional_local_pattern.tex
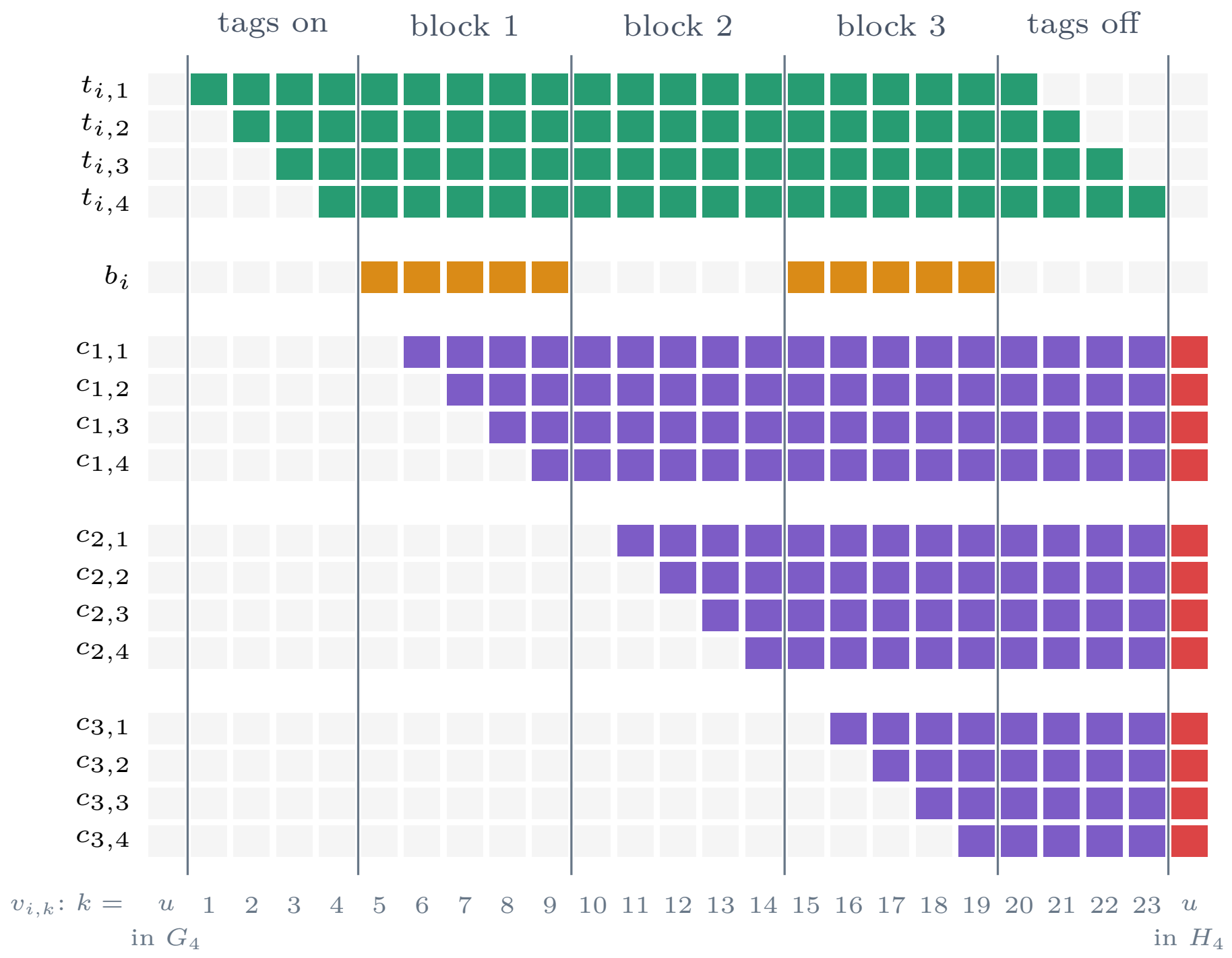
\begin{figure}[t]
    \centering
    \resizebox{0.723\textwidth}{!}{\input{tikz/fractional_local_pattern}}
    \caption{Visualisation of the adjacency patterns $\chi_{G_4}(u), \chi(v_{i, 1}), \ldots, \chi(v_{i, 23}), \chi_{H_4}(u)$ obtained by following the cyclic permutation $\sigma_i$ of \cref{lemma:local_gap_fractional} in $G_4$ and $H_4$.}
    \label{figure:local_unbounded_gap_pattern}
\end{figure}

%% file: tikz/fractional_local_pattern.tex
\begingroup
\input{figures/palette}%
\colorlet{HfourData}{figBlue}%
\colorlet{HfourClock}{figPurple}%
\colorlet{HfourTag}{figGreen}%
\colorlet{HfourBit}{figOrange}%
\colorlet{HfourAdded}{figRed}%
\colorlet{HfourText}{figInk}%
\colorlet{HfourMuted}{figMuted}%
\colorlet{HfourEmpty}{figEmpty}

\begin{tikzpicture}[
  x=0.34cm, y=-0.3cm,
  every node/.style={inner sep=0pt, outer sep=0pt},
  h4 label/.style={font=\fontsize{6}{7}\selectfont, anchor=east},
  h4 col label/.style={font=\fontsize{5.5}{6}\selectfont, text=HfourMuted},
  h4 block label/.style={font=\fontsize{7}{9}\selectfont, text=HfourText},
]
  \def\cell#1#2#3{\fill[#3] ({#1+0.08},{#2+0.08}) rectangle ({#1+0.92},{#2+0.92});}

  \foreach \k in {0,...,24} {
    \foreach \r in {0,1,2,3,5,7,8,9,10,12,13,14,15,17,18,19,20} {
      \fill[HfourEmpty] ({\k+0.08},{\r+0.08}) rectangle ({\k+0.92},{\r+0.92});
    }
  }

  \foreach \k in {0,...,24} {
    \foreach \h in {1,...,4} {
      \pgfmathtruncatemacro{\on}{(\k>=\h) && (\k<20+\h) && (\k>0) && (\k<24)}
      \ifnum\on=1 \cell{\k}{\h-1}{HfourTag} \fi
    }
    \pgfmathtruncatemacro{\on}{((\k>=5) && (\k<=9)) || ((\k>=15) && (\k<=19))}
    \ifnum\on=1 \cell{\k}{5}{HfourBit} \fi
    \foreach \j in {1,...,3} {
      \foreach \h in {1,...,4} {
        \pgfmathtruncatemacro{\on}{\k>=5*\j+\h}
        \pgfmathtruncatemacro{\r}{7+(\j-1)*5+\h-1}
        \ifnum\on=1
          \ifnum\k=24 \cell{\k}{\r}{HfourAdded} \else \cell{\k}{\r}{HfourClock} \fi
        \fi
      }
    }
  }

  \foreach \x in {1,5,10,15,20,24} {
    \draw[HfourMuted, line width=0.5pt] (\x,-0.4) -- (\x,21.4);
  }
  \foreach \a/\b/\name in {1/5/tags on, 5/10/block $1$, 10/15/block $2$, 15/20/block $3$, 20/24/tags off} {
    \node[h4 block label] at ({(\a+\b)/2},-1.2) {\name};
  }

  \foreach \h in {1,...,4} {\node[h4 label] at (-0.3,{\h-0.5}) {$t_{i,\h}$};}
  \node[h4 label] at (-0.3,5.5) {$b_i$};
  \foreach \j in {1,...,3} {
    \foreach \h in {1,...,4} {
      \node[h4 label] at (-0.3,{7+(\j-1)*5+\h-0.5}) {$c_{\j,\h}$};
    }
  }
  \node[h4 col label] at (0.5,22.2) {$u$};
  \node[h4 col label] at (0.5,23.1) {in $G_4$};
  \node[h4 col label] at (24.5,22.2) {$u$};
  \node[h4 col label] at (24.5,23.1) {in $H_4$};
  \foreach \k in {1,...,23} {
    \node[h4 col label] at ({\k+0.5},22.2) {\k};
  }
  \node[h4 col label, anchor=east] at (-0.3,22.2) {$v_{i,k}$:\ $k={}$};
\end{tikzpicture}%
\endgroup

%% file: appendices/05_cut_distance_bounds.tex
\section{The Integrality Gap of the Cut Distance}
\label{appendix:cut_factor_two_gap}
\subsection{The Lower Bound}

\noindent We give a formal construction of the graphs depicted in \cref{figure:cut_factor_two_graphs}. The graph $G$ has vertex set $V$ and the graph $H$ has vertex set $W$, each consisting of a core of $10$ vertices in five pairs, and $135$ markers.
We denote the core vertices of $G$ by $v_i^{\sigma}$ and the core vertices of $H$ by $w_i^{\sigma}$ with $i \in \Z_5$ and $\sigma \in \{+1, -1\}$, and write $C_V \subseteq V$ and $C_W \subseteq W$ for the two cores.
Core vertices are connected by edges based on the following matrix
\[
    Z \ceq \begin{pmatrix}
         0 &  +1 & -1 & -1 &  +1\\
         +1 &  0 &  +1 & -1 & -1\\
        -1 &  +1 &  0 &  +1 & -1\\
        -1 & -1 &  +1 &  0 &  +1\\
         +1 & -1 & -1 &  +1 &  0
    \end{pmatrix} \in \{-1, 0, +1\}^{\Z_5 \times \Z_5},
\]
that is, $Z_{ij} = 1$ for neighbouring corners of a pentagon ($j = i \pm 1$) and $Z_{ij} = -1$ for corners two apart ($j = i \pm 2$) where addition is modulo 5.
For $i \neq j$, we let
\[
    v_i^\sigma v_j^\tau \in E(G) \iff \sigma\tau = Z_{ij},
    \qquad
    w_i^\sigma w_j^\tau \in E(H) \iff \sigma\tau = -Z_{ij}.
\]
So in $G$, along a side of the pentagon, $+$ joins $+$ and $-$ joins $-$, while along a diagonal, $+$ joins $-$, while in $H$ it is exactly opposite (compare \cref{figure:cut_factor_two_graphs}).
Thus, each core vertex has exactly four core neighbours.
For each $i \in \Z_5$, where we identify $\Z_5$ with $\{0, \ldots, 4\}$, the vertices $v_i^+$ and $v_i^-$ are connected to a set $L_i \subseteq V$ of $9(i+1)$ markers, and likewise $w_i^+$ and $w_i^-$ are connected to a set $L_i' \subseteq W$ of $9(i+1)$ markers. 
As a result, the pairs have degrees $13, 22, 31, 40, 49$ respectively and hence moving a core vertex to a different pair yields a degree mismatch of at least $9$.
Moreover, as markers are connected identically to core vertices of the same pair, mixing within a pair of core vertices incurs no cost (relative to the marker edges).

\begin{lemma}
    \label{lemma:cut_bijective}
    $\d{\cut}(G, H) = 8$.
\end{lemma}
\begin{proof}
Let $\pi\in\Bij(V,W)$ and $A := A^\pi_G - A_H$, so that
$\Cost{\cut}(\pi) = \norm{A}_\cut$. Taking $S := \{\pi(v)\}$ and $T := W$
shows $\Cost{\cut}(\pi) \geq |\deg_G(v) - \deg_H(\pi(v))|$. Since the pairs have
degrees $13, 22, 31, 40, 49$ and the markers have degree $2$, we get
$\Cost{\cut}(\pi) \geq 9$ unless $\pi$ maps every pair $\{v_i^+, v_i^-\}$ onto
$\{w_i^+, w_i^-\}$. We may therefore assume that $\pi(v_i^\sigma) =
w_i^{r_i\sigma}$ for signs $r_i \in \{-1,+1\}$.
Hence, each pair is either kept as-is or swapped under $\pi$.
Let
$K := \{i \in \Z_5 \mid r_i = +1\}$ be the set of indices of pairs that are kept.

For $i \neq j$ we have $Z_{ij} \in \{-1,+1\}$, so
$A_G(v_i^\sigma, v_j^\tau) = \frac12(1 + Z_{ij}\sigma\tau)$ and
$A_H(w_i^\sigma, w_j^\tau) = \frac12(1 - Z_{ij}\sigma\tau)$. As
$\pi^{-1}(w_i^\sigma) = v_i^{r_i\sigma}$, it follows that
\[
    A(w_i^\sigma, w_j^\tau) = \tfrac12 Z_{ij}\,\sigma\tau\,(1 + r_i r_j)
  = M_{ij}\,\sigma\tau,
\]
where $M := Z[K] \oplus Z[\overline{K}]$. This also holds for $i = j$, since
there are no edges within a pair and $Z_{ii} = 0$. Now let $S, T \subseteq C_W$,
and define vectors $\x,\y \in \{-1,0,+1\}^{\Z_5}$ with entries $x_i := \sum_\sigma \sigma\ \indicator[w_i^\sigma \in S]$ and
$y_j := \sum_\tau \tau\indicator[w_j^\tau \in T]$ for $i, j \in \Z_5$. Then
\[
  \sum_{s \in S,\, t \in T} A(s,t)
  = \sum_{i,j \in \Z_5} M_{ij}
    \bigg(\sum_{\sigma \colon w_i^\sigma \in S} \sigma\bigg)
    \bigg(\sum_{\tau \colon w_j^\tau \in T} \tau\bigg)
  = \x^\top M \y,
\]
and every pair $\x, \y \in \{-1,0,+1\}^{\mathbb{Z}_5}$ arises in this way. In
particular, $S$ and $T$ realise all pairs of sign vectors. 
Hence $\Cost{\cut}(\pi) \geq \norm{M}_{\infty \to 1}$ and since $M$ is block diagonal we can compute its norm as the sum of the norms of its blocks.
The four remaining cases (up to symmetries) are $|K| = 0$, $|K| = 1$ and $|K|=2$ (once with two neighbouring pairs in $K$ and once two non-adjacent pairs) and one easily checks that in each case $\norm{M}_{\infty \to 1} \geq 8$.
To show that equality holds, let $\id \in \Bij(V, W)$ denote the natural bijection between the two graphs with $v_i^\sigma \mapsto
w_i^\sigma$ and $L_i \to L_i'$ for each $i \in \Z_5$.
For this alignment, we have $M = Z$ and $A$ is zero outside
$C_W \times C_W$ because $G$ and $H$ agree outside their cores. Every cut value
is thus exactly of the form $\x^\top Z \y$ with $\x, \y \in \{-1,0,+1\}^{\Z_5}$. The maximum of this bilinear expression is attained at sign vectors, and one can show that $\Cost{\cut}(\id) = \norm{Z}_{\infty\to1} = 8$.
\end{proof}

\begin{lemma}
    \label{lemma:cut_fractional}
    $\dOT{\cut}(G, H) \leq 4$.
\end{lemma}
\begin{proof}
    Let $Q \in \DS(V, W)$ map each marker set $L_i$ bijectively onto $L_i'$ and match each pair $\{v_i^+, v_i^-\}$ uniformly to $\{w_i^+, w_i^-\}$, that is
    \[
        Q(v_i^\sigma, w_i^\tau) \ceq \tfrac12 \qquad \text{for all } i \in \Z_5 \text{ and } \sigma, \tau \in \{+1,-1\}.
    \]
    Since each marker is adjacent to both vertices of its pair, every discrepancy involving a marker is zero.
    The remaining $20$ atoms $(v_i^\sigma, w_i^\tau)$ in $\supp(Q)$ each carry mass $\frac12$, with discrepancy
    \[
        \Disc(v_i^\sigma, w_i^\tau, v_j^{\sigma'}, w_j^{\tau'}) = \tfrac12 Z_{ij} (\sigma\sigma' + \tau\tau').
    \]
    Call an atom \emph{straight} if $\sigma = \tau$ and \emph{crossed} if $\sigma = -\tau$.
    Between a straight and a crossed atom, the discrepancy is $0$, while between two atoms of the same kind it is $Z_{ij}\sigma\sigma'$.
    As a result, the discrepancy decomposes into two disjoint copies of the pentagon pattern, each with weight of $\frac{1}{2}$.
    In each copy, a cut has value $\frac{1}{4}\x^\top Z\y$ with $\x, \y \in \{-1, 0, +1\}^{\Z_5}$, hence at most $\frac14 \cdot 8 = 2$ in absolute value 
    since $\Cost{\cut}(\id) = 8$.
    Combining both copies then shows $\dOT{\cut}(G, H) \leq \Cost{\cut}(Q) \leq 4$.
\end{proof}

\begin{proof}[Proof of \cref{theorem:cut_factor_two_gap}]
    This follows from \cref{lemma:cut_bijective,lemma:cut_fractional}.
\end{proof}

\subsection{The Upper Bound}
\label{appendix:cut_almost_linear}

Throughout this section, $G$ and $H$ are graphs of the same order $n\ge 2$. For simplicity, we assume that $V(G)=V(H)=[n]$. The adjacency matrices of $G,H$ are $A,B\in\{0,1\}^{n\times n}$. 
For $Q\in\DS(n)$, we let
\begin{align}
    \label{eq:epsQ1}
    \cst{Q} \, \coloneqq& \mathmakebox[55pt][c]{\max_{S, T \subseteq [n]^2}} \frac{1}{n^2}\Bigl| \sum_{\substack{(i, j) \in S\\ (k, \ell) \in T}} Q_{ij} Q_{k\ell} \bigl( A_{ik} - B_{j\ell} \bigr) \Bigr|\\
    \label{eq:epsQ2}
    =& \mathmakebox[55pt][c]{\max_{f, g:[n]^2\to[0,1]}} \frac{1}{n^2}\Bigl| \sum_{i, j, k, \ell \in [n]} f(i, j)\, Q_{ij} Q_{k\ell}\bigl( A_{ik} - B_{j\ell} \bigr)\, g(k, \ell) \Bigr|,
\end{align}
where equality holds because the expression is bilinear in $(f, g)$, so its absolute value is maximised at vertices of the cube $[0, 1]^{[n]^2}$, \ie{} at indicator functions of sets $S,T\subseteq[n]^2$.
Note that
\[
    \ndOT{\cut}(G, H) = \min_{Q \in \DS(n)} \cst{Q}
    \qquad\text{and}\qquad
    \nd{\cut}(G, H) = \min_{P \in \Perm(n)} \cst{P}.
\]
As explained in the proof overview, we prove two upper bounds, a multiplicative and an additive one, using completely different techniques. We start with the multiplicative upper bound.

\subsubsection{Multiplicative Rounding via Thin Matchings}
\label{appendix:cutmid_thin_v2}

Let $\alpha>0$. A permutation matrix $P\in\Perm(n)$ is \emph{$\alpha$-thin} \wrt{} $Q \in \DS(n)$ if 
\begin{equation}
    \label{equation:cutmid_v2_thin}
    \sum_{(i, j) \in (X\times \overline{Y})\cup(\overline{X} \times Y)} P_{ij} \leq \alpha \sum_{(i, j) \in (X\times \overline{Y})\cup(\overline{X} \times Y)} Q_{ij}
    \qquad\text{for all } X,Y\subseteq[n].
\end{equation}
Here we write $\overline X$, $\overline Y$ to denote the complements $[n]\setminus X,[n]\setminus Y$ of the sets.
If we view $P$ as describing a matching of the complete bipartite graph $K_{n,n}$ and $Q$ as a fractional matching, then for all cuts $(X\cup Y,\overline X\cup \overline Y)$ the number of matching edges across the cut is bounded from above by $\alpha$ times the total weight of the fractional edges across the cut.
Haqi and Oveis Gharan showed that thin permutations exist where $\alpha$ only grows polylogarithmically in $n$.
\begin{theorem}[{\cite[Theorem~1.4]{haqi2026thin}}]
    \label{theorem:cutmid_v2_haqi}
    There is an absolute constant $C_{\mathrm{thin}} > 0$ such that every $Q \in \DS(n)$ admits an $\alpha$-thin permutation matrix \wrt{} $Q$ with $\alpha \leq C_{\mathrm{thin}} \ln(2n) \ln\ln(4n)$.
\end{theorem}

The following lemma turns thinness (\cref{equation:cutmid_v2_thin}) into a cost bound for line metrics based on a level-set decomposition.
A similar but different argument was used in the thin-tree approach to ATSP \cite{asadpour2017atsp}.

\begin{lemma}
    \label[lemma]{lemma:cutmid_v2_transport}
    Let $\pi$ be a permutation of $[n]$ whose permutation matrix $P = P_\pi$ is $\alpha$-thin \wrt{} $Q \in \DS(n)$. Then, for all $\x, \y \in \R^n$, we have
    \[
        \sum_{i \in [n]} \bigl| x_i - y_{\pi(i)} \bigr| \leq \alpha \sum_{i, j \in [n]} Q_{ij} \bigl| x_i - y_j \bigr|.
    \]
\end{lemma}
\begin{proof}
    For $s \in \R$, consider the level sets $X_s \ceq \{i \in [n] : x_i > s\}$ and $Y_s \ceq \{j \in [n] : y_j > s\}$.
    For $i, j \in [n]$, we have $\bigl| \indicator[x_i > s] - \indicator[y_j > s] \bigr| = 1$ if $(i, j) \in (X_s \times \overline{Y_s}) \cup (\overline{X_s} \times Y_s)$, and the expression is $0$ otherwise.
    Hence \cref{equation:cutmid_v2_thin}, applied to $X \ceq X_s$ and $Y \ceq Y_s$, yields
    \[
        \sum_{i \in [n]} \bigl| \indicator[x_i > s] - \indicator[y_{\pi(i)} > s] \bigr|
        = \sum_{i, j \in [n]} P_{ij} \bigl| \indicator[x_i > s] - \indicator[y_j > s] \bigr|
        \leq \alpha \sum_{i, j \in [n]} Q_{ij} \bigl| \indicator[x_i > s] - \indicator[y_j > s] \bigr|.
    \]
    For $x, y \in \R$, the function $s \mapsto |\indicator[x > s] - \indicator[y > s]|$ is the indicator of the interval between $x$ and $y$, so $\int_\R |\indicator[x > s] - \indicator[y > s]|\, ds = |x - y|$.
    Integrating the inequality over $s \in \R$ and exchanging the integral with the finite sums therefore proves the claim.
\end{proof}

The above lemma allows us to transfer guarantees about the pairs of vertices that are matched by $Q$ to the pairs that are matched by $\pi$.

\begin{lemma}[Local Agreement]
    \label[lemma]{lemma:cutmid_v2_local}
    Let $Q \in \DS(n)$ and $\vec y \in [0, 1]^n$, and let $\vec x \ceq A\vec y \in \R^n$.
    Then
    \[
        \frac{1}{n^2} \sum_{i, j \in [n]} Q_{ij} \bigl| x_i - (Q^\top \vec x)_j \bigr| \leq 4 \cst{Q}.
    \]
\end{lemma}

To understand the intuition behind this lemma, consider the case where $\vec y=\indicator_Y$ for a subset $Y\subseteq [n]$. Then the entry $x_i$ of $\vec x$ is exactly the number of neighbours vertex $i$ has in the set $Y$ in the graph $G$. 
For $j \in [n]$, the entry $(Q^\top \vec x)_j = \sum_i Q_{ij} x_i$ is a convex combination of the entries $x_i$ of the vertices $i$ that $Q$ matches to $j$. Intuitively, $x_i$ and $(Q^\top \vec x)_j$ are close if on average the vertices $i'$ that $Q$ matches to $j$ have about the same number of neighbours in $Y$ as $i$. 
The lemma states that $x_i$ and $(Q^\top \vec x)_j$ are close for most pairs $(i, j)$ in the support of $Q$: the sum of $Q_{ij} |x_i - (Q^\top \vec x)_j|$ over all pairs $(i, j)$ is small.
With a bit of a stretch, this means that $Q$ only matches different vertices $i, i'$ of $G$ to the same vertex $j$ of $H$ if $i$ and $i'$ have similar neighbourhoods.

\begin{proof}
    For $i, j \in [n]$, let
    \[
        F_{ij} \ceq \sum_{k, \ell \in [n]} Q_{k\ell} \bigl( A_{ik} - B_{j\ell} \bigr) y_k = x_i - r_j,
        \qquad\text{where } \vec r \ceq B Q^\top \vec y,
    \]
    and the identity uses that the rows of $Q$ sum to one.
    We first show that
    \begin{equation}
        \label{equation:cutmid_v2_local_F}
        \frac{1}{n^2} \sum_{i, j \in [n]} Q_{ij} |F_{ij}| \leq 2 \cst{Q}.
    \end{equation}
    Let $S_+ \ceq \{(i, j) \in [n]^2 \mid F_{ij} > 0\}$ and $S_- \ceq \{(i, j) \in [n]^2 \mid F_{ij} < 0\}$, and let $S \in \{S_+, S_-\}$.
    Then
    \[
        \frac{1}{n^2} \sum_{(i, j) \in S} Q_{ij} F_{ij}
        = \frac{1}{n^2} \sum_{(i, j) \in S} \sum_{k, \ell \in [n]} Q_{ij} Q_{k\ell} \bigl( A_{ik} - B_{j\ell} \bigr)\, y_k
    \]
    is the expression in \cref{eq:epsQ2} for the functions $f, g\colon [n]^2 \to [0, 1]$ with $f(i, j) \ceq 1$ if $(i, j) \in S$ and $f(i, j) \ceq 0$ otherwise, and $g(k, \ell) \ceq y_k$.
    Hence it is at most $\cst{Q}$ in absolute value, for $S = S_+$ as well as for $S = S_-$.
    As $|F_{ij}| = F_{ij}$ on $S_+$, $|F_{ij}| = -F_{ij}$ on $S_-$, and $F_{ij} = 0$ otherwise, this proves \cref{equation:cutmid_v2_local_F}.

    Since the columns of $Q$ sum to one, $(Q^\top \vec x)_j = \sum_{i'} Q_{i'j} (F_{i'j} + r_j) = r_j + \sum_{i'} Q_{i'j} F_{i'j}$, and hence
    \[
        x_i - (Q^\top \vec x)_j = F_{ij} - \sum_{i' \in [n]} Q_{i'j} F_{i'j}
        \qquad\text{for all } i, j \in [n].
    \]
    By the triangle inequality and, once more, the fact that the columns of $Q$ sum to one, we have
    \begin{align*}
        \frac{1}{n^2} \sum_{i, j \in [n]} Q_{ij} \bigl| x_i - (Q^\top \vec x)_j \bigr|
        &\leq \frac{1}{n^2} \sum_{i, j \in [n]} Q_{ij} |F_{ij}| + \frac{1}{n^2} \sum_{j \in [n]} \Bigl( \sum_{i \in [n]} Q_{ij} \Bigr) \sum_{i' \in [n]} Q_{i'j} |F_{i'j}|\\
        &= \frac{2}{n^2} \sum_{i, j \in [n]} Q_{ij} |F_{ij}|
        \leq 4 \cst{Q}.
        \qedhere
    \end{align*}
\end{proof}

It is well known and easy to see from the definition that the cut norm 
\[\norm{K}_\cut = \max_{X, Y \subseteq [n]} \Bigl|\sum_{i \in X} \sum_{k \in Y} K_{ik}\Bigr|\] of a matrix $K \in \R^{n \times n}$ is invariant under transpositions and under permuting rows and columns. That is, $\norm{K^\top}_\cut = \norm{K}_\cut$ and $\norm{PKP^\top}_\cut = \norm{K}_\cut$ for $P\in\Perm(n)$. Furthermore, for $R\in\DS(n)$ we have
\begin{equation}\label{eq:ds-cutnorm}
    \norm{RK}_\cut \leq \norm{K}_\cut.
\end{equation}
To see this, let $X, Y \subseteq [n]$ and $\vec x \ceq R^\top \indicator_X$. Then $\vec x \in [0, 1]^n$ and
    \[
        \sum_{i \in X} \sum_{k \in Y} (RK)_{ik} = \sum_{j \in [n]} x_j \sum_{k \in Y} K_{jk}.
    \]
    The right-hand side is linear in $\vec x$, so its absolute value over $\vec x \in [0, 1]^n$ is maximised at some indicator vector $\indicator_{X'}$, where it equals $|\sum_{j \in X'} \sum_{k \in Y} K_{jk}| \leq \norm{K}_\cut$.

With this, we are ready to prove the result of this section.

\begin{theorem}
    \label{theorem:cutmid_v2_thin}
    Let $\alpha \ceq C_{\mathrm{thin}} \ln(2n) \ln\ln(4n)$, where $C_{\mathrm{thin}}$ is the constant of \cref{theorem:cutmid_v2_haqi}. Then for every $Q\in\DS(n)$ there is a $P\in\Perm(n)$ such that $\cst{P} \leq (1 + 4\alpha) \cst{Q}$.
    %
\end{theorem}
\begin{proof}
    Let $Q\in\DS(n)$. By \cref{theorem:cutmid_v2_haqi}, there is a permutation matrix $P \in \Perm(n)$ that is $\alpha'$-thin \wrt{} $Q$ for some $\alpha' \leq \alpha$, and since the entries of $Q$ are non-negative, $P$ is then also $\alpha$-thin \wrt{} $Q$. 
    Let $\pi$ be the permutation of $[n]$ with $P = P_\pi$.
    We show that 
    \begin{equation}\label{eq:cutmid_v2_thin1}
        \cst{P} \leq (1 + 4\alpha) \cst{Q}.
    \end{equation}
    We first express $\cst{P}$ as a cut norm.
    Since $P_{ij} = 1$ if and only if $j = \pi(i)$, only pairs of the form $(i, \pi(i))$ contribute to the sum in \cref{eq:epsQ1} for $\cst{P}$.
    Writing $X \ceq \{i : (i, \pi(i)) \in S\}$ and $Y \ceq \{k : (k, \pi(k)) \in T\}$, the sum equals $\sum_{i \in X} \sum_{k \in Y} (A_{ik} - B_{\pi(i)\pi(k)}) = \sum_{i \in X} \sum_{k \in Y} (A - PBP^\top)_{ik}$, because $(PBP^\top)_{ik} = B_{\pi(i)\pi(k)}$.
    As $X, Y$ range over all subsets of $[n]$, we obtain
    \[
        \cst{P} = n^{-2} \norm{A - PBP^\top}_\cut.
    \]

    Let $R \ceq PQ^\top$, which is doubly stochastic and satisfies $(R\vec x)_i = (Q^\top \vec x)_{\pi(i)}$ for all $\vec x \in \R^n$ and $i \in [n]$.
    Since $RAR^\top = P(Q^\top A Q)P^\top$, we split
    \[
        A - PBP^\top = \bigl( A - RAR^\top \bigr) + P \bigl( Q^\top A Q - B \bigr) P^\top
    \]
    and bound the norm of the two terms of the right-hand side separately.
    Since the cut norm is invariant under conjugation with $P$, by the triangle inequality, to prove \eqref{eq:cutmid_v2_thin1} it suffices to show
    \begin{equation}\label{eq:cutmid_v2_thin2}
        n^{-2} \norm{Q^\top A Q - B}_\cut \leq \cst{Q}
        \qquad\text{and}\qquad
        n^{-2} \norm{A - RAR^\top}_\cut \leq 4\alpha \cst{Q}.
    \end{equation}

    For the first bound, let $X, Y \subseteq [n]$.
    Since the columns of $Q$ sum to one, $\sum_{i, k} Q_{ij} Q_{k\ell} B_{j\ell} = B_{j\ell}$, and together with $\sum_{i, k} Q_{ij} Q_{k\ell} A_{ik} = (Q^\top A Q)_{j\ell}$ this gives
    \[   
        \frac{1}{n^2} \sum_{j \in X} \sum_{\ell \in Y} \bigl( (Q^\top A Q)_{j\ell} - B_{j\ell} \bigr)
        =
        \frac{1}{n^2} \sum_{\substack{(i, j) \in [n] \times X \\ (k, \ell) \in [n] \times Y}} Q_{ij} Q_{k\ell} \bigl( A_{ik} - B_{j\ell} \bigr),
    \]
    which is the expression in \cref{eq:epsQ1} for $S = [n] \times X$ and $T = [n] \times Y$.
    Hence the left-hand side is at most $\cst{Q}$ in absolute value for all $X, Y \subseteq [n]$.

    For the second bound in \eqref{eq:cutmid_v2_thin2}, let $E \ceq I - R$.
    Let $Y \subseteq [n]$ and put $\vec x \ceq A \indicator_Y \in \R^n$ and $\vec f \ceq EA\indicator_Y = \vec x - R\vec x$, so that
    \[
        f_i = x_i - (Q^\top \vec x)_{\pi(i)}
        \qquad\text{for all } i \in [n].
    \]
    Applying \cref{lemma:cutmid_v2_transport} to the vectors $\vec x$ and $Q^\top \vec x$, followed by \cref{lemma:cutmid_v2_local} with $\vec y \ceq \indicator_Y$, gives
    \[
        \frac{1}{n^2} \sum_{i \in [n]} |f_i|
        \leq \frac{\alpha}{n^2} \sum_{i, j \in [n]} Q_{ij} \bigl| x_i - (Q^\top \vec x)_j \bigr|
        \leq 4\alpha \cst{Q}.
    \]
    As the columns of $R$ sum to one, $\sum_i (R\vec x)_i = \sum_k x_k \sum_i R_{ik} = \sum_k x_k$. So $\sum_i f_i = \sum_i x_i - \sum_i (R\vec x)_i = 0$,
and hence the positive and the negative parts of $\vec f$ have the same sum. It follows that $|\sum_{i \in X} f_i| \leq \frac12 \sum_{i \in [n]} |f_i|$ for every $X \subseteq [n]$.
    Therefore,
    \[
        \frac{1}{n^2} \Bigl| \sum_{i \in X} \sum_{k \in Y} (EA)_{ik} \Bigr|
        = \frac{1}{n^2} \Bigl| \sum_{i \in X} f_i \Bigr|
        \leq \frac{1}{2n^2} \sum_{i \in [n]} |f_i|
        \leq 2\alpha \cst{Q}
    \]
    for all $X, Y \subseteq [n]$, that is, $n^{-2} \norm{EA}_\cut \leq 2\alpha \cst{Q}$.
    Finally, since $A$ is symmetric, we have
    \begin{align}
        \begin{aligned}
        \label{equation:arar_to_ea}
        A - RAR^\top &= (A - RA) + (RA - RAR^\top) \\
        &= (A - RA) + RA(I-R^\top) \\
        &= (A - RA) + R(EA)^\top \\
        &= EA + R(EA)^\top,            
        \end{aligned}
    \end{align}
    and \cref{eq:ds-cutnorm} together with the invariance of the cut norm under transposition bounds the cut norms of $EA$ and $R(EA)^\top$ by $\norm{EA}_\cut$.
    This gives $n^{-2} \norm{A - RAR^\top}_\cut \leq 4\alpha \cst{Q}$ and concludes the proof.
\end{proof}

\subsubsection{Additive Rounding}
\label{appendix:cutmid_additive}

In this section, we shall prove the following additive bound.

\begin{theorem}
    \label{theorem:cut_additive_rounding}
    Let $d_n \ceq \lceil 2\log n \rceil$.
    For every $Q \in \DS(n)$ there is a permutation matrix $P \in \Perm(n)$ such that
    \[
        \cst{P} \leq \cst{Q} + 36 \sqrt{d_n}\, n^{-1/6}.
    \]
\end{theorem}

This is going to be significantly harder than the multiplicative bound of Theorem~\ref{theorem:cutmid_v2_thin} (which, admittedly, used the thin matchings for free). We start by fixing notations and collecting some tools from geometry and high-dimensional probability. 

\subparagraph*{The Space $\Ltwo$.}
We write $\Ltwo$ for $\R^n$ equipped with the inner product and norm
\[
    \ip{\vec x}{\vec y} \ceq \frac1n \sum_{i \in [n]} x_i y_i
    \qquad\text{and}\qquad
    \norm{\vec x} \ceq \sqrt{\ip{\vec x}{\vec x}},
\]
that is, the $L^2$-space of the uniform distribution on $[n]$.
We may view the elements of $\Ltwo$ as vectors in $\R^n$ or as functions $[n] \to \R$, whichever is more convenient.
In this section, a matrix $M \in \R^{n \times n}$ acts on $\Ltwo$ as an \emph{averaging operator}, $(M\vec x)_i \ceq \frac1n \sum_{j \in [n]} M_{ij} x_j$, and its \emph{operator norm} is
\[
    \norm{M}_{\Ltwo} \ceq \max_{\norm{\vec x} = 1} \norm{M\vec x} = \max_{\norm{\vec x} = \norm{\vec y} = 1} \bigl| \ip{\vec x}{M\vec y} \bigr| = \frac1n \norm{M}_{2 \to 2}.
\]
For symmetric $M$, the averaging operator is self-adjoint \wrt{} $\ip{\cdot}{\cdot}$, so by the spectral theorem $\Ltwo$ has an orthonormal basis of eigenvectors of $M$, and $\norm{M}_{\Ltwo}$ is the largest absolute value of an eigenvalue.
Note that the eigenvalues of the averaging operator are those of the matrix $M/n$.

\subparagraph*{Kernels and Weighted Cut Norm.}
For a finite probability space $(\Omega, \mu)$, a \emph{kernel} is a function $K\colon \Omega \times \Omega \to \R$, i.e.\ a matrix indexed by $\Omega$, which we always pair with the measure $\mu$ as in \cite{lovasz_large_networks}. Its \emph{weighted cut norm} is
\[
    \norm{K}_{\cut, \mu} \ceq \max_{S, T \subseteq \Omega} \Bigl| \sum_{x \in S} \sum_{y \in T} \mu(x) \mu(y) K(x, y) \Bigr|.
\]
Note that if $\Omega=[n]$ and $\mu$ is the uniform distribution we have $\norm{K}_{\cut, \mu}=n^{-2} \norm{K}_\cut$.

\subparagraph*{Couplings and Discrepancy.}
For $Q \in \DS(n)$, let $\Omega_Q \ceq [n] \times [n]$ carry the probability weights $\mu_Q(i, j) \ceq Q_{ij}/n$; thus $\mu_Q \in \DSS(n)$ is a coupling of the uniform distributions on $[n]$. 
We view the \emph{discrepancy matrix} $\Disc_{A', B'}$ with entries $\Disc_{A', B'}(i, j, k, \ell) = A'_{ik} - B'_{j\ell}$ (cf.~\cref{equation:def_disc_matrix}) as a kernel on $\Omega_Q$.
For $A', B' \in \R^{n \times n}$, we let
\[
    \cst{Q}(A', B') \ceq \norm{\Disc_{A', B'}}_{\cut, \mu_Q}.
\]
Note that for the adjacency matrices $A, B$ of our graphs $G, H$, we have $\cst{Q}(A, B) = \cst{Q}$ by \cref{eq:epsQ1}.
We need the following elementary fact about these weighted cut norms.
\begin{lemma}[Lifted Kernels]
    \label[lemma]{lemma:cutmid_cut_facts}
    Let $Q \in \DS(n)$, let $K \in \R^{n \times n}$ with $\norm{K}_{\Ltwo} \leq t$, and let $K^{(1)}$ and $K^{(2)}$ be the kernels on $\Omega_Q$ with $K^{(1)}\bigl( (i, j), (k, \ell) \bigr) \ceq K_{ik}$ and $K^{(2)}\bigl( (i, j), (k, \ell) \bigr) \ceq K_{j\ell}$.
    Then $\norm{K^{(1)}}_{\cut, \mu_Q}, \norm{K^{(2)}}_{\cut, \mu_Q} \leq t$.
\end{lemma}
\begin{proof}
    Let $S, T \subseteq \Omega_Q$ and define $\vec s, \vec \sigma \in \R^n$ by $s_i \ceq \sum_{j \in [n]} Q_{ij} \indicator_S(i, j)$ and $\sigma_k \ceq \sum_{\ell \in [n]} Q_{k\ell} \indicator_T(k, \ell)$.
    Since the rows of $Q$ sum to one, $\vec s, \vec \sigma \in [0, 1]^n$, hence $\norm{\vec s}, \norm{\vec \sigma} \leq 1$, and
    \[
        \sum_{(i, j) \in S} \sum_{(k, \ell) \in T} \mu_Q(i, j)\, \mu_Q(k, \ell)\, K^{(1)}\bigl( (i, j), (k, \ell) \bigr)
        = \frac{1}{n^2} \sum_{i, k \in [n]} s_i K_{ik} \sigma_k
        = \ip{\vec s}{K \vec \sigma}
    \]
    has absolute value at most $\norm{\vec s} \norm{K}_{\Ltwo} \norm{\vec \sigma} \leq t$ by Cauchy--Schwarz.
    Maximising over $S, T$ gives $\norm{K^{(1)}}_{\cut, \mu_Q} \leq t$.
    For $K^{(2)}$, the same argument applies with $s_j \ceq \sum_{i \in [n]} Q_{ij} \indicator_S(i, j)$ and $\sigma_\ell \ceq \sum_{k \in [n]} Q_{k\ell} \indicator_T(k, \ell)$, which lie in $[0, 1]$ since the columns of $Q$ sum to one.
\end{proof}

\subparagraph*{Tools from Geometry.}
In the remainder of this section, $\Hs \ceq \R^m$ for some $m \in \Npos$, equipped with the standard inner product $\ip{\vec u}{\vec w} \ceq \sum_{r \in [m]} u_r w_r$ and norm $\norm{\vec u} \ceq \sqrt{\ip{\vec u}{\vec u}}$; in our application, $m \leq 2n$ is fixed in the proof of \cref{lemma:cutmid_spectral}.
For a linear operator $T$ on $\Hs$, we write $\norm{T}_{2 \to 2} \ceq \max_{\norm{\vec u} = 1} \norm{T\vec u}$ for its operator norm; if $T$ is diagonal, this is the largest absolute value of a diagonal entry.
We use the same symbols $\ip{\cdot}{\cdot}$ and $\norm{\cdot}$ as for $\Ltwo$; the two spaces only meet in the proof of \cref{lemma:cutmid_spectral}, where we say which one is meant.
For a non-empty compact convex set $Z \subseteq \Hs$, let
\[
    h_Z(\vec w) \ceq \max_{\vec u \in Z} \ip{\vec w}{\vec u} \qquad \text{for } \vec w \in \Hs
\]
be its \emph{support function}, and let $\ball$ be the closed unit ball of $\Hs$.
\begin{lemma}[Close Support Functions]
    \label[lemma]{lemma:cutmid_support}
    Let $Z, Z' \subseteq \Hs$ be non-empty compact convex sets contained in the ball of radius $M$ with $|h_Z(\vec w) - h_{Z'}(\vec w)| \leq \eta$ for all $\vec w \in \ball$, and let $T$ be a linear operator with $\norm{T}_{2 \to 2} \leq \tau$.
    Then
    \[
        \Bigl| \max_{\vec u, \vec u' \in Z} |\ip{\vec u}{T\vec u'}| - \max_{\vec u, \vec u' \in Z'} |\ip{\vec u}{T\vec u'}| \Bigr| \leq 2M\eta\tau.
    \]
\end{lemma}
\begin{proof}
    The hypothesis gives $|h_Z(\vec w) - h_{Z'}(\vec w)| \leq \eta \norm{\vec w}$ for all $\vec w \in \Hs$.
    For non-empty compact convex $Y, Y' \subseteq \Hs$, let $m_T(Y, Y') \ceq \max_{\vec u \in Y, \vec u' \in Y'} \ip{\vec u}{T\vec u'}$.
    Maximising over $\vec u$ first gives $m_T(Y, Y') = \max_{\vec u' \in Y'} h_Y(T\vec u')$, and maximising over $\vec u'$ first gives $m_T(Y, Y') = \max_{\vec u \in Y} h_{Y'}(T^\ast \vec u)$ for the adjoint operator $T^\ast$, which satisfies $\norm{T^\ast}_{2 \to 2} = \norm{T}_{2 \to 2} \leq \tau$.
    For $\vec u\in Z$ we have $\norm{T\vec u} \leq \tau M$, so $|h_Z(T\vec u) - h_{Z'}(T\vec u)| \leq M\eta\tau$, and hence $|m_T(Z, Z) - m_T(Z', Z)| \leq M\eta\tau$.
    Likewise, $\norm{T^\ast \vec u} \leq \tau M$ for $\vec u \in Z'$ gives $|m_T(Z', Z) - m_T(Z', Z')| \leq M\eta\tau$.
    Thus $|m_T(Z, Z) - m_T(Z', Z')| \leq 2M\eta\tau$, and the same holds for $-T$.
    Since $\max_{\vec u, \vec u' \in Y} |\ip{\vec u}{T\vec u'}| = \max\{m_T(Y, Y), m_{-T}(Y, Y)\}$ and $|\max\{a, b\} - \max\{a', b'\}| \leq \max\{|a - a'|, |b - b'|\}$, the claim follows.
\end{proof}

\subparagraph*{Tools from Probability.}

The following lemma uses a standard argument giving a concentration bound in terms of Rademacher complexity (cf.\ \cite[Lemma~2.3.1]{vandervaart1996weak} or \cite[Lemma 26.2]{shalevshwartz2014understanding}).

\begin{lemma}
    \label[lemma]{lemma:cutmid_symmetrisation}
    Let $W$ be a set, let $d_1, \dots, d_b\colon W \to \R$ be bounded, and let $X_1, \dots, X_b$ be independent $\{0, 1\}$-valued random variables with $\theta_j \ceq \Expected X_j$.
    Then
    \[
        \Expected_X \sup_{w \in W} \Bigl| \sum_{j \in [b]} (X_j - \theta_j)\, d_j(w) \Bigr|
        \leq 2\, \Expected_X \Expected_{\vec\sigma \sim \Unif\{\pm1\}^b} \sup_{w \in W} \Bigl| \sum_{j \in [b]} \sigma_j X_j\, d_j(w) \Bigr|.
    \]
\end{lemma}

\begin{proof}
    Let $X'$ be an independent copy of $X$.
    Since $\theta_j = \Expected X'_j$ and the supremum of absolute values is convex, Jensen's inequality bounds the left-hand side by $\Expected_{X, X'} \sup_{w} |\sum_j (X_j - X'_j)\, d_j(w)|$.
    The differences $X_j - X'_j$ are independent and symmetric, so $(X_j - X'_j)_j$ has the same distribution as $(\sigma_j (X_j - X'_j))_j$ for every fixed $\vec\sigma \in \{\pm1\}^b$.
    Averaging over $\vec\sigma$ and applying the triangle inequality, the left-hand side is at most
    \[
        \Expected_{X, X'} \Expected_{\vec\sigma} \sup_{w} \Bigl| \sum_j \sigma_j X_j\, d_j(w) \Bigr| + \Expected_{X, X'} \Expected_{\vec\sigma} \sup_{w} \Bigl| \sum_j \sigma_j X'_j\, d_j(w) \Bigr|,
    \]
    and both terms equal the expectation on the right-hand side of the lemma.
\end{proof}

Next, we need a standard lemma stating that applying a contraction, that is, a $1$-Lipschitz function $\gamma\colon\R\to\R$ with $\gamma(0)=0$, increases the Rademacher complexity by at most a factor of $2$ (see \cite[Theorem~4.12]{ledoux1991probability} for a proof; \cite[Lemma~5.7]{mohri2018foundations} and \cite[Lemma~26.9]{shalevshwartz2014understanding} give the version without absolute values).

\begin{lemma}[Talagrand Contraction Lemma]
    \label[lemma]{lemma:cutmid_contraction}
    Let $S \subseteq \R^b$ be bounded, and let $\gamma_1, \dots, \gamma_b\colon \R \to \R$ be contractions. Then
    \[
        \Expected_{\vec\sigma \sim \Unif\{\pm1\}^b} \sup_{\vec x \in S} \Bigl| \sum_{j \in [b]} \sigma_j\, \gamma_j(x_j) \Bigr|
        \leq 2\, \Expected_{\vec\sigma \sim \Unif\{\pm1\}^b} \sup_{\vec x \in S} \Bigl| \sum_{j \in [b]} \sigma_j\, x_j \Bigr|.
    \]
\end{lemma}

The following lemma gives the main estimate we need later. Recall that for $x\in\R$ we let $(x)_+\coloneqq\max\{x,0\}$.

\begin{lemma}
    \label[lemma]{lemma:cutmid_rademacher}
    Let $B_1, \dots, B_b \subseteq [n]$ be pairwise disjoint and non-empty, with $\ell_j = |B_j|$, and let $\vec u_1, \dots, \vec u_n \in \Hs$ with $\norm{\vec u_i} \leq M$.
    Then
    \[
        \Expected_{\vec\sigma \sim \Unif\{\pm1\}^b} \sup_{\vec w \in \ball} \Bigl| \sum_{j \in [b]} \sigma_j \sum_{i \in B_j} \pos{\ip{\vec w}{\vec u_i}} \Bigr| \leq 2M \Bigl( \sum_{j \in [b]} \ell_j^2 \Bigr)^{1/2}.
    \]
\end{lemma}

\begin{proof}
    Let $I_j \sim \Unif(B_j)$ independently for $j \in [b]$, so that
    \[
        \sum_{i \in B_j} \pos{\ip{\vec w}{\vec u_i}} = \Expected_{I_j} \pos{\ell_j \ip{\vec w}{\vec u_{I_j}}}.
    \]
    Since the supremum of absolute values is convex, Jensen's inequality bounds the left-hand side of the lemma by the expectation over $I$ of the same expression with $\pos{\ell_j \ip{\vec w}{\vec u_{I_j}}}$ in place of the block sums.
    So it suffices to show that, for every fixed $I = (I_1, \dots, I_b)$,
    \begin{equation}
        \label{equation:cutmid_rademacher_fixed}
        \Expected_{\vec\sigma} \sup_{\vec w \in \ball} \Bigl| \sum_{j \in [b]} \sigma_j \pos{\ell_j \ip{\vec w}{\vec u_{I_j}}} \Bigr|
        \leq 2M \Bigl( \sum_{j \in [b]} \ell_j^2 \Bigr)^{1/2}.
    \end{equation}
    Fix $I$ and let $S \ceq \{ (\ell_j \ip{\vec w}{\vec u_{I_j}})_{j \in [b]} \mid \vec w \in \ball \} \subseteq \R^b$.
    This set is bounded, since $|\ell_j \ip{\vec w}{\vec u_{I_j}}| \leq \ell_j M$ for $\vec w \in \ball$ by Cauchy--Schwarz.
    The map $s \mapsto \pos{s}$ is a contraction, so \cref{lemma:cutmid_contraction} with $\gamma_j(s) \ceq \pos{s}$ for all $j \in [b]$ gives
    \[
        \Expected_{\vec\sigma} \sup_{\vec w \in \ball} \Bigl| \sum_{j \in [b]} \sigma_j \pos{\ell_j \ip{\vec w}{\vec u_{I_j}}} \Bigr|
        \leq 2\, \Expected_{\vec\sigma} \sup_{\vec w \in \ball} \Bigl| \sum_{j \in [b]} \sigma_j \ell_j \ip{\vec w}{\vec u_{I_j}} \Bigr|.
    \]
    For fixed $\vec\sigma$, let $\vec g_{\vec\sigma} \ceq \sum_{j \in [b]} \sigma_j \ell_j \vec u_{I_j} \in \Hs$.
    By linearity of the inner product, $\sum_{j \in [b]} \sigma_j \ell_j \ip{\vec w}{\vec u_{I_j}} = \ip{\vec w}{\vec g_{\vec\sigma}}$, and by Cauchy--Schwarz, $|\ip{\vec w}{\vec g_{\vec\sigma}}| \leq \norm{\vec g_{\vec\sigma}}$ for all $\vec w \in \ball$, with equality for $\vec w = \vec g_{\vec\sigma} / \norm{\vec g_{\vec\sigma}}$ if $\vec g_{\vec\sigma} \neq \vec 0$.
    Hence $\sup_{\vec w \in \ball} |\sum_{j} \sigma_j \ell_j \ip{\vec w}{\vec u_{I_j}}| = \norm{\vec g_{\vec\sigma}}$, and the right-hand side above equals $2\, \Expected_{\vec\sigma} \norm{\vec g_{\vec\sigma}}$.
    Since the square root is concave, Jensen's inequality yields $\Expected_{\vec\sigma} \norm{\vec g_{\vec\sigma}} \leq (\Expected_{\vec\sigma} \norm{\vec g_{\vec\sigma}}^2)^{1/2}$.
    Expanding the square and using $\Expected_{\vec\sigma} \sigma_j \sigma_{j'} = 1$ if $j = j'$ and $\Expected_{\vec\sigma} \sigma_j \sigma_{j'} = 0$ if $j \neq j'$, we obtain
    \[
        \Expected_{\vec\sigma} \norm{\vec g_{\vec\sigma}}^2
        = \sum_{j, j' \in [b]} \ell_j \ell_{j'} \ip{\vec u_{I_j}}{\vec u_{I_{j'}}}\, \Expected_{\vec\sigma} \sigma_j \sigma_{j'}
        = \sum_{j \in [b]} \ell_j^2 \norm{\vec u_{I_j}}^2
        \leq M^2 \sum_{j \in [b]} \ell_j^2.
    \]
    Combining the three displays proves \cref{equation:cutmid_rademacher_fixed}, and taking the expectation over $I$ proves the lemma.
\end{proof}

\subparagraph*{Spectral Approximation of the Cut Norm.}

The following lemma is the spectral heart of the argument.
It shows that, up to an additive error $2t$, we may replace the entries $A_{ik} - B_{j\ell}$ of the discrepancy matrix in the definition \cref{eq:epsQ1} of $\cst{Q}$ by inner products $\ip{\vec z_{ij}}{T_t \vec z_{k\ell}}$ of vectors in a Euclidean space, where the vectors are bounded and the operator $T_t$ has norm at most $1/t$.
The parameter $t$ thus controls a trade-off between the error made by discarding spectral information and the norm of $T_t$; it will be chosen at the very end, in the proof of \cref{theorem:cut_additive_rounding}.
\begin{lemma}
    \label[lemma]{lemma:cutmid_spectral}
    For every $t > 0$ there are $m$ with $1 \leq m \leq 2n$, vectors $(\vec z_{ij})_{i, j \in [n]}$ in $\Hs = \R^m$ with $\norm{\vec z_{ij}} \leq \sqrt2$, and a linear operator $T_t$ on $\Hs$ with $\norm{T_t}_{2 \to 2} \leq 1/t$ such that every $Q \in \DS(n)$ satisfies
    \begin{equation}
        \label{equation:zonotope_cut_approximation}
        \Bigl| \cst{Q} - \max_{S, T \subseteq [n]^2} \frac{1}{n^2} \Bigl| \sum_{(i, j) \in S} \sum_{(k, \ell) \in T} Q_{ij} Q_{k\ell} \ip{\vec z_{ij}}{T_t \vec z_{k\ell}} \Bigr| \Bigr| \leq 2t.
    \end{equation}
 \end{lemma}
\begin{proof}
    By the spectral theorem for the averaging operators $A$ and $B$ on $\Ltwo$, there are eigenvectors $(\vec\phi_r)_{r \in [n]}$ and $(\vec\psi_s)_{s \in [n]}$, orthonormal in $\Ltwo$, and eigenvalues $(\lambda_r)_{r \in [n]}$ and $(\nu_s)_{s \in [n]}$ with
    \[
        A_{ik} = \sum_{r \in [n]} \lambda_r \phi_{ri} \phi_{rk}
        \qquad\text{and}\qquad
        B_{j\ell} = \sum_{s \in [n]} \nu_s \psi_{sj} \psi_{s\ell}.
    \]
    Let $I_A \ceq \{r \in [n] : |\lambda_r| > t\}$ and $I_B \ceq \{s \in [n] : |\nu_s| > t\}$, and let
    \[
        (A_t)_{ik} \ceq \sum_{r \in I_A} \lambda_r \phi_{ri} \phi_{rk}
        \qquad\text{and}\qquad
        (B_t)_{j\ell} \ceq \sum_{s \in I_B} \nu_s \psi_{sj} \psi_{s\ell}
    \]
    be the spectral truncations of $A$ and $B$.
    The averaging operators $A - A_t$ and $B - B_t$ on $\Ltwo$ have all their eigenvalues in $[-t, t]$, so $\norm{A - A_t}_{\Ltwo}, \norm{B - B_t}_{\Ltwo} \leq t$.
    We have
    \[
        \Disc_{A, B} = \Disc_{A_t, B_t} + (A - A_t)^{(1)} - (B - B_t)^{(2)}
    \]
    in the notation of \cref{lemma:cutmid_cut_facts}, where the kernels $(A - A_t)^{(1)}$ and $(B - B_t)^{(2)}$ depend only on the first and only on the second coordinates, respectively.
    \Cref{lemma:cutmid_cut_facts} bounds their weighted cut norms $\norm{\cdot}_{\cut, \mu_Q}$ by $t$ for every $Q \in \DS(n)$.
    By the triangle inequality for $\norm{\cdot}_{\cut, \mu_Q}$,
    \begin{equation}
        \label{equation:cutmid_truncation}
        \bigl| \cst{Q}(A, B) - \cst{Q}(A_t, B_t) \bigr| \leq 2t
        \qquad\text{for every } Q \in \DS(n).
    \end{equation}
    If $I_A = I_B = \emptyset$, then $A_t = B_t = 0$, so $\cst{Q} \leq 2t$ for every $Q \in \DS(n)$ by \cref{equation:cutmid_truncation}, and the lemma holds with $m = 1$, $\vec z_{ij} \ceq \vec 0$ and $T_t \ceq 0$.
    Otherwise, let $m \ceq |I_A| + |I_B| \leq 2n$ and identify $\Hs = \R^m$ with $\R^{I_A} \oplus \R^{I_B}$, with the standard inner product; let
    \[
        \vec z_{ij} \ceq \Bigl( \bigl(\lambda_r \phi_{ri}\bigr)_{r \in I_A}, \bigl(\nu_s \psi_{sj}\bigr)_{s \in I_B} \Bigr)
        \qquad\text{and}\qquad
        T_t \ceq \diag\Bigl( \bigl(\lambda_r^{-1}\bigr)_{r \in I_A}, \bigl(-\nu_s^{-1}\bigr)_{s \in I_B} \Bigr).
    \]
    The $i$-th row of $A$, as an element of $\Ltwo$, equals $\sum_{r \in [n]} \lambda_r \phi_{ri} \vec\phi_r$, so by Parseval's identity in $\Ltwo$, the first component of $\vec z_{ij}$ has squared norm $\sum_{r \in I_A} \lambda_r^2 \phi_{ri}^2 \leq \frac1n \sum_{k \in [n]} A_{ik}^2 \leq 1$; similarly for the second component.
    Hence
    \[
        \norm{\vec z_{ij}} \leq \sqrt2,
        \qquad
        \norm{T_t}_{2 \to 2} \leq \frac1t,
        \qquad\text{and}\qquad
        \ip{\vec z_{ij}}{T_t \vec z_{k\ell}} = (A_t)_{ik} - (B_t)_{j\ell}.
    \]
    Now let $Q \in \DS(n)$.
    By the definition of $\cst{Q}(A_t, B_t)$ as the weighted cut norm of $\Disc_{A_t, B_t}$ on $(\Omega_Q, \mu_Q)$, and since $\mu_Q(i, j) \mu_Q(k, \ell) = Q_{ij} Q_{k\ell} / n^2$,
    \begin{align*}
        \cst{Q}(A_t, B_t)
        &= \max_{S, T \subseteq [n]^2} \frac{1}{n^2} \Bigl| \sum_{(i, j) \in S} \sum_{(k, \ell) \in T} Q_{ij} Q_{k\ell} \bigl( (A_t)_{ik} - (B_t)_{j\ell} \bigr) \Bigr| \\
        &= \max_{S, T \subseteq [n]^2} \frac{1}{n^2} \Bigl| \sum_{(i, j) \in S} \sum_{(k, \ell) \in T} Q_{ij} Q_{k\ell} \ip{\vec z_{ij}}{T_t \vec z_{k\ell}} \Bigr|.
    \end{align*}
    Together with \cref{equation:cutmid_truncation} and $\cst{Q}(A, B) = \cst{Q}$, this proves \cref{equation:zonotope_cut_approximation}.
\end{proof}

\subparagraph*{Zonotopes and a Pseudometric on Doubly Stochastic Matrices.}

For the remainder of this section, we fix $t > 0$ and, with it, $m \leq 2n$, vectors $(\vec z_{ij})_{i, j \in [n]}$ in $\Hs = \R^m$ and an operator $T_t$ on $\Hs$ as in \cref{lemma:cutmid_spectral}.
We let $M \ceq \sqrt2$, so that $\norm{\vec z_{ij}} \leq M$ for all $i, j \in [n]$.
For $Q \in \DS(n)$, we let
\begin{equation}
    \label{equation:cutmid_zonotope}
    Z_Q \ceq \Bigl\{ \frac1n \sum_{i, j \in [n]} Q_{ij} a_{ij} \vec z_{ij} : a_{ij} \in [0, 1]\text{ for }i,j \Bigr\}.
\end{equation}
Then $Z_Q$ is a zonotope, the Minkowski sum of the line segments $[0, \frac1n Q_{ij} \vec z_{ij}]$. Note that $Z_Q$ is a non-empty compact convex set contained in the ball of radius $M$, because $\frac1n \sum_{i, j} Q_{ij} = 1$. Its support function $h_Q\coloneqq h_{Z_Q}$ satisfies
\begin{equation}\label{eq:support_zonotope}
h_Q(\vec w) = \max_{\vec u \in Z_Q} \ip{\vec w}{\vec u} = \frac{1}{n} \sum_{i, j \in [n]} Q_{ij}\, \pos{\ip{\vec w}{\vec z_{ij}}}
\end{equation}
for all $\vec w\in\Hs$, because the function $a \mapsto \ip{\vec w}{\frac1n \sum_{i, j} Q_{ij} a_{ij} \vec z_{ij}}$ on $[0,1]^{n\times n}$ is maximised by $a_{ij} = \indicator[\ip{\vec w}{\vec z_{ij}} > 0]$.

For $Q, Q' \in \DS(n)$, let
\begin{equation}
    \label{equation:cutmid_h}
    \rho(Q, Q') \ceq \sup_{\vec w \in \ball} \bigl| h_Q(\vec w) - h_{Q'}(\vec w) \bigr|.
\end{equation}
Then $\rho$ is a pseudometric on $\DS(n)$. Furthermore, since by \eqref{eq:support_zonotope}, $h_Q$ depends linearly on $Q$, for all $Q_0, Q_1, Q'_0, Q'_1 \in \DS(n)$ and $\theta \in [0, 1]$ we have
\begin{equation}
    \label{equation:cutmid_convex}
    \rho\bigl( (1 - \theta) Q_0 + \theta Q_1, (1 - \theta) Q'_0 + \theta Q'_1 \bigr) \leq (1 - \theta) \rho(Q_0, Q'_0) + \theta \rho(Q_1, Q'_1).
\end{equation}

The maximum in \cref{equation:zonotope_cut_approximation} can be written as the maximum of a bilinear form over the zonotope $Z_Q$: for every $Q \in \DS(n)$,
\begin{equation}
    \label{equation:cutmid_zonotope_cut}
    \max_{\vec u, \vec u' \in Z_Q} \bigl| \ip{\vec u}{T_t \vec u'} \bigr|
    = \max_{S, T \subseteq [n]^2} \frac{1}{n^2} \Bigl| \sum_{(i, j) \in S} \sum_{(k, \ell) \in T} Q_{ij} Q_{k\ell} \ip{\vec z_{ij}}{T_t \vec z_{k\ell}} \Bigr|,
\end{equation}
and hence, by \cref{equation:zonotope_cut_approximation},
\begin{equation}
    \label{equation:cutmid_zonotope_eps}
    \Bigl| \cst{Q} - \max_{\vec u, \vec u' \in Z_Q} \bigl| \ip{\vec u}{T_t \vec u'} \bigr| \Bigr| \leq 2t.
\end{equation}
Indeed, for $a, a' \in [0, 1]^{n \times n}$, the points $\vec u_a \ceq \frac1n \sum_{i, j} Q_{ij} a_{ij} \vec z_{ij}$ and $\vec u_{a'} \ceq \frac1n \sum_{k, \ell} Q_{k\ell} a'_{k\ell} \vec z_{k\ell}$ of $Z_Q$ satisfy
\[
    \ip{\vec u_a}{T_t \vec u_{a'}} = \frac{1}{n^2} \sum_{i, j, k, \ell \in [n]} Q_{ij} Q_{k\ell}\, a_{ij}\, a'_{k\ell}\, \ip{\vec z_{ij}}{T_t \vec z_{k\ell}}
\]
by bilinearity of the inner product.
The right-hand side is bilinear in $(a, a')$, so for fixed $a'$ its absolute value is a convex function of $a$ and attains its maximum over the cube $[0, 1]^{n \times n}$ at a vertex, that is, at an indicator $a = \indicator_S$ for some $S \subseteq [n]^2$; likewise for $a'$.
Hence the maximum over $Z_Q$ is attained at indicators $a = \indicator_S$, $a' = \indicator_T$, which proves \cref{equation:cutmid_zonotope_cut}.

\begin{corollary}\label[corollary]{cor:cutmid_spectral}
       Let $P,Q \in \DS(n)$. Then
    \[
        \cst{P} \leq \cst{Q} + 4t + \frac{2\sqrt2}{t}\, \rho(P, Q).
    \]
\end{corollary}

\begin{proof}
    The zonotopes $Z_P$ and $Z_Q$ are non-empty, compact and convex, and they lie in the ball of radius $M = \sqrt{2}$.
    By the definition \cref{equation:cutmid_h} of $\rho$, their support functions $h_P$ and $h_Q$ differ by at most $\rho(P, Q)$ on $\ball$.
    Hence \cref{lemma:cutmid_support}, applied with $M = \sqrt{2}$, $\eta = \rho(P, Q)$ and $\tau = 1/t$, shows that
    \[
        \max_{\vec u, \vec u' \in Z_P} \bigl| \ip{\vec u}{T_t \vec u'} \bigr|
        \leq \max_{\vec u, \vec u' \in Z_Q} \bigl| \ip{\vec u}{T_t \vec u'} \bigr| + \frac{2\sqrt2}{t} \rho(P, Q).
    \]
    Applying \cref{equation:cutmid_zonotope_eps} to $P$ and to $Q$, we obtain
    \[
        \cst{P}
        \leq \max_{\vec u, \vec u' \in Z_P} \bigl| \ip{\vec u}{T_t \vec u'} \bigr| + 2t
        \leq \max_{\vec u, \vec u' \in Z_Q} \bigl| \ip{\vec u}{T_t \vec u'} \bigr| + \frac{2\sqrt2}{t} \rho(P, Q) + 2t
        \leq \cst{Q} + 4t + \frac{2\sqrt2}{t} \rho(P, Q).
        \qedhere
    \]
\end{proof}

\subparagraph*{Approximating Doubly Stochastic Matrices by Permutation Matrices.}
The next lemma is the crucial step of the proof. It shows that, with respect to $\rho$, every convex combination of two permutation matrices is close to a single permutation matrix.
By \cref{cor:cutmid_spectral}, the cut norm then cannot increase by much.

\begin{lemma}[Merging Two Permutations]
    \label[lemma]{lemma:cutmid_merge}
    For all permutation matrices $P_0, P_1$ and $\theta \in [0, 1]$, there is a permutation matrix $P$ with
    \[
        \rho\bigl( P, (1 - \theta) P_0 + \theta P_1 \bigr) \leq 20 M n^{-1/3}.
    \]
\end{lemma}
\begin{proof}
    Let $P_0 = P_{\pi_0}$ and $P_1 = P_{\pi_1}$ for permutations $\pi_0, \pi_1$ of $[n]$.
    The proof below uses no property of the vectors $\vec z_{ij}$ other than $\norm{\vec z_{ij}} \leq M$, so we may relabel the columns along $\pi_0$, that is, replace $\vec z_{ij}$ by $\vec z'_{ij} \ceq \vec z_{i, \pi_0(j)}$ for all $i, j \in [n]$.
    By \cref{eq:support_zonotope}, the support function of $Q \in \DS(n)$ \wrt{} the vectors $\vec z'_{ij}$ equals $h_{Q P_0}$, so the pseudometric $\rho'$ defined by the vectors $\vec z'_{ij}$ satisfies $\rho(Q, Q') = \rho'(Q P_0^\top, Q' P_0^\top)$ for all $Q, Q' \in \DS(n)$.
    Since $Q \mapsto Q P_0^\top$ is a linear bijection of $\DS(n)$ that maps $\Perm(n)$ onto itself, $P_0$ to $I$ and $P_1$ to $P_1 P_0^\top$, it suffices to prove the lemma for the vectors $\vec z'_{ij}$ and the pair $I, P_1 P_0^\top$.
    Hence we may \WLOG{} assume that $P_0 = I = P_\id$ and $P_1 = P_\tau$ for some permutation $\tau$ of $[n]$. 
    We merge $P_0$ and $P_1$ into a new permutation $P$ by selecting for each row $i$, either column $i$ or $\tau(i)$.
    This will generally result in conflicts, so the resulting maps from rows to columns need not be injective. For a map $T \colon [n] \to [n]$, let
    \[
        h_T(\vec w) \ceq \frac1n \sum_{i \in [n]} \pos{\ip{\vec w}{\vec z_{i, T(i)}}}
        \qquad\text{and}\qquad
        \Delta(T) \ceq \sup_{\vec w \in \ball} \bigl| h_T(\vec w) - (1 - \theta) h_{\mathrm{id}}(\vec w) - \theta h_{\tau}(\vec w) \bigr|,
    \]
    so that $h_{P_\pi} = h_\pi$ for permutations $\pi$, and $\Delta(\pi) = \rho(P_\pi, (1 - \theta) I + \theta P_\tau)$ by linearity of $Q \mapsto h_Q$.

    Our goal is to construct a map $T$ that is \enquote{close} to being bijective while $\Delta(T)$ remains small. 
    We can then \enquote{repair} $T$ to obtain a bijection (permutation matrix) while only increasing the distance under $\rho$ slightly.

    Let $L \ceq \lceil n^{1/3} \rceil$, so that $n^{1/3} \leq L \leq 2n^{1/3}$.
    We partition $[n]$ into blocks $B_1, \dots, B_b$ of consecutive elements $i, \tau(i), \tau^2(i), \dots$ of the cycles of $\tau$.
    Cycles of length at most $L$ form a block, whereas cycles of length $\ell > L$ are cut into $\lfloor \ell/L \rfloor$ blocks. The resulting blocks are all of length at most $L$, with (possibly) the exception of the last block, which has length less than $2L$. 
    The following properties hold:
    \begin{enumerate}[(i)]
        \item every block has length less than $2L$, so $\sum_{a \in [b]} |B_a|^2 \leq 2L \sum_{a \in [b]} |B_a| = 2Ln$;
        \item if $i$ and $\tau(i)$ lie in different blocks, then $i$ is the last element of a block of length at least $L$, so there are at most $n/L$ such $i$.
    \end{enumerate}
    For $i \in [n]$, let $a(i)$ denote the index of the block containing $i$.

    For independent $X_1, \dots, X_b \sim \Ber(\theta)$, let every block either stay in place or follow $\tau$, that is,
    \[
        T_X(i) \ceq
        \begin{cases}
            i, & X_{a(i)} = 0,\\
            \tau(i), & X_{a(i)} = 1,
        \end{cases}
    \]
    for all $i \in [n]$.
    Switching block $a$ from the identity to $\tau$ changes the quantity $n \cdot h_{T_X}(\vec w)$ by
    \[
        d_a(\vec w) \ceq \sum_{i \in B_a} \Bigl[ \pos{\ip{\vec w}{\vec z_{i, \tau(i)}}} - \pos{\ip{\vec w}{\vec z_{i, i}}} \Bigr].
    \]
    Hence $h_{T_X} = h_{\mathrm{id}} + \frac{1}{n}\sum_a X_a d_a$ and $h_{\tau} = h_{\mathrm{id}} + \frac{1}{n}\sum_a d_a$, and therefore
    \[
        n \cdot \bigl( h_{T_X} - (1 - \theta) h_{\mathrm{id}} - \theta h_{\tau} \bigr) = \sum_{a \in [b]} (X_a - \theta)\, d_a.
    \]
    By \cref{lemma:cutmid_symmetrisation} with $W = \ball$ and $\theta_j = \theta$ for all $j \in [b]$,
    \[
        n\, \Expected_X\, \Delta(T_X)
        \leq 2\, \Expected_X \Expected_{\vec\sigma} \sup_{\vec w \in \ball} \Bigl| \sum_{a : X_a = 1} \sigma_a d_a(\vec w) \Bigr|.
    \]
    Splitting $d_a$ into its two sums and applying \cref{lemma:cutmid_rademacher} to the blocks $\{B_a \mid a \in [b], X_a = 1\}$, once with $\vec u_i = \vec z_{i, \tau(i)}$ and once with $\vec u_i = \vec z_{i, i}$, bounds the right-hand side by 
    \[
        2 \cdot (2M + 2M) \Big(\sum_{a \in [b]} |B_a|^2\Big)^{1/2} \leq 8M \sqrt{2Ln},
    \]
    by~(i).
    In particular, there is an $X \in \{0, 1\}^b$ with
    \begin{equation}
        \label{equation:cutmid_merge_T}
        \Delta(T_X) \leq 8M \sqrt{\mfrac{2L}{n}} \leq 16 M n^{-1/3}.
    \end{equation}

    It remains to transform $T_X$ into a permutation matrix. By (ii), at most $\kappa \leq n/L$ columns have two preimages.
    Removing the row $\tau^{-1}(c)$ for each of them leaves an injective map on $n - \kappa$ rows.
    Matching the $\kappa$ removed rows to the $\kappa$ unused columns arbitrarily then gives a permutation $\pi$ that differs from $T_X$ in at most $n/L$ rows.
    Since $s \mapsto s_+$ is $1$-Lipschitz, changing the column of one row $i$ from $j$ to $j'$ changes $h_T(\vec w)$ for $\vec w \in \ball$ by at most
    \[
        \frac1n \Bigl| \pos{\ip{\vec w}{\vec z_{ij}}} - \pos{\ip{\vec w}{\vec z_{ij'}}} \Bigr|
        \leq \frac1n \norm{\vec z_{ij} - \vec z_{ij'}}
        \leq \frac{2M}{n}.
    \]
    Hence 
    \[
        \sup_{\vec w \in \ball} |h_\pi(\vec w) - h_{T_X}(\vec w)| \leq \frac{n}{L} \cdot \frac{2M}{n} \leq 2Mn^{-1/3},
    \]
    and together with \cref{equation:cutmid_merge_T} the triangle inequality gives 
    \[
        \rho(P_\pi, (1 - \theta) I + \theta P_\tau) = \Delta(\pi) \leq 18 M n^{-1/3},
    \]
    concluding the proof.
\end{proof}

We now apply \cref{lemma:cutmid_merge} to the permutations in the support of $Q$. To keep the error small we merge permutations along a binary tree. In doing so the error only scales logarithmically with the number of permutations.

\begin{lemma}
    \label[lemma]{lemma:cutmid_tree}
    Recall that $n \geq 2$ and $d_n = \lceil 2 \log n \rceil = \lceil \log(n^2) \rceil$.
    For every $Q \in \DS(n)$ there is a permutation matrix $P$ with
    \[
        \rho(P, Q) \leq 20 Md_nn^{-1/3}.
    \]
\end{lemma}
\begin{proof}
    Let $\eta_0 \ceq 20 M n^{-1/3}$ be the bound of \cref{lemma:cutmid_merge}.
    By \cref{equation:birkhoff_decomposition} and Carath\'eodory's theorem~\cite{schrijver_linear_programming}, we can write $Q = \sum_{s \in [N]} \lambda_s P_s$ as a convex combination of $N \leq n^2$ permutation matrices.
    Place the terms at the leaves of a binary tree of height $\lceil \log N \rceil \leq d_n$, padding with leaves of weight zero.
    For a node $\nu$ with leaf weight $\Lambda_\nu > 0$, let
    \[
        Q_\nu \ceq \frac{1}{\Lambda_\nu} \sum_{s \in \mathrm{leaves}(\nu)} \lambda_s P_s,
    \]
    so that $Q_{\mathrm{root}} = Q$.
    If both children $\nu_0, \nu_1$ of $\nu$ have positive weight, then $Q_\nu = (1 - \theta_\nu) Q_{\nu_0} + \theta_\nu Q_{\nu_1}$ with $\theta_\nu = \Lambda_{\nu_1}/\Lambda_\nu$.
    Bottom-up, give each leaf its $P_s$, let a node with one child of positive weight copy that child's matrix, and otherwise let $P_\nu$ be the matrix provided by \cref{lemma:cutmid_merge} for $P_{\nu_0}, P_{\nu_1}$ and $\theta_\nu$.
    By induction on the height $k(\nu)$, the triangle inequality and \cref{equation:cutmid_convex}, we have
    \begin{align*}
        \rho(P_\nu, Q_\nu)
        &\leq \eta_0 + (1 - \theta_\nu) \rho(P_{\nu_0}, Q_{\nu_0}) + \theta_\nu \rho(P_{\nu_1}, Q_{\nu_1}) \\
        &\leq \eta_0 + (k(\nu) - 1) \eta_0 \\
        &= k(\nu) \eta_0.
    \end{align*}
    Since the root has height at most $d_n$, the matrix $P = P_{\mathrm{root}}$ satisfies $\rho(P, Q) \leq d_n \eta_0$.
\end{proof}

We can now prove \cref{theorem:cut_additive_rounding} by choosing $t$ so as to balance the truncation error $4t$ against the transfer error $2\sqrt2\, \rho(P, Q) / t$ in \cref{cor:cutmid_spectral}.
\begin{proof}[Proof of \cref{theorem:cut_additive_rounding}]
    Recall that $M = \sqrt2$ and that $t > 0$ was arbitrary, so we may choose it below.
    By \cref{lemma:cutmid_tree}, there is a permutation matrix $P$ with $\rho(P, Q) \leq 20\sqrt2\, d_n n^{-1/3}$.
    \Cref{cor:cutmid_spectral} gives
    \[
        \cst{P} \leq \cst{Q} + 4t + \frac{2\sqrt2 \cdot 20\sqrt2\, d_n n^{-1/3}}{t} = \cst{Q} + 4t + \frac{80\, d_n n^{-1/3}}{t}.
    \]
    For $t = \sqrt{20 d_n}\, n^{-1/6}$ both error terms equal $4t$, and $8t = 8\sqrt{20} \sqrt{d_n}\, n^{-1/6} < 36 \sqrt{d_n}\, n^{-1/6}$.
\end{proof}
\subsubsection{Proof of \headingref{Theorem}{theorem:cut_almost_linear}}
\label{appendix:cutmid_main}
\begin{proof}
    The lower bound holds because permutation matrices are doubly stochastic.
    Let $\eps \ceq \ndOT{\cut}(G, H)$ and let $Q \in \DS(n)$ be optimal, so that $\cst{Q} = \eps$.
    Since $\ln(2n) \ln\ln(4n) \geq \ln 4 \cdot \ln\ln 8 > 1$ for all $n \geq 2$, \cref{theorem:cutmid_v2_thin} gives, with $C_1 \ceq 1 + 4 C_{\mathrm{thin}}$,
    \begin{equation}
        \label{equation:cutmid_thin_bound}
        \nd{\cut}(G, H) \leq C_1 \ln(2n) \ln\ln(4n)\, \eps.
    \end{equation}
    This settles $\eps = 0$.
    If $\eps \geq 1/2$, the claim holds for $C \geq 2$, since $\nd{\cut}(G, H) \leq 1$ and both logarithmic factors are at least one.
    So let $0 < \eps < 1/2$ and set $L_\eps \ceq \ln(e/\eps) = 1 + \ln(1/\eps)$.

    We distinguish two cases:
    If $n \leq \eps^{-12}$, then
    \[
        \ln(2n) \leq \ln 2 + 12 \ln(1/\eps) \leq 13 L_\eps
        \qquad\text{and}\qquad
        \ln\ln(4n) \leq \ln(14 L_\eps) \leq 3 \ln(e L_\eps),
    \]
    where the last step uses $\ln 14 < 3$.
    Hence \cref{equation:cutmid_thin_bound} gives the claim with $C = 39 C_1$.

    Now suppose that $n > \eps^{-12}$.
    Then $n > 2^{12}$ and 
    \[
        d_n \leq 2\log n + 1 \leq 3\log n = \frac{3}{\ln 2} \ln n.
    \]
    The function $x \mapsto x^{-1/6} \sqrt{\ln x}$ is decreasing for $x \geq e^3$, so the error term of \cref{theorem:cut_additive_rounding} satisfies
    \begin{align*}
        36 \sqrt{d_n}\, n^{-1/6}
        \leq 36 \sqrt{\tfrac{3}{\ln 2}}\, (\eps^{-12})^{-1/6} \sqrt{12 \ln(1/\eps)} 
        = \frac{216}{\sqrt{\ln 2}}\, \eps \cdot \eps \sqrt{\ln(1/\eps)} 
        \leq \frac{216}{\sqrt{2e \ln 2}}\, \eps 
        < 112\, \eps,
    \end{align*}
    using $\eps \sqrt{\ln(1/\eps)} \leq (2e)^{-1/2}$.
    So $\nd{\cut}(G, H) \leq 113 \eps$, and the claim holds with $C = \max\{2, 39C_1, 113\}$.
    
    Since $\eps^{1 - \alpha} \ln(e/\eps) \ln(e \ln(e/\eps))$ is bounded on $(0, 1]$ for every $\alpha < 1$, it follows that
    \[
        \nd{\cut}(G, H) \leq C_\alpha\, \ndOT{\cut}(G, H)^\alpha
    \]
    for a constant $C_\alpha > 0$ depending only on $\alpha$.
    This concludes the proof of \cref{theorem:cut_almost_linear}.
\end{proof}

%% file: appendices/06_spectral_distance_bounds.tex
\section{The Integrality Gap of the Spectral Distance}
\label{appendix:spectral}
Before considering the integrality gap of $\nd{2}$, we first prove the relation between $\nd{2}$ and $\nd{\cut}$.
\begin{proof}[Proof of \cref{proposition:cut_spectral_equivalence}]
    Let $V, W$ be finite sets, $m \ceq |V|$ and $n \ceq |W|$. We show that for all real matrices $A \in \R^{V \times W}$, it holds that
    \[
        \frac{1}{\sqrt{mn}}\norm{A}_\cut \leq \norm{A}_2 \leq 2 \sqrt{\norm{A}_{(\infty)}\norm{A}_\cut}
    \]
    where $\norm{A}_{(\infty)} \ceq \max_{v \in V, w \in W} |A_{vw}|$.
    This implies the proposition by applying it to the matrices $A_G^\pi - A_H$, which have entrywise maximum norm at most one, and observing the different normalisation factors of $\nd{\cut}$ and $\nd{2}$.

    Let $A \in \R^{V \times W}$.
    For every set $U \subseteq V$, we denote by $\indicator_U \in \mathbb{R}^V$ the
    characteristic vector of $U$, that is, $\indicator_U(i) = 1$ if $i \in U$ and
    $\indicator_U(i) = 0$ otherwise and similarly for subsets of $W$. Observe that $\norm{\indicator_U}_2 \le \sqrt{|U|}$.
    
    For the first inequality, we have
    \begin{align*}
      \|A\|_\square
        &= \max_{S \subseteq V,\, T \subseteq W}
           \left| \sum_{s \in S,\, t \in T} A_{st} \right| \\
        &= \max_{S \subseteq V,\, T \subseteq W}
           \bigl| \langle \indicator_S, A \indicator_T \rangle \bigr| \\
        &\leq \max_{S\subseteq V,\, T \subseteq W}
           \|\indicator_S\|_2 \cdot \|A \indicator_T\|_2
           \tag*{(Cauchy--Schwarz)}\\
        &\leq \sqrt{mn} \, \max_{T \subseteq W, T \neq \emptyset}
           \frac{\|A \indicator_T\|_2}{\|\indicator_T\|_2} \\
        &\leq \sqrt{mn} \, \|A\|_2.
    \end{align*}
    The second inequality was shown by Nikiforov \cite[Theorem 3]{nikiforov2009cutnormsspectramatrices}.
\end{proof}

\subsection{The Lower Bound}
\label{appendix:spectral_lower}
The construction is a simplified version of the one in \cref{appendix:gaps_local_lower}, without the bits and without the need for additional vertices to fix markers.
Similarly to before, bijections incur a large cost compared to fractional alignments, since the latter can spread error across many branches, reducing the spectral norm of the resulting matrix.

Fix an integer $r \geq 1$.
The graphs $G_r$ and $H_r$ have the common vertex set $V \ceq X \cup M$ with the \emph{core vertices} and \emph{markers}
\[
    X \ceq \{u\} \cup \{v_{i, k} \mid i \in [2r], 0 \leq k \leq r\}
    \qquad\text{and}\qquad
    M \ceq C \cup T_1 \cup \cdots \cup T_{2r},
\]
where $C \ceq \{c_1, \ldots, c_r\}$ are the \emph{clock markers} and $T_i \ceq \{t_{i, 1}, \ldots, t_{i, r}\}$ are the \emph{tags} of branch $i \in [2r]$.
Thus $|X| = 2r(r + 1) + 1$, $|M| = r + 2r^2$, and let $n \ceq 4r^2 + 3r + 1$.
In both graphs, $N(v_{i, k}) = T_i \cup \{c_1, \ldots, c_k\}$.
In $G_r$, the vertex $u$ is isolated, while in $H_r$ it is adjacent to all clock markers.
There are no further edges, so $X$ and $M$ are independent sets in both graphs, and $G_r$ and $H_r$ are bipartite.
Along branch $i$, the tags $T_i$ are present throughout, while the clock markers are switched on one by one (\cref{figure:spectral_lower_bound}).
Note that $u$ is isolated in $G_r$ and $H_r$ has minimum degree $r$.

\input{figures/spectral_lower_bound}

It is easy to see that $\d{2}(G_r, H_r) = \sqrt{r}$. Every bijection matches the all-zeros row of $u$ with a row with at least $r$ non-zero entries since $\deg_{H_r}(\pi(u)) \geq r$, so this row has Euclidean norm at least $\sqrt{r}$, and as the spectral norm of a matrix is at least the Euclidean norm of each of its rows, hence $\norm{A_G^\pi - A_H}_2 \geq \sqrt{r}$. Taking the identity alignment, $A_G - A_H$ is the adjacency matrix of a star with $r$ leaves (up to entries being negative) whose spectral norm is $\sqrt{r}$.

\begin{lemma}
    \label{lemma:spectral_gap_fractional}
    $\frac{r}{\sqrt{n}} \leq \dOT{2}(G_r, H_r) \leq 1$.
\end{lemma}
\begin{proof}
    For the upper bound, for $i \in [2r]$, 
    let $\sigma_i$ be the bijection from $G_r$ to $H_r$ that shifts the core vertices of the $i$-th branch along one step, that is $\sigma_i(u) = v_{i,0}$, $\sigma_i(v_{i,k}) = v_{i,k+1}$ for $k \in \{0, \ldots, r-1\}$ and $\sigma_i(v_{i,r}) = u$.
    Then define
    \[
        Q \ceq \frac{1}{2r} \sum_{i = 1}^{2r} P_{\sigma_i}.
    \]
    Similarly to the proof of \cref{lemma:local_gap_fractional}, $Q$ fixes all markers and spreads $u$ uniformly across the first vertices of the $2r$ branches, moves mass $\frac{1}{2r}$ one step along each branch, and moves mass $\frac{1}{2r}$ from the end of each branch to $u$ again.
    For a pair $(x, y) \in \supp(Q)$ of core vertices, let $\vec{d}_{xy} \ceq \indicator_{N_{G_r}(x)} - \indicator_{N_{H_r}(y)} \in \{-1, 0, 1\}^M$.
    The nonzero vectors of this kind are
    \[
        \vec{d}_{u v_{i, 0}} = -\indicator_{T_i},
        \qquad
        \vec{d}_{v_{i, k} v_{i, k + 1}} = -\indicator_{\{c_{k + 1}\}} \text{ for } 0 \leq k < r,
        \qquad\text{and}\qquad
        \vec{d}_{v_{i, r} u} = \indicator_{T_i},
    \]
    and for each such pair $(x, y)$, we have $Q(x, y) = \frac{1}{2r}$.
    As $X$ and $M$ are independent sets in both graphs, $\Disc \ceq \Disc_{G_r, H_r}$ is zero for pairs of atoms that both consist of core vertices or both consist of markers. Moreover, $\Disc\big((x, y), (z, z)\big) = \vec{d}_{xy}(z)$ for a pair $(x, y) \in \supp(Q)$ of core vertices and a marker $z \in M$.
    Hence the matrix $B_{2 \to 2}(G_r, H_r, Q)$ has the block form $\left(\begin{smallmatrix} 0 & K \\ K^\top & 0 \end{smallmatrix}\right)$, where $K$ has rows $\sqrt{Q(x, y)} \vec{d}_{xy}^\top$, and therefore
    \[
        \Cost{2}(Q)^2
        = \norm{K}_2^2 
        = \norm{K^\top K}_2 
        = \Bigl\| \sum_{(x, y) \in \supp(Q)} Q(x, y)\, \vec{d}_{xy} \vec{d}_{xy}^\top \Bigr\|_2 
        = \Bigl\| \frac{1}{r} \sum_{i = 1}^{2r} \indicator_{T_i} \indicator_{T_i}^\top + \sum_{j = 1}^r \indicator_{\{c_j\}} \indicator_{\{c_j\}}^\top \Bigr\|_2.
    \]
    Here every tag set contributes twice with mass $\frac{1}{2r}$, and every clock marker contributes once on each of the $2r$ branches with mass $\frac{1}{2r}$.
    The sets $C, T_1, \ldots, T_{2r}$ are disjoint, so the matrix on the right-hand side is block-diagonal with blocks $\frac1r \indicator_{T_i}\indicator_{T_i}^\top$ of norm $\frac{|T_i|}{r} = 1$ and an identity block for $C$.
    Thus $\Cost{2}(Q) = 1$ and $\dOT{2}(G_r, H_r) \leq 1$.

    For the lower bound, let $R \in \DS(V, W)$ be arbitrary and
    let $\one_R \in \R^{V \times W}$ be the unit vector with $\one_R(x, y) \ceq \sqrt{R(x, y)/n}$ for $(x, y) \in \supp(R)$ and zero otherwise.
    Then, writing $B \ceq B_{2\to2}(G_r, H_r, R)$, we have
    
    \begin{align*}
        \big(B\one_R\big)(x, y)
        &= \sqrt{R(x, y)} \sum_{(x', y') \in \supp(R)} D\big((x, y), (x', y')\big) \frac{R(x', y')}{\sqrt{n}} \\
        &= \one_R(x, y) \big(\deg_{G_r}(x) - \deg_{H_r}(y)\big)
    \end{align*}
    for every $(x, y) \in V \times W$ since $R$ has row and column sums one.
    
    As $u$ is isolated in $G_r$ and $H_r$ has minimum degree $r$, we have $|\deg_{G_r}(u) - \deg_{H_r}(y)| \geq r$ for all pairs $(u, y) \in \supp(R)$, and 
    these pairs satisfy $\sum_{y} \one_R(u, y)^2 = \frac1n \sum_y R(u, y) = \frac1n$.
    Hence
    \begin{align*}
        \Cost{2}(R) &\geq \norm{B\one_R}_2 
        \geq \bigg(\sum_{(u,y) \in \supp(R)} \big( B\one_R(u, y) \big)^2 \bigg)^{1/2}
        \geq \bigg( r^2 \sum_{(u,y) \in \supp(R)} \one_R(u,y)^2 \bigg)^{1/2}
        = \frac{r}{\sqrt{n}},
    \end{align*}
    and thus $\dOT{2}(G_r, H_r) \geq \frac{r}{\sqrt{n}}$.
\end{proof}

\begin{proof}[Proof of \cref{theorem:spectral_lower_bound}]
    With \cref{lemma:spectral_gap_fractional} together with the fact that $\d{2}(G_r, H_r) = \sqrt{r}$, it remains to observe that $\frac{r}{\sqrt{n}} \geq \frac{1}{2\sqrt{2}}$ since $8r^2 \geq 4r^2 + 3r + 1 = n$.
\end{proof}

\subsection{The Upper Bound}
Throughout this section, $G$ and $H$ are graphs of the same order $n$. For simplicity, we assume that $V(G) = V(H) = [n]$. The adjacency matrices of $G, H$ are $A, B \in \{0, 1\}^{n \times n}$.
As in \cref{appendix:cutmid_thin_v2}, matrices act on vectors by ordinary matrix multiplication.
We recall that $\norm{K}_2 \leq \norm{K}_F$ and $\norm{K}_F^2 = \sum_{r \in [n]} \norm{K\vec\phi_r}_2^2$ for every orthonormal basis $(\vec\phi_r)_{r \in [n]}$ of $\R^n$.

We show that the spectral distance is bounded by the square root of its optimal transport relaxation, that is, we prove \cref{theorem:spectral_upper_bound_hard}.
The proof is mostly an adaptation of the multiplicative rounding for the cut distance (\cref{theorem:cutmid_v2_thin}), however a crucial step based on the transport via thin matchings breaks down here and needs to be repaired.
We do this by means of spectral truncation similar to the one that was used in the additive cut estimate (compare \cref{lemma:cutmid_spectral}).
We therefore establish the necessary results for arbitrary matrices $A'$ and $B'$, however they will be chosen to be \enquote{close} to $A$ and $B$ later in the proof.

We first translate \cref{eq:epsQ2} to the setting of the spectral distance. There, the test functions $f, g$ represented cuts in the cut norm. Here, similar functions are used to encode the vectors in the optimisation within the spectral norm.
Let $Q \in \DS(n)$.
For functions $f, g\colon [n]^2 \to \mathbb{R}$, we define the
\emph{$Q$-weighted inner product} and the \emph{$Q$-weighted norm} by
\[
    \langle f, g\rangle_Q := \frac{1}{n}\sum_{i,j\in[n]} Q_{ij} f(i,j) g(i,j)
    \qquad\text{and}\qquad
    \norm{f}_Q := \sqrt{\ip{f}{f}_Q}.
\]
Strictly speaking, $\langle\cdot,\cdot\rangle_Q$ is only a semi-inner product,
as $\norm{f}_Q = 0$ whenever $f$ vanishes on the support of $Q$, however, this does not
affect any of our arguments.
Then, for $A', B' \in \R^{n \times n}$, we let
\begin{equation}
    \label{equation:spectral_v2_eta}
    \scst{Q}(A', B') \ceq 
    \max_{f, g} 
    \frac{1}{n^2} \Bigl| \sum_{i, j, k, \ell \in [n]} f(i, j)\, Q_{ij} Q_{k\ell} \bigl( A'_{ik} - B'_{j\ell} \bigr)\, g(k, \ell) \Bigr|
    \qquad\text{and}\qquad
    \scst{Q} \ceq \scst{Q}(A, B),
\end{equation}
where the maximum is over all $f, g\colon [n]^2 \to \R$ with $\norm{f}_Q, \norm{g}_Q = 1$.
Note that
\[
    \ndOT{2}(G, H) = \min_{Q \in \DS(n)} \scst{Q}
    \qquad\text{and}\qquad
    \nd{2}(G, H) = \min_{P \in \Perm(n)} \scst{P}.
\]

We recall the proof of \cref{theorem:cutmid_v2_thin}. For a given optimal $Q \in \DS(n)$ and a permutation matrix $P = P_\pi$, we let $R \ceq PQ^\top$ so that $(R\vec x)_i = (Q^\top \vec x)_{\pi(i)}$ for all $\vec x \in \R^n$.
Using $RAR^\top = P(Q^\top AQ)P^\top$, we split
\begin{align}
    \label{equation:spectral_v2_split}
    A - PBP^\top &= \bigl( A - RAR^\top \bigr) + P \bigl( Q^\top A Q - B \bigr) P^\top
\end{align}
and bound the norm of the two terms of the right-hand side separately as for \cref{eq:cutmid_v2_thin2}.

The second term can be bounded by $\scst{Q}$ using the following lemma and the fact that the spectral norm is invariant under conjugation with $P$.
\begin{lemma}
    \label[lemma]{lemma:spectral_v2_compress}
    Let $Q \in \DS(n)$ and $A', B' \in \R^{n \times n}$. Then $n^{-1} \norm{Q^\top A' Q - B'}_2 \leq \scst{Q}(A', B')$.
\end{lemma}
The proof is essentially the same as for the first bound in \cref{eq:cutmid_v2_thin2}: testing $\scst{Q}(A',B')$
with $f(i,j) := \sqrt{n}\,x_j$ and $g(k,\ell) := \sqrt{n}\,y_\ell$ for unit
vectors $\x, \y \in \mathbb{R}^n$ yields $n^{-1}\bigl|\x^\top(Q^\top A' Q - B')\y\bigr|$,
and since the columns of $Q$ sum to one, these lifts satisfy the norm constraint
in \cref{equation:spectral_v2_eta}, just as the indicator of $[n]\times X$ is an admissible test set for the cut norm. 
Then, maximising over $\x, \y$ and applying \cref{lemma:norm_bilinear_identity} proves the claim.

Expanding $A - RAR^\top = EA + R(EA)^\top$ with $E \ceq I - R$ (see \cref{equation:arar_to_ea}), our goal is to show that $\norm{EA}_2$ is small. 
This suffices because $\norm{RK}_2 \leq \norm{R}_2 \norm{K}_2 \leq \norm{K}_2$ for $R \in \DS(n)$, as $R$ is a convex combination of permutation matrices, which have spectral norm one.
We bound $\norm{EA'}_2$ where $A'$ is a suitable approximation of $A$.
For a test vector $\y$ and $\x \ceq A'\y$ it holds that $(EA'\y)_i = \x_i - (Q^\top \x)_{\pi(i)}$, so we need the value of $\x$ at each $i$ to be close to its $Q$-weighted average over the vertices coupled with $\pi(i)$.

We do this in two steps.
The first is a local agreement property similar to the one that was shown for the cut distance:
We show that
$x_i \approx (Q^\top x)_j$ on average over the pairs
$(i,j)$ weighted by $Q$.
\begin{lemma}[Local Agreement]
    \label[lemma]{lemma:spectral_v2_local}
    Let $Q \in \DS(n)$, $A', B' \in \R^{n \times n}$ and $\vec y \in \R^n$, and let $\vec x \ceq A'\vec y \in \R^n$.
    Then
    \[
        \frac{1}{n^2} \sum_{i, j \in [n]} Q_{ij} \bigl( x_i - (Q^\top \vec x)_j \bigr)^2 \leq \scst{Q}(A', B')^2\, \norm{\vec y}_2^2.
    \]
\end{lemma}
\begin{proof}
As in the proof of \cref{lemma:cutmid_v2_local}, for $i, j \in [n]$, let
\[
  F_{ij} := \sum_{k,\ell\in[n]} Q_{k\ell}\bigl(A'_{ik} - B'_{j\ell}\bigr)\, y_k
  = x_i - r_j, \qquad \text{where } \vec r := B'Q^\top \y,
\]
and the identity uses that the rows of $Q$ sum to one. We first show that
\begin{align}
    \label{equation:spectral_v2_local_F}
    \frac{1}{n^2}\sum_{i,j\in[n]} Q_{ij} F_{ij}^2 \leq \scst{Q}(A',B')^2\, \norm{\y}_2^2.
\end{align}
To this end, define $S \ceq \big(\sum_{i,j\in[n]} Q_{ij} F_{ij}^2\big)^{1/2}$.
Then
\[
  \frac{S^2}{n^2} = \frac{1}{n^2}\sum_{i,j\in[n]} F_{ij}\, Q_{ij} F_{ij}
  = \frac{1}{n^2}\sum_{i,j,k,\ell\in[n]} F_{ij}\, Q_{ij} Q_{k\ell}\bigl(A'_{ik} - B'_{j\ell}\bigr)\, y_k
\]
is the expression in \cref{equation:spectral_v2_eta} for the functions $f, g\colon [n]^2 \to \mathbb{R}$
with $f(i,j) := F_{ij}$ and $g(k,\ell) := y_k$. As this expression is bilinear
in $f$ and $g$, it is at most $\scst{Q}(A',B')\,\norm{f}_Q\,\norm{g}_Q$, where
$\norm{f}_Q = \frac{1}{\sqrt{n}} S$ and, since the rows of $Q$ sum to one,
$\norm{g}_Q = \frac{1}{\sqrt{n}}\norm{\y}_2$. Hence $S^2 \le n\,\scst{Q}(A',B')\,\norm{\y}_2\, S$,
which proves \cref{equation:spectral_v2_local_F}, the case $S = 0$ being trivial.

Since the columns of $Q$ sum to one, we have
$(Q^\top \x)_j = \sum_{i'} Q_{i'j}(F_{i'j} + r_j) = r_j + m_j$ with
$m_j := \sum_{i'\in[n]} Q_{i'j} F_{i'j}$, and hence
\[
  x_i - (Q^\top \x)_j = F_{ij} - m_j \qquad \text{for all } i, j \in [n].
\]
Expanding the square and using once more that the columns of $Q$ sum to one,
we have
\[
  \frac{1}{n^2}\sum_{i,j\in[n]} Q_{ij}\bigl(x_i - (Q^\top \x)_j\bigr)^2
  = \frac{1}{n^2}\sum_{i,j\in[n]} Q_{ij} F_{ij}^2 - \frac{1}{n^2}\sum_{j\in[n]} m_j^2
  \le \scst{Q}(A',B')^2\, \norm{\y}_2^2. \qedhere
\]
\end{proof}

\noindent Compared to \cref{lemma:cutmid_v2_local}, the factors of two disappear since we can test with $F$
itself rather than with its positive and negative parts.

In a second step, we now want to transfer this guarantee from the pairs $(i, j)$ weighted by $Q$ to the pairs $(i, \pi(i))$.
However, here we cannot rely on a thinness argument as for the cut distance (\cref{lemma:cutmid_v2_transport}). Indeed, the graphs we constructed in \cref{appendix:spectral_lower} show that the loss of an $\ell_2$-version of such a transfer can, in general, grow with the square root of the order of the graphs.

We therefore follow a different strategy, again rounding to a permutation in a Birkhoff decomposition of $Q$ by making a random choice, see \cref{equation:birkhoff_decomposition}.

\begin{lemma}
    \label[lemma]{lemma:spectral_v2_random}
    Let $Q \in \DS(n)$, let $\boldsymbol P = P_{\boldsymbol\pi}$ be a random permutation matrix with $\E{\boldsymbol P} = Q$, and let $\boldsymbol R \ceq \boldsymbol P Q^\top$.
    Then for every $\vec x \in \R^n$,
    \[
        \Expected \norm{(I - \boldsymbol R)\vec x}_2^2 = \sum_{i, j \in [n]} Q_{ij} \bigl( x_i - (Q^\top \vec x)_j \bigr)^2.
    \]
\end{lemma}
\begin{proof}
    We have $((I - \boldsymbol R)\vec x)_i = x_i - (Q^\top \vec x)_{\boldsymbol\pi(i)}$ and $\Pr{\boldsymbol\pi(i) = j} = \E{\boldsymbol P_{ij}} = Q_{ij}$ for all $i, j \in [n]$.
    Linearity of expectation proves the claim.
\end{proof}
Thus, a random permutation satisfies this \enquote{$\ell_2$-version} of \cref{lemma:cutmid_v2_transport} with constant one on average for each fixed test vector.
Of course, the permutation we finally choose has to be good for all test vectors simultaneously.
This is the central obstacle in the following proof.

\begin{proof}[Proof of \cref{theorem:spectral_upper_bound_hard}]
    Let $Q \in \DS(n)$ be optimal, so that $\scst{Q} = \ndOT{2}(G, H)$.
    We start by \enquote{centering} the matrices $A$ and $B$, which allows us to control parts of the rounding error via the Frobenius norm later in the proof.
    Let $p_G \ceq 2|E(G)|/n^2$ and $p_H \ceq 2|E(H)|/n^2$, and define
    $h \ceq \min\big\{ \sqrt{p_G(1 - p_G)}, \sqrt{p_H(1 - p_H)} \big\}$.
    By symmetry, we may assume that $h = \sqrt{p_G(1-p_G)}$.
    Letting $J$ be the all-ones matrix, we define $A' \ceq A - p_G J$ and $B' \ceq B - p_G J$.
    Then $\Disc_{A', B'} = \Disc_{A, B}$, so $\scst{Q}(A', B') = \scst{Q}$, and $A' - PB'P^\top = A - PBP^\top$ for every $P \in \Perm(n)$, as $PJP^\top = J$.
    Since $A_{ik}^2 = A_{ik}$ and $\sum_{i, k \in [n]} A_{ik} = n^2 p_G$, we have $\norm{A'}_F^2 = n^2 (p_G - p_G^2) = n^2 h^2$.

    Since $A'$ is symmetric, by the spectral theorem, we can write $A' = \sum_{r \in [n]} \lambda_r \vec\phi_r \vec\phi_r^\top$ where $(\vec\phi_r)_{r \in [n]}$ is an  orthonormal basis.
    For $t > 0$, let
    \[
        A'_t \ceq \sum_{r \in I_t} \lambda_r \vec\phi_r \vec\phi_r^\top
        \qquad\text{with}\qquad
        I_t \ceq \{r \in [n] : |\lambda_r| > nt\}
    \]
    be its spectral truncation. (Note the similarity to the spectral truncation used in the additive cut estimate, compare \cref{lemma:cutmid_spectral}.)
    This truncation reduces the size of the orthonormal basis, however the resulting simplified representation is lossy. This yields two different sources of error which we bound in the following.
    For the first kind of error, as $A' - A'_t$ has all its eigenvalues in $[-nt, nt]$, we have $\norm{A' - A'_t}_2 \leq nt$.

    Turning to the second kind of error, we first show that $|I_t|$ can be bounded by a function of $t$. Recalling that $n^2h^2 = \norm{A'}_F^2$, we have
    \begin{align*}
        n^2h^2 = \norm{A'}_F^2 = \sum_{r \in [n]} \lambda_r^2 \geq \sum_{r \in I_t} \lambda_r^2 \geq \sum_{r \in I_t} (nt)^2 = n^2t^2|I_t|,
    \end{align*}
    so $|I_t| \leq h^2/t^2$.
    Our goal is to now show that there is a permutation $P$ within the support of $Q$ such that its spectral norm restricted to optimisation over vectors in the span of the eigenvectors of eigenvalues in $I_t$ is bounded by a quantity depending on $\scst{Q}$.
    Intuitively, we consider the expected sum of costs for all eigenvectors in $I_t$ where the expectation is over the choice of $\boldsymbol P$. 
    Hence if the total cost is small, there must also exist some choice of $P$ with small cost.
    
    Since the maximum does not commute with expectation, we first bound the spectral norm by the Frobenius norm and then use linearity of expectation to bound the spectral norm for each of the $|I_t|$ eigenvectors individually.

    Let $\boldsymbol P$ and $\boldsymbol R \ceq \boldsymbol P Q^\top$ as in \cref{lemma:spectral_v2_random}, and define $\boldsymbol E \ceq I - \boldsymbol R$.
    Then, we have
    \begin{align}
        \label{equation:expectation_spectral}
        \Expected \norm{\boldsymbol E A'_t}_2^2
        \leq \Expected \norm{\boldsymbol E A'_t}_F^2
        = \sum_{r \in I_t} \Expected \norm{\boldsymbol E A'\vec\phi_r}_2^2
        \leq n^2 \scst{Q}^2 |I_t|
        \leq \frac{n^2 \scst{Q}^2 h^2}{t^2},
    \end{align}
    where the first inequality holds since $\|\cdot\|_2 \le \|\cdot\|_F$.
    For the equality, let $\Phi$ be the orthogonal matrix with columns
    $\boldsymbol\phi_r$. Since the Frobenius norm is invariant under
    multiplication with orthogonal matrices, every 
    matrix $M \in \mathbb{R}^{n\times n}$ satisfies
    $\|M\|_F^2 = \|M\Phi\|_F^2 = \sum_{r\in[n]} \|M\boldsymbol\phi_r\|_2^2$.
    We apply this to $M := \boldsymbol{E}A'_t$ and use that
    $A'_t\boldsymbol\phi_r = \lambda_r\boldsymbol\phi_r = A'\boldsymbol\phi_r$
    for $r \in I_t$ and $A'_t\boldsymbol\phi_r = \boldsymbol{0}$ for $r \notin I_t$;
    linearity of expectation then gives the equality.
    The next inequality follows by chaining \cref{lemma:spectral_v2_random} and \cref{lemma:spectral_v2_local} with $\x \ceq  A'\vec\phi_r$ and $\y \ceq \vec \phi_r$, respectively, using that $\norm{\vec\phi_r}_2 = 1$.
    The last inequality then follows from the size bound on $I_t$.

    Taking the square root of both sides in \cref{equation:expectation_spectral} therefore shows that there is a value $P$ of $\vec P$ that satisfies $\norm{EA'_t}_2 \leq n\scst{Q} h/t$ for $R \ceq PQ^\top$ and $E \ceq I - R$.

    Splitting $A' = (A' - A'_t) + A'_t$, we have
    \begin{align*}
        \norm{A' - RA'R^\top}_2
        &= \norm{(A' - A'_t) - R(A' - A'_t)R^\top + A'_t - RA'_tR^\top}_2 \\
        &\leq \norm{(A' - A'_t) - R(A' - A'_t)R^\top}_2  + \norm{A'_t - RA'_tR^\top}_2 \\
        &\leq \norm{(A' - A'_t) - R(A' - A'_t)R^\top}_2  + 2\norm{EA'_t}_2\tag{a} \\
        &\leq \norm{(A' - A'_t) - R(A' - A'_t)R^\top}_2  + \mfrac{2n\scst{Q}h}{t} \\
        &\leq \norm{A' - A'_t}_2 + \norm{R(A' - A'_t)R^\top}_2  + \mfrac{2n\scst{Q}h}{t} \\
        &\leq 2nt + \mfrac{2n\scst{Q} h}{t}
    \end{align*}
    where we used \cref{equation:arar_to_ea} in step (a) with $A'_t$ instead of $A$ and the fact that $\norm{R}_2 \leq 1$ for the last inequality.
   
    Hence, together with \cref{equation:spectral_v2_split} for $A', B'$ and \cref{lemma:spectral_v2_compress}, we get
    \[
        \nd{2}(G, H) \leq \scst{P} = \mfrac1n \norm{A' - PB'P^\top}_2 \leq \scst{Q} + 2t + \mfrac{2\scst{Q} h}{t}.
    \]
    If $\scst{Q} h > 0$, we choose $t \ceq \sqrt{\scst{Q} h}$ to balance the two error terms optimally and obtain $\nd{2}(G, H) \leq \scst{Q} + 4\sqrt{\scst{Q} h}$.
    Otherwise, letting $t \to 0$ shows that $\nd{2}(G, H) \leq \scst{Q}$ as the second error term is always zero.
    Finally, we always have $h \leq \frac12$, since $x(1 - x) \leq \frac14$ for all $x \in \R$, so $4\sqrt{h\scst{Q}} \leq 2\sqrt{2\scst{Q}}$.
\end{proof}

%% file: figures/spectral_lower_bound.tex
\input{figures/palette}
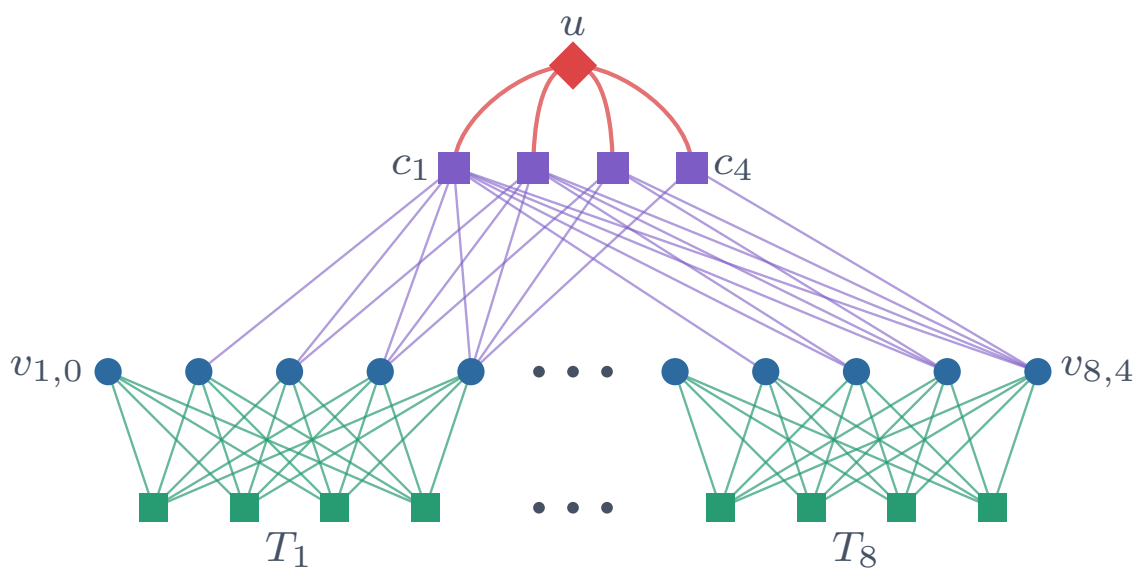
\begin{figure}[t]
    \centering
    \begingroup
    \colorlet{SpecData}{figBlue}
    \colorlet{SpecClock}{figPurple}
    \colorlet{SpecTag}{figGreen}
    \colorlet{SpecAdded}{figRed}
    \colorlet{SpecText}{figInk}
    \begin{tikzpicture}[
        x=0.75cm, y=0.75cm,
        line cap=round, line join=round,
        every node/.style={inner sep=0pt, outer sep=0pt},
        spec data/.style={circle, fill=SpecData, minimum size=1.8mm},
        spec clock/.style={rectangle, fill=SpecClock, minimum size=2.1mm},
        spec tag/.style={rectangle, fill=SpecTag, minimum size=1.9mm},
        spec source/.style={diamond, aspect=1, fill=SpecAdded, minimum size=3.2mm},
        spec label/.style={font=\scriptsize, text=SpecText},
        spec clock edge/.style={draw=SpecClock, line width=0.45pt, opacity=0.6},
        spec tag edge/.style={draw=SpecTag, line width=0.45pt, opacity=0.7},
        spec added edge/.style={draw=SpecAdded, line width=0.8pt, opacity=0.75}
    ]
        \foreach \i/\x in {1/0, 8/5} {
            \foreach \k in {0,...,4} {
                \coordinate (v-\i-\k) at ({\x + \k * 0.8}, 0);
            }
            \foreach \h in {1,...,4} {
                \coordinate (t-\i-\h) at ({\x + 1.6 + (\h - 2.5) * 0.8}, -1.2);
            }
        }
        \foreach \j in {1,...,4} {
            \coordinate (c-\j) at ({4.1 + (\j - 2.5) * 0.7}, 1.8);
        }
        \coordinate (u) at (4.1, 2.7);

        \foreach \i in {1,8} {
            \foreach \k in {0,...,4} {
                \foreach \h in {1,...,4} {
                    \draw[spec tag edge] (v-\i-\k) -- (t-\i-\h);
                }
            }
            \foreach \j in {1,...,4} {
                \foreach \k in {\j,...,4} {
                    \draw[spec clock edge] (v-\i-\k) -- (c-\j);
                }
            }
        }
        \foreach \j/\out in {1/180, 2/180, 3/0, 4/0} {
            \draw[spec added edge] (u) to[out=\out, in=90, looseness=0.8] (c-\j);
        }

        \foreach \i in {1,8} {
            \foreach \k in {0,...,4} {
                \node[spec data] at (v-\i-\k) {};
            }
            \foreach \h in {1,...,4} {
                \node[spec tag] at (t-\i-\h) {};
            }
            \node[spec label, below=1.6mm] at ($(t-\i-2)!0.5!(t-\i-3)$) {$T_{\i}$};
        }
        \foreach \y in {0, -1.2} {
            \foreach \d in {-0.3, 0, 0.3} {
                \fill[SpecText] ({4.1 + \d}, \y) circle (1.1pt);
            }
        }
        \node[spec label, left=1.5mm] at (v-1-0) {$v_{1, 0}$};
        \node[spec label, right=1.5mm] at (v-8-4) {$v_{8, 4}$};
        \foreach \j in {1,...,4} {
            \node[spec clock] at (c-\j) {};
        }
        \node[spec label, left=1.4mm] at (c-1) {$c_1$};
        \node[spec label, right=1.4mm] at (c-4) {$c_4$};
        \node[spec source] at (u) {};
        \node[spec label, above=2mm] at (u) {$u$};
    \end{tikzpicture}
    \endgroup
    \caption{The graph $H_4$ of \cref{theorem:spectral_lower_bound}, only the first and last branch are drawn.
    The vertices $v_{i,0}, \ldots, v_{i, 4}$ of a branch $i$ are arranged horizontally and fully connected to its corresponding tag set $T_i$.
    Every vertex $v_{i, k}$ is adjacent to the clock markers $c_1, \ldots, c_k$.
    The graph $G_4$ is obtained by removing the red edges incident to $u$.}
    \label{figure:spectral_lower_bound}
\end{figure}

%% file: appendices/07_hardness.tex
\section{Proofs Omitted from \headingref{Section}{section:hardness}}
\label{appendix:hardness}
\begin{proof}[Proof of \cref{theorem:ot_edit_dist_hardness}]
    We give a reduction from the following \NP-hard variant of the \threepartition{} problem \cite{garey1979computers}.
    The input consists of integers $A$ and $a_1, \ldots, a_{3m}$ in unary representation satisfying $\sum_{i=1}^{3m} a_i = mA$ and $A/4 < a_i < A/2$ for $1 \leq i \leq 3m$.
    The question then is to decide whether $a_1, \ldots, a_{3m}$ can be partitioned into $m$ triples such that the sum of each triple is exactly $A$.
    Given an instance $(A, a_1, \ldots, a_{3m})$, we construct a forest $F_1$ which is the disjoint union of paths on $a_i$ vertices for each $i \in [3m]$.
    The forest $F_2$ is the disjoint union of $m$ paths on $A$ vertices.
    Moreover, we set $t \ceq 4m$. Denote the vertex sets of $F_1$ and $F_2$ by $V$ and $W$, respectively.
   
    We claim that the resulting instance $(F_1, F_2, t)$ is a YES-instance of \GSim{\dOT{\edit}} if and only if $(A, a_1, \ldots, a_{3m})$ is a YES-instance of \threepartition{}.
    By \cite[Theorem 3.10]{grohe2025thoughtsgraphsimilarity}, $\dOT{\edit}(F_1, F_2) = \dBU{\edit}(F_1, F_2)$, so we may argue with blowups.
    In the following, we assume that $a_i \geq 4$ for all $i \in [3m]$. It is easy to see that we can always add 3 to every $a_i$ and $9$ to $A$ to ensure that this is the case.

    Given a YES-instance of \threepartition{}, let $\sigma: [3m] \to [m]$ be the assignment of the $a_i$ to the sums they occur in, \ie{} $|\sigma^{-1}(\{\ell\})| = 3$ for every $\ell \in [m]$.
    Clearly, we can map the paths of $F_1$ to the paths of $F_2$ according to $\sigma$, incurring 2 mismatches per triple, let $\rho \in \Bij(V, W)$ be the corresponding mapping.
    Note that $\rho$ maps every edge of $F_1$ to an edge of $F_2$, so the mismatches are exactly the $2m$ edges of $F_2$ that are not hit by $\rho$.
    For any $k \geq 1$, 
    consider the bijection $\rho^{\blowup k} \in \Bij(V \times [k], W \times [k])$ between $F_1^{\blowup k}$ and $F_2^{\blowup k}$ defined by
    $\rho^{\blowup k}\big( (v, j) \big) := (\rho(v), j)$
    for $v \in V$ and $j \in [k]$. 
    In words, the clouds of the blowups get mapped to each other as if they were the same vertex.
    The number of mismatches in the $k$-blowup increases by a factor of $k^2$ compared to the underlying graphs, so there are $2mk^2$ mismatches, and as $\norm{\cdot}_{(1)}$ counts every mismatched edge twice we obtain
    $\d{\edit}(F^{\blowup k}_1, F^{\blowup k}_2) \leq 4mk^2$.
    Since this holds for all $k \geq 1$ it follows that $\dBU{\edit}(F_1, F_2) \leq t$.

    Conversely, suppose that $\d{\edit}^{\blowup}(F_1,F_2)\leq t$,
    and set $\gamma \ceq \frac{1}{2mA}$. Then, there exist $k\geq 1$ and an alignment
    $\pi$ of the $k$-blowups $F_1' \ceq F_1^{\blowup k}$ and
    $F_2'\ceq F_2^{\blowup k}$ such that
    $\norm{A_{F_1'}^\pi-A_{F_2'}}_{(1)}\leq (4m+4\gamma)k^2$.
    Let $b$ be the number of edges of $F_1'$ that $\pi$ maps to
    non-edges of $F_2'$. Then exactly $|E(F_2')|-|E(F_1')|+b$ edges
    of $F_2'$ are not hit. Since each edge discrepancy contributes
    twice, the edge counts give
    \[
        \norm{A_{F_1'}^\pi-A_{F_2'}}_{(1)}=4mk^2+4b.
    \]
    Thus $b\leq\gamma k^2$.
    For $v\in V$ and $j\in[m]$, let $n_j(v)$ be the number of
    vertices in the cloud of $v$ that $\pi$ maps into the blowup of
    the $j$-th path of $F_2$. In particular, $\sum_j n_j(v)=k$.
    If $vv'\in E(F_1)$, an edge between their clouds can be preserved
    only if both endpoints are mapped into the same path of $F_2$.
    Hence at least
    \[
    k^2-\sum_j n_j(v)n_j(v')
    \ge k\bigl(k-\max_j n_j(v)\bigr)
    \]
    of the edges between these clouds are not preserved. These edges
    are counted by $b$, so $\max_j n_j(v)\geq(1-\gamma)k$.
    Every vertex in $V$ has a neighbour, so this condition holds for every $v \in V$.
    As $\gamma = \frac{1}{2mA} \leq 1/4$, each $v \in V$ has
    a unique index $j(v)$ satisfying $n_{j(v)}(v)\ge(1-\gamma)k$.
    
    We claim that $j(v)=j(v')$ whenever $vv'\in E(F_1)$.
    Otherwise, writing $a:=j(v)$, we have $n_a(v')\le\gamma k$,
    because $j(v')\ne a$. Also,
    $\sum_{j\ne a}n_j(v)=k-n_a(v)\le\gamma k$. Consequently,
    \[
    \sum_j n_j(v)n_j(v')
    \le k\,n_a(v')+k\sum_{j\ne a}n_j(v)
    \le2\gamma k^2.
    \]
    At least $(1-2\gamma)k^2$ edges between these clouds are therefore
    not preserved. This implies
    $b\ge(1-2\gamma)k^2>\gamma k^2$, since $\gamma\le1/4$,
    contradicting the bound on $b$.
    Thus $j(\cdot)$ is constant on every path $P_i$ of $F_1$. We
    denote this common value by $j(i)$ for each $i \in [3m]$.

    Finally, let $j \in [m]$ and $S_j \ceq \sum_{i \,:\, j(i) = j} a_i$. As $\pi$ is a bijection, $\sum_v n_j(v) = Ak$, where the $S_j$ vertices of paths $P_i$ with $j(i) = j$ contribute more than $(1 - \gamma)k$ each and all other vertices less than $\gamma k$ each.
    Thus $|A - S_j| < \gamma mA = \frac{1}{2}$, so $S_j = A$, and as $A/4 < a_i < A/2$, exactly three paths satisfy $j(i) = j$.
    Hence we obtain a valid solution for the \threepartition{}-instance.
\end{proof}

\begin{proof}[Proof of \cref{theorem:ot_local_edit_dist_hardness}]
    We again reduce from \textsc{Three-Partition}. This time $F_2$ is the collection of $m$ disjoint cycles on $A$ vertices each.
    As in the proof of \cref{theorem:ot_edit_dist_hardness}, we may assume that $a_i \geq 3$ for all $i \in [3m]$: adding $3$ to every $a_i$ and $9$ to $A$ yields an equivalent instance that still satisfies $A/4 < a_i < A/2$. In particular every path of $F_1$ has at least three vertices.
    We show that $\dBU{1}(F_1, F_2) \leq 1$ if $\mathcal I = (A, a_1, \ldots, a_{3m})$ is a YES-instance, and $\dBU{1}(F_1, F_2) \geq 2$ otherwise.
    
    A YES-instance of \textsc{Three-Partition} gives a mapping of the triples into the cycles such that the vertices at the endpoints of each path in $F_1$ get an additional neighbour in $F_2$. Let $\pi \in \Sym([mA])$ be the corresponding permutation.
    Then for each $k \geq 1$, define $\pi^{\blowup k} \in \Sym([mA]\times[k])$ by letting $\pi^{\blowup k}(v, j) := (\pi(v), j)$, that is $\pi^{\blowup k}$ maps clouds in $F^{\blowup k}_1$ to clouds in $F^{\blowup k}_2$.
    Clearly, the vertices in the clouds corresponding to endpoints of paths of $F_1$ gain $k$ new neighbours under $\pi^{\blowup k}$ whereas (non-)adjacency is preserved for all other vertices of $F^{\blowup k}_1$.
    Hence $\d{1}(F^{\blowup k}_1, F^{\blowup k}_2) \leq k$ which proves $\dBU{1}(F_1, F_2) \leq 1$.

    For the other direction, assume that $(A, a_1, \ldots, a_{3m})$ is a NO-instance, let $k \geq 1$ and let $\pi \in \Sym([mA] \times [k])$ be an arbitrary alignment of $F^{\blowup k}_1$ and $F^{\blowup k}_2$.
    We show that some vertex incurs a mismatch of at least $2k$ under $\pi$.
    For $i \in [3m]$, we write $P^{\blowup k}_i$ for the $k$-blowup of the path $P_i$ of $F_1$. Similarly, for each $j \in [m]$, we write $C^{\blowup k}_j$ for the blowup of the cycle $C_j$ in $F_2$.

    Note that it cannot hold that for each $i \in [3m]$ there is a $j \in [m]$ such that
    \begin{equation}
        \label{eq:subclaim-rev}
        \pi\left( V\big(P^{\blowup k}_i\big) \right) \subseteq V\big(C^{\blowup k}_j\big).
    \end{equation}
    Indeed, otherwise, since $\pi$ is a bijection, and since $\big|C^{\blowup k}_j\big| = Ak$ for each $j \in [m]$, and $\big|P^{\blowup k}_i\big| = a_ik$ and $A/4 < a_i < A/2$ for each $i \in [3m]$, it follows that exactly three paths $P^{\blowup k}_i$ are mapped to each $C^{\blowup k}_j$.
    Hence, the corresponding partition mapping $\sigma: [3m] \to [m]$ yields a solution for the original instance $(A, a_1, \ldots, a_{3m})$, contradicting that $\mathcal I$ is a NO-instance.

    Hence there is a path $P$ whose blowup is mapped to multiple blown-up cycles.
    There are two cases to distinguish.
    \subparagraph*{Case 1.}
    $\pi$ maps all vertices of each cloud of $P^{\blowup k}$ to the same cycle.
    Since $\pi$ does not map all vertices to the same cycle, there are vertices $v, w \in V(P)$ with $vw \in E(P)$ such that $\{v\} \times [k]$ is mapped to a different cycle (blowup) in $F^{\blowup k}_2$ than $\{w\} \times [k]$ under $\pi$.
    Since each path has at least three vertices, at least one of $v$ and $w$ has degree two in $P$, \WLOG{} we assume this to be $v$. Then clearly $\pi(v, j)\pi(w, j') \notin E(F^{\blowup k}_2)$ for any choice $j, j' \in [k]$, so the vertices $(v, j)$ are missing $k$ adjacencies from $F^{\blowup k}_1$. Moreover since $\deg_{F^{\blowup k}_2}(\pi(v, j)) = 2k$ there are at least $k$ vertices that $\pi(v, j)$ is adjacent to in $F^{\blowup k}_2$ whose preimages under $\pi$ are not adjacent to $(v, j)$ in $F^{\blowup k}_1$.
    Hence the mismatch count for $(v, j)$ is at least $2k$.

    \subparagraph*{Case 2.}
    A cloud is split across multiple cycles.
    In particular, there is $v \in V(P)$ and $j, j' \in [k]$ such that $N_{F^{\blowup k}_2}(\pi(v, j)) \cap N_{F^{\blowup k}_2}(\pi(v, j')) = \emptyset$ and $|N_{F^{\blowup k}_2}(\pi(v, j)) \cup N_{F^{\blowup k}_2}(\pi(v, j'))| = 4k$.
    Letting $d := \deg_{F_1}(v)$, the set $B := N_{F^{\blowup k}_1}(v, j) = N_{F^{\blowup k}_1}(v, j')$ has size $dk$. Let $\ell, \ell'$ be the number of neighbours of $\pi(v, j)$ (respectively $\pi(v, j')$) in $F^{\blowup k}_2$ that are also neighbours of $(v, j)$ in $F^{\blowup k}_1$ (respectively $(v, j')$).
    Clearly, $\ell + \ell' \leq dk$.
    Then, since $(v, j)$ only keeps $\ell$ adjacencies, equivalently it loses $dk - \ell$ adjacencies.
    However since $\deg_{F^{\blowup k}_2}(\pi(v, j)) = 2k$ it also gains adjacency to $2k - \ell$ other vertices whose preimage under $\pi$ is not adjacent to $(v, j)$. The same holds for $(v, j')$.
    Thus, the total number of mismatches of $(v, j)$ under $\pi$ is
    \[
        (2k - \ell) + (dk - \ell) = 2k + (dk - 2\ell).
    \]
    Since $\ell + \ell' \leq dk$, either $dk - 2\ell \geq 0$ or $dk - 2\ell' \geq 0$ which implies that the mismatch count of at least one of $(v, j)$ and $(v, j')$ under $\pi$ is at least $2k$.
    \smallskip\newline
    In either case, it follows that there is a vertex incurring a mismatch of at least $2k$ under $\pi$.
    Since $\norm{A^\pi_{F^{\blowup k}_1} - A_{F^{\blowup k}_2}}_1$ is the maximum column sum, \ie{} the maximum mismatch count of a vertex, and $\pi$ was arbitrary, we obtain $\d{1}(F^{\blowup k}_1, F^{\blowup k}_2) \geq 2k$ for every $k \geq 1$, and hence $\dBU{1}(F_1, F_2) \geq 2$.

    By \cref{theorem:blowup_ot_identity}, $\dOT{1} = \dBU{1}$, so YES-instances satisfy $\dOT{1}(F_1, F_2) \leq 1$ and NO-instances satisfy $\dOT{1}(F_1, F_2) \geq 2$.
    A polynomial-time algorithm approximating $\dOT{1}$ (equivalently, $\ndOT{1}$) within a factor of $2 - \eps$ would distinguish the two cases and thus decide \threepartition{}.
    In particular, \GSim{\dOT{1}} is \NP-hard.
    This proves correctness of our reduction and concludes the proof.
\end{proof}

Our main tool for proving \cref{theorem:ot_op_p_general_hardness}
is a lower bound that reduces the analysis of
$\norm{B_p(G,H,Q)}_p$ to that of a smaller matrix which averages the discrepancies according to $Q$.
For $Q \in \DS(V, W)$, define the $V \times V$ matrix $A_{G,H}^Q \ceq A_G - QA_HQ^\top$ with entries
\begin{equation}
    \label{equ:entrywise_def}
    A_{G,H}^Q(v,v') \ceq
    \sum_{w,w' \in W}
    Q(v,w)\,\Disc_{G,H}(v,w,v',w')\,Q(v',w')
\end{equation}
for all $v,v' \in V$.
The following lemma allows us
to derive a lower bound from $A_{G,H}^Q$.

\begin{lemma}
    \label{lemma:lower_bound_operator_p_blowup}
    For all graphs $G, H$ on the same number of nodes, all $Q \in \DS(V, W)$, and all $1 \leq p \leq \infty$ it holds that
    \[
        \norm{A_{G, H}^Q}_p \leq \norm{B_p(G, H, Q)}_p.
    \]
\end{lemma}

\begin{proof}[Proof of \cref{lemma:lower_bound_operator_p_blowup}]
    By \cref{lemma:norm_bilinear_identity}, let $\x, \y \in \R^V$ with $\norm{\x}_p = 1$ and $\norm{\y}_{p^\ast} = 1$ such that $\norm{A_{G, H}^Q}_p = \left| \y^\top A_{G, H}^Q \x \right|$.
    We define vectors $\Tilde{\x}, \Tilde{\y} \in \R^{V \times W}$ by letting
    \begin{align*}
        \Tilde{\y}(v, w) &\ceq Q(v, w)^{1/p^\ast}\y(v) = Q(v, w)^{1-1/p}\y(v), \quad\text{and}\\
        \Tilde{\x}(v, w) &\ceq Q(v, w)^{1/p}\x(v)
    \end{align*}
    for all $v \in V, \, w \in W$.
    We claim that $\norm{\Tilde{\y}}_{p^\ast} = 1$ and $\norm{\Tilde{\x}}_p = 1$.
    Let $(\z, r) \in \{ (\x, p), (\y, p^\ast) \}$.
    If $r < \infty$, then, as $\sum_{w \in W} Q(v, w) = 1$ for all $v \in V$,
    \[
        \norm{\Tilde{\z}}_r
        = \left( \sum_{(v, w) \in V \times W} Q(v, w) \left| \z(v) \right|^r \right)^{1/r}
        = \left( \sum_{v \in V} |\z(v)|^r \sum_{w \in W} Q(v, w) \right)^{1/r} = \norm{\z}_r = 1.
    \]
    If $r = \infty$, then by the convention $0^0 = 0$ we have $\Tilde{\z}(v, w) = \z(v)$ if $Q(v, w) > 0$ and $\Tilde{\z}(v, w) = 0$ otherwise; since every row of $Q$ has a positive entry, $\norm{\Tilde{\z}}_\infty = \norm{\z}_\infty = 1$.

    Moreover, $Q(v, w) = Q(v, w)^{1/p} \cdot Q(v, w)^{1/p^\ast}$ and $Q(v', w') = Q(v', w')^{1 - 1/p} \cdot Q(v', w')^{1/p}$ for all entries (if $Q(v, w) = 0$, both sides vanish by the convention $0^0 = 0$).
    Letting $B \ceq B_p(G, H, Q)$, we hence obtain
    \begin{align*}
        &\quad\, \left| \y^\top A_{G, H}^Q \x \right| \\
        &=
        \left| \sum_{v, v' \in V} \y(v) \left( 
            \sum_{w, w' \in W} \Disc(v, w, v', w') Q(v, w) Q(v',w')
        \right) \x(v') \right| \\
        &=
        \left| \sum_{\substack{v, v' \in V\\ w, w' \in W}} \Disc(v, w, v', w') 
        Q(v, w)\y(v) 
        Q(v',w')\x(v') \right| \\
        &=
        \left| \sum_{\substack{v, v' \in V\\ w, w' \in W}} \Disc(v, w, v', w') 
        Q(v, w)^{1/p} \underbrace{Q(v, w)^{1-1/p}\y(v)}_{\Tilde{\y}(v, w)} 
        Q(v',w')^{1-1/p} \underbrace{Q(v',w')^{1/p}\x(v')}_{\Tilde{\x}(v', w')} \right| \\
        &=
        \left| \sum_{(v, w), (v', w') \in V \times W} 
        \Tilde{\y}(v, w) 
        \underbrace{ 
            \Disc(v, w, v', w') 
            Q(v, w)^{1/p}
            Q(v',w')^{1-1/p}
        }_{= \, B((v, w), (v', w'))}
        \Tilde{\x}(v', w') 
        \right| \\
        &\leq
        \max_{\norm{\Tilde{\y}}_{p^\ast} = 1, \norm{\Tilde{\x}}_p = 1} \left| \Tilde{\y}^\top B \Tilde{\x} \right|
        = \norm{B}_p
    \end{align*}
    which completes the proof.
\end{proof}

The above lemma has the following simple consequence.
\begin{corollary}
    \label{claim:lower_bounding_A_GHQ}
    Let $1 \leq p \leq \infty$.
    Suppose the matrix $A^Q_{G,H}$ has a column with at least $k$ entries of absolute value at least $\eps$.
    Then $\norm{A_{G, H}^Q}_p \geq k^{b(p)} \eps$.
    In particular, if $k > (c/\eps)^{1/b(p)}$ for some $c > 0$, then $\norm{B_p(G, H, Q)}_p > c$.
\end{corollary}

\begin{proof}
    Let $\mathbf e_v \in \R^V$ denote the unit vector with $\mathbf e_v(v) = 1$ and $\mathbf e_v(u) = 0$ for $u \neq v$, so that $\norm{\mathbf e_v}_s = 1$ for all $1 \leq s \leq \infty$.
    As $A_{G, H}^Q \mathbf e_v$ is the column $A_{G, H}^Q(\cdot, v)$, we get
    \[
        \norm{A_{G, H}^Q}_p \geq \norm{A_{G, H}^Q \mathbf e_v}_p \geq \left(k \cdot \eps^p\right)^{1/p} = k^{1/p} \eps,
    \]
    where for $p = \infty$ the right-hand side equals $\eps$.
    Moreover, $A_{G, H}^Q$ is symmetric (exchange $(v, w)$ and $(v', w')$ in \cref{equ:entrywise_def}), and $\norm{A}_p = \norm{A^\top}_{p^\ast}$ for every matrix $A$ (see \cref{appendix:extended_preliminaries}).
    Hence, analogously,
    \[
        \norm{A_{G, H}^Q}_p = \norm{A_{G, H}^Q}_{p^\ast} \geq \norm{A_{G, H}^Q \mathbf e_v}_{p^\ast} \geq k^{1/p^\ast} \eps = k^{1 - 1/p} \eps.
    \]
    Together, $\norm{A_{G, H}^Q}_p \geq k^{b(p)}\eps$ with $b(p) = \max\{1/p, 1 - 1/p\}$.
    If $k > (c/\eps)^{1/b(p)}$, then $k^{b(p)} \eps > c$, and \cref{lemma:lower_bound_operator_p_blowup} yields $\norm{B_p(G, H, Q)}_p \geq \norm{A_{G, H}^Q}_p > c$ for this fixed $Q$.
\end{proof}

\begin{proof}[Proof of \cref{theorem:ot_op_p_general_hardness}]
    The proof is by reduction from \threepartition{}.
    Specifically, for a given instance $\mathcal I = (A, a_1, \ldots, a_{3m})$, the graph $G'$ consists of $3m$ paths where the $i$-th path, denoted $P_i$ has $a_i$ vertices.
    We assume \WLOG{} that $a_i \geq 3$ for all $i \in [3m]$. 
    Consequently, $A > 2a_i \geq 6$, so $A \geq 7$.
    The graph $H'$ consists of $m$ disjoint cycles on $A$ vertices each, which are denoted $C_j$.
    The graphs $G$ and $H$ are obtained from $G'$ and $H'$ respectively, by attaching at each vertex $u$ a total of $M$ leaves.
    The value of $M$ will be specified later in the proof.
    We refer to the vertex set of $G'$ and $H'$ by $V$ and $W$, respectively. The vertex sets of $G$ and $H$ are referred to as $X$ and $Y$ and we have $V \subseteq X$ and $W \subseteq Y$.
    For each graph, we partition the vertex set into the base vertex sets $\mathcal B_G, \mathcal B_H$, which correspond to the vertices in $G$ and $H$ that were already present in $G'$ and $H'$, and the leaf sets $\mathcal L_G, \mathcal L_H$ containing the vertices that were added in the augmentation.
    For a base vertex $v \in \mathcal B_G$, let $\mathcal L(v) \ceq N_G(v) \cap \mathcal L_G$ denote the set of leaves attached to $v$. We use the same notation for vertices in $H$.

    For a fixed $Q \in \DS(X, Y)$, we allow both arguments to $Q$ to also be sets, letting $Q(S, T) \ceq \sum_{s \in S, t\in T} Q(s, t)$ for $S \subseteq X, T \subseteq Y$. For simplicity we omit the set notation for input sets that are singletons. If $S = \emptyset$ or $T = \emptyset$, then $Q(S, T) = 0$.

    Finally we make the following useful observation that will be used throughout the proof.
    If $xx' \in E(G)$, then
    \[
        A_{G, H}^Q(x, x') = \sum_{yy' \notin E(H)} Q(x,y)Q(x', y').
    \]

    It remains to show correctness of the proposed reduction.
    Fix $0 < \eps^\ast < \sqrt 2 - 1$.
    We show that YES-instances satisfy $\dOT{p}(G, H) \leq 1$, whereas NO-instances satisfy $\dOT{p}(G, H) > 2^{b(p)} - \eps^\ast > 1$, where the last inequality holds as $b(p) \geq 1/2$.
    This proves both \NP-hardness and the claimed inapproximability (a $(2^{b(p)} - \eps^\ast)$-approximation would return a value of at most $2^{b(p)} - \eps^\ast$ on YES-instances and of more than $2^{b(p)} - \eps^\ast$ on NO-instances). 

    If $\mathcal I$ is a YES-instance, then the witnessing partition function $\sigma: [3m] \to [m]$ yields a natural way of linking up three paths into a cycle by inserting three edges, yielding a bijection $\tau: \mathcal B_G \to \mathcal B_H$.
    We extend $\tau$ to a bijection $\pi: X \to Y$ by choosing an arbitrary bijection $\mu_v \colon \mathcal L(v) \to \mathcal L(\tau(v))$ for each $v \in \mathcal B_G$ and setting $\pi \ceq \tau \cup \bigcup_{v \in \mathcal B_G} \mu_v$.
    Since $A_H - A_G^\pi$ corresponds to the adjacency matrix of a graph consisting of $3m$ disjoint edges and isolated vertices, we have $\norm{A_G^\pi - A_H}_p = 1$.

    In the following, let $\mathcal I$ be a NO-instance.
    Fix an arbitrary $Q \in \DS(X, Y)$.
    We start with two claims that allow us to conclude that any admissible $Q$ is approximately a bijection between $\mathcal B_G$ and $\mathcal B_H$.
    \begin{claim}
        \label{subclaim:fraction_to_base_vertex-rev}
        Let $v \in \mathcal B_G$ and suppose that $Q(v, \mathcal L_H) \geq \eta$ for some $\eta > 0$.
        If $M > 2\left( \frac{4}{\eta} \right)^2 + 1$, then $\norm{B_p(G, H, Q)}_p > 2^{b(p)}$.
    \end{claim}
    \begin{claimproof}
        Let $N \ceq N_G(v) \subseteq X$ and note that $|N| \geq M+1$.
        Let $L \ceq \{ y \in \mathcal L_H \mid Q(v, y) > 0 \}$ and denote by $z_y$ the unique neighbour of $y$ in $H$ and let $Z \ceq \{ z_y \mid y \in L\} \subseteq \mathcal B_H$.
        We distinguish the following two cases.
        \subparagraph*{Case 1.} For all  $w \in \mathcal B_H$, $Q(v, \mathcal L(w)) \leq \eta/2$.\\
        Then, for all $x \in N$, we have
        \begin{align*}
            A_{G,H}^Q(x, v) &\geq \sum_{y \in L} Q(v, y) \sum_{y' \in Y, yy' \notin E(H)} Q(x, y') \\
            &\geq \sum_{y \in L} Q(v, y) \sum_{y' \in Y \setminus \{z_y\} } Q(x, y') \\
            &= \sum_{y' \in Y} Q(x, y') \sum_{y \in L: z_y \neq y'} Q(v, y) \\
            &\geq \mfrac{\eta}{2} \sum_{y' \in Y} Q(x, y') \\
            &\geq \mfrac{\eta}{2}.
        \end{align*}

        \subparagraph*{Case 2.} There is $w \in \mathcal B_H$ with $Q(v, \mathcal L(w)) > \eta/2$.\\
        For any $0 < \delta < 1$, at most $\lfloor 1/\delta \rfloor$ vertices $x$ can satisfy $Q(x, w) > \delta$.
        Conversely, at least $M+1 - \lfloor 1/\delta \rfloor$ vertices $x \in N$ have $Q(x, w) \leq \delta$, or equivalently $Q(x, Y \setminus \{w\}) \geq 1 - \delta$.
        For any such $x$ it follows that
        \begin{align*}
            A_{G, H}^Q(x, v) &\geq \left(\sum_{y \in L: z_y = w} Q(v, y)\right) \left(\sum_{y' \in Y \setminus \{w\}} Q(x, y')\right)
            \geq \mfrac{\eta}{2} (1-\delta).
        \end{align*}

        Choosing $\delta \ceq 1/2$, in either case we find at least $M-1$ entries in column $v$ of $A_{G, H}^Q$ that are at least $\eta/4$ each.
        Hence, if $M-1 > 2\left( \frac{4}{\eta} \right)^2$, then $\norm{B_p(G, H, Q)}_p > 2^{b(p)}$ 
        by \cref{claim:lower_bounding_A_GHQ}.
    \end{claimproof}

    \begin{claim}
        \label{subclaim:no_base_vertex_split-rev}
        Let $v \in \mathcal B_G$ and $w, w' \in \mathcal B_H$ be distinct with $Q(v, w), Q(v, w') \geq \eps$ for some $\eps > 0$.
        If $M > 2\left(\frac{2}{\eps}\right)^2 + 1$, then $\norm{B_p(G, H, Q)}_p > 2^{b(p)}$.
    \end{claim}
    \begin{claimproof}
        Let $I \ceq N_H(w) \cap N_H(w')$ and observe that $|I| \leq 1$ since $|V(C_j)| = A \geq 5$.
        Let $N \ceq N_G(v)$.
        Since $|N| \geq M+1$, a total mass of at least $M$ of the vertices in $N$ is mapped such that it is adjacent to at most one of $w$ and $w'$.
        For $0 < \delta < 1$, there are at most $\lfloor 1/\delta \rfloor$ vertices $x \in N$ with $Q(x, I) > \delta$, so for at least $|N| - \lfloor 1/\delta \rfloor$ vertices we have
        that $Q(x, Y \setminus I) \geq 1-\delta$.
        In particular for each $y \in Y \setminus I$ there is a $w_y \in \{w, w'\}$ such that $yw_y \notin E(H)$.
        Thus for at least $M+1 - \lfloor 1/\delta \rfloor$ vertices $x \in N$,
        it holds that
        \begin{align*}
            A_{G, H}^Q(x, v) &\geq \sum_{y \in Y \setminus I} Q(x, y) \left( \sum_{y' \in Y, yy' \notin E(H)} Q(v, y') \right) \\
            &\geq \sum_{y \in Y \setminus I} Q(x, y) Q(v, w_y) \\
            &\geq \eps \cdot \sum_{y \in Y \setminus I} Q(x, y) \\
            &\geq \eps(1-\delta).
        \end{align*}
        Choosing $\delta \ceq 1/2$ it follows that the $v$-th column of $A_{G,H}^Q$ has at least $M-1$ entries that are at least $\frac{\eps}{2}$ each.
        Then $\norm{B_p(G, H, Q)}_p > 2^{b(p)}$ by \cref{claim:lower_bounding_A_GHQ}.
    \end{claimproof}

    Let $\eta \ceq \frac{\eps^\ast}{8}$, $\eps \ceq \frac{\eps^\ast}{8}$ and set $\eps' \ceq \frac{\eps}{mA-1}$ and choose
    \begin{equation}
        \label{equation:condition_M-rev}
        M > \max\left\{ 2\left(\mfrac{4}{\eta}\right)^2 + 1,\, 2\left(\mfrac{2}{\eps'}\right)^2 + 1 \right\}.
    \end{equation}
    Since $\eps^\ast$ is a fixed constant, $M$ can be chosen as $O\big((mA)^2\big)$, independently of $p$.
    If the precondition of \cref{subclaim:fraction_to_base_vertex-rev} or \cref{subclaim:no_base_vertex_split-rev} holds for some $v \in \mathcal B_G$, then $\norm{B_p(G, H, Q)}_p > 2^{b(p)} > 2^{b(p)} - \eps^\ast$ and we are done. So assume otherwise.
    Invoking \cref{subclaim:no_base_vertex_split-rev} with $\eps'$, for every $v \in \mathcal B_G$, there are no two distinct vertices $w, w' \in \mathcal B_H$ with $Q(v, w), Q(v, w') \geq \eps'$.
    Since $|\mathcal B_H| = mA$, in particular, there is a $w \in \mathcal B_H$ with $Q(v, \mathcal L_H \cup \{w\}) \geq 1 - (mA-1) \eps' = 1 - \eps$.
    Additionally, by \cref{subclaim:fraction_to_base_vertex-rev}, $Q(v, \mathcal B_H) \geq 1-\eta$ or equivalently $Q(v, \mathcal L_H) \leq \eta$ which implies that $Q(v, w) \geq 1-\eps-\eta$.

    As $1 - \eps - \eta = 1 - \eps^\ast/4 > 1/2$, $Q$ induces a bijection $\tau: \mathcal B_G \to \mathcal B_H$, by letting $\tau(v) \ceq \arg\max_{w \in \mathcal B_H} Q(v, w)$.

    Now since $\mathcal I$ is a NO-instance, for all partition functions $\sigma: [3m] \to [m]$ with $|\sigma^{-1}(j)| = 3$ for each $j \in [m]$, there is a $j \in [m]$ with $\sum_{i \in \sigma^{-1}(j)} a_i \geq A + 1$.
    We claim that there exist $i \in [3m]$ and $v, v' \in V(P_i)$ with $vv' \in E(P_i)$ such that $\tau(v)$ and $\tau(v')$ lie in distinct cycles of $H'$.
    Otherwise, by connectivity of the paths, for each $i$ there is $j(i)$ with $\tau(V(P_i)) \subseteq V(C_{j(i)})$. Since $\tau$ is a bijection $\mathcal B_G \to \mathcal B_H$, each cycle $C_j$ receives entire paths of total size $\sum_{i : j(i) = j} a_i = A$, and the assumption $A/4 < a_i < A/2$ forces exactly three paths to be mapped into each cycle. Hence, this yields a valid 3-partition, contradicting that $\mathcal I$ is a NO-instance.
    Specifically, $Q(v, \tau(v)), Q(v', \tau(v')) \geq 1 - \eps - \eta$ and $\tau(v)\tau(v') \notin E(H)$.
    This implies that $A_{G, H}^Q(v', v) \geq (1 - \eps - \eta)^2$.

    As $a_i \geq 3$ and hence $|V(P_i)| \geq 3$, we may assume that $\deg_{P_i}(v) = 2$ and denote by $v''$ the second neighbour of $v$ in $P_i$.
    If $\tau(v'') \notin N_H(\tau(v))$, then similarly to before $A_{G, H}^Q(v'', v) \geq (1 - \eps - \eta)^2$.
    So suppose that $\tau(v'') \in N_H(\tau(v))$. Since $\deg_{H[\mathcal B_H]}(\tau(v)) = 2$, there exists $w^\ast \in (N_H(\tau(v)) \cap \mathcal B_H) \setminus \{\tau(v'')\}$ with $\tau^{-1}(w^\ast) \notin N_G(v)$.
    It follows that $A_{G, H}^Q(\tau^{-1}(w^\ast), v) \leq - (1-\eps-\eta)^2$.
    Intuitively, we have shown that $v$ either loses both its neighbours in $P_i$, or loses adjacency to one neighbour and gains adjacency to at least one non-neighbour under $\tau$.

    In either case, $A_{G, H}^Q(\cdot, v)$ contains at least two entries that are at least $(1-\eps-\eta)^2$ in absolute value, and therefore \cref{claim:lower_bounding_A_GHQ} (with $k = 2$) and \cref{lemma:lower_bound_operator_p_blowup} yield
    \[
        \norm{B_p(G, H, Q)}_p \geq \norm{A_{G, H}^Q}_p \geq 2^{b(p)}(1 - \eps - \eta)^2.
    \]
    By the choice $\eps = \eta = \eps^\ast/8$, we have $(1 - \eps - \eta)^2  > 1 - \eps^\ast/2$, and hence, using $b(p) \leq 1$,
    \[
        \norm{B_p(G, H, Q)}_p > 2^{b(p)}(1 - \eps^\ast/2) \geq 2^{b(p)} - \eps^\ast.
    \]
    Since $Q$ was arbitrary, we obtain $\dOT{p}(G, H) > 2^{b(p)} - \eps^\ast$. This shows correctness of our reduction, thus concluding the proof.
\end{proof}

\begin{proposition}
    \label{proposition:ot_edit_local_containment_in_np}
    Let $G, H$ be graphs of orders $n, m$ and let $\normsubscript \in \{1, \edit\}$.
    \begin{enumerate}
        \item \GSim{\ndOT{1}} and \GSim{\ndOT{\edit}} are \NP-complete.
        \item $\ndOT{1}(G, H)$ can be computed in time $2^{nm} \cdot \poly(n + m)$.
        \item There is $k \in \Npos$ with $\log k \leq \poly(n + m)$ such that $\ndOT{\normsubscript}(G, H) = \nd{\normsubscript}\big(G^{\blowup mk}, H^{\blowup nk}\big)$.
    \end{enumerate}
\end{proposition}

\begin{proof}
    \NP-hardness follows from \cref{theorem:ot_edit_dist_hardness,theorem:ot_local_edit_dist_hardness} since $\dOT{\normsubscript}$ and $\ndOT{\normsubscript}$ differ by a factor only depending on the order of the graphs.
    By \cref{equ:local_transport_identity}, $\ndOT{1}(G, H)$ is the optimal value of the mixed integer linear program
    \[
        \begin{aligned}
            \min \quad & z \\
            \text{s.t.} \quad
            & & \sum_{(v,w) \in V \times W} \big|\Disc(v, w, v', w')\big|\, x_{v,w}
              & {} \leq z + 1 - y_{v',w'}
              && \forall\, (v', w') \in V \times W, \\
            & & \x & {} \leq \y, \\
            & & \x & {} \in \DSS(V, W), \\
            & & \y & {} \in \{0, 1\}^{V \times W}, \\
            & & z & {} \geq 0.
        \end{aligned}
    \]

    Indeed, $x_{v', w'} > 0$ forces $y_{v', w'} = 1$, and for $y_{v', w'} = 0$ the first constraint is void as column sums are at most $\sum_{v, w} x_{v, w} = 1$; so an optimal choice is $y = \indicator_{\supp(x)}$.
    For fixed $\y$, this is a linear program of polynomial size, solvable in polynomial time with an optimal vertex solution of polynomial encoding length \cite[Theorem~10.2]{schrijver_linear_programming}.
    Hence guessing $\y$ shows \GSim{\ndOT{1}} $\in \NP$, enumerating all $2^{nm}$ choices of $\y$ proves the second statement, and $\ndOT{1}$ has an optimal $Q^\ast$ of polynomial encoding length.
    Similarly, $\ndOT{\edit}(G, H) = \min_{\x \in \DSS(V, W)} \sum |\Disc(v, w, v', w')|\, x_{v, w} x_{v', w'}$ is a quadratic program, which by \cite{vavasis_qp_in_np} has an optimal solution $Q^\ast$ of polynomial encoding length; it serves as the certificate.

    For the third statement, let $k$ be a common denominator of the entries of $Q^\ast$ and $N \ceq nmk$.
    For an alignment $\pi$ of $G^{\blowup mk}$ and $H^{\blowup nk}$, let $N Q_\pi(v, w)$ be the number of vertices of $G^{\blowup mk}$ mapped from the cloud of $v$ to the cloud of $w$; then $Q_\pi \in \DSS(V, W)$.
    Counting mismatches, $\norm{A_{G^{\blowup mk}}^\pi - A_{H^{\blowup nk}}}_{\normsubscript} = N^{\kappa_\normsubscript} \norm{B_{\normsubscript}(G, H, Q_\pi)}_{\normsubscript}$ where $\kappa_1 = 1$ and $\kappa_\edit = 2$, so $\nd{\normsubscript}(G^{\blowup mk}, H^{\blowup nk}) \geq \ndOT{\normsubscript}(G, H)$, with equality for $\pi$ satisfying $Q_\pi = Q^\ast$, which exists as $NQ^\ast$ is integral.
\end{proof}

\begin{proof}[Proof of \cref{theorem:ot_cut_hardness}]
    We follow the proof of \cite[Theorem 23]{gervens2022graphsimilaritybasedmatrix}, in which they showed the hardness of $\GSim{\d{\cut}}$.
    Given a graph $F$ on $n$ vertices and $m$ edges, let $G', H'$ be the graphs constructed in their proof (before they additionally introduce leaves attached to each vertex), which satisfy $\d{\cut}(G', H') = \Cost{\cut}(\pi) = 2\cdot \maxcut(F)$ where $\maxcut(F)$ denotes the size of a maximum cut in $F$ and $\pi$ is the canonical bijection between the two graphs.
    Let $n' \ceq 2m+n$ denote the order of $G'$ and $H'$ and let $v_1, \ldots, v_{n'}$ and $w_1, \ldots, w_{n'}$ be enumerations of their respective vertex sets.
    Recall that in the final step of their construction, $i \cdot (n^2 + n)$ leaves are attached to the $i$-th vertex in each graph.
    Our graphs $G, H$ are obtained by adding $i \cdot (n^2 + n) \cdot k$ leaves to the $i$-th vertex in both graphs, where $k$ will be chosen later.
    Note that each non-leaf vertex has at least $M \ceq k(n^2+n)$ adjacent leaves.
    Let $N$ denote the order of the resulting graphs.
    We write $\pi$ for the canonical bijection $v_i \mapsto w_i$ between the base vertices (both carry the same number $\ell_i$ of leaves).
    We claim that
    \begin{equation}
        \label{equation:cut_approx_bound-rev}
        \mfrac{33}{17} \cdot \maxcut(F) \leq \dOT{\cut}(G, H) \leq 2 \cdot \maxcut(F).
    \end{equation}
    The second inequality follows from $\dOT{\cut}(G, H) \leq \d{\cut}(G, H)$ for all graphs $G, H$ of the same order and the original proof in \cite{gervens2022graphsimilaritybasedmatrix}.
    Hence, we only need to show the first inequality.
    Note that \cref{equation:cut_approx_bound-rev} implies that exact computation of $\dOT{\cut}$ is \NP-hard, because for any $\eps > 0$, unless $\P = \NP$, \textsc{MaxCut} does not admit a polynomial-time approximation algorithm with multiplicative guarantee $\frac{16}{17}+\eps$ \cite{hastad_inapproximability_results}.
    We fix $\eps$ in the previous equation to be $\eps \ceq \frac{1}{34}$.

    The reduction of \cite{gervens2022graphsimilaritybasedmatrix} leverages the quantities of leaves attached at each vertex to force the optimal bijection between the graphs $G$ and $H$ to be the identity.

    We employ a similar strategy, however we first have to rule out fractional mappings that split base vertices, that is, the vertices that are not added as leaves, too much.
    Let $\mathcal B_G, \mathcal B_H$ denote the base vertices of the construction and $\mathcal L_G, \mathcal L_H$ the leaf vertices.
    For a vertex $v \in \mathcal B_G$, we define $\mathcal L(v) \ceq \mathcal L_G \cap N_G(v)$, the definition for vertices in $H$ is analogous. Note that this notation suppresses the subscript indicating the graph, however it will be clear which graph we refer to by vertex naming.
    In the remainder of the proof, $V \ceq V(G)$ and $W \ceq V(H)$ denote the vertex sets of the graphs.
    Fix an arbitrary $Q \in \DS(V, W)$ and write $B \ceq B(G, H, Q)$. We show that $\norm{B}_\cut \geq \frac{33}{17}\maxcut(F)$; as $Q$ is arbitrary, this proves the first inequality of \cref{equation:cut_approx_bound-rev}.

    \begin{claim}
        \label{subclaim:cut_base_to_leaf_bound-rev}
        Let $v \in \mathcal B_G$ and suppose that $Q(v, \mathcal L_H) \geq \eta$ for some $\eta > 0$.
        If $k \geq \frac{4}{\eta}$, then $\norm{B(G, H, Q)}_\cut \geq 2\cdot\maxcut(F)$.
    \end{claim}
    \begin{claimproof}
        Let $U \ceq \mathcal L(v) \subseteq V$ and note that $|U| \geq M$. Let $L \ceq \{ y \in \L_H \mid Q(v, y) > 0\}$ and denote by $z_y$ the unique neighbour of $y$ in $H$ and let $Z \ceq \{ z_y \mid y \in L \} \subseteq \mathcal B_H$.

        Write $B \ceq B(G, H, Q)$.
        We distinguish two cases.

        \subparagraph*{Case 1.} For all $w \in \mathcal B_H$, $Q(v, \mathcal L(w)) \leq \eta/2$. \\
        Set $S' \ceq \{v\} \times L$ and $T' \ceq U \times W$.
        Then
        \begin{align*}
             B(S', T') &= \sum_{\substack{v' \in U\\w \in L, w' \in W}} \Big( \underbrace{A_G(v, v')}_{=1} - A_H(w, w') \Big)Q(v, w)Q(v', w') \\
             &= \sum_{\substack{v' \in U\\w \in L, w' \in W \setminus \{z_w\} }} Q(v, w)Q(v', w') \\
             &= \sum_{v' \in U, w' \in W} Q(v', w') \underbrace{\sum_{w \in L: z_w \neq w'} Q(v, w)}_{\geq \frac{\eta}{2}} \\
             &\geq \mfrac{\eta}{2} \sum_{v' \in U, w' \in W} Q(v', w') \\
             &\geq M \cdot \mfrac{\eta}{2}
        \end{align*}
        and thus $\norm{B}_\cut \geq \frac{1}{2} \eta M$.

        \subparagraph*{Case 2.} There is $w^\ast \in \mathcal B_H$ with $Q(v, \mathcal L(w^\ast)) > \eta/2$.\\
        For any $0 < \delta < 1$, at most $\lfloor 1/\delta \rfloor$ vertices $x$ can satisfy $Q(x, w^\ast) > \delta$.
        Conversely, at least $M - \lfloor 1/\delta \rfloor$ vertices $x \in U$ have $Q(x, w^\ast) \leq \delta$, or equivalently $Q(x, W \setminus \{w^\ast\}) \geq 1 - \delta$.
        Let $U' \ceq \{x \in U \mid Q(x, W \setminus \{w^\ast\}) \geq 1 - \delta \}$.
        By the previous observation $|U'| \geq M - \lfloor 1/\delta \rfloor$.
        Choose $S' \ceq \{v\} \times \mathcal L(w^\ast)$ and $T' \ceq U' \times (W \setminus \{w^\ast\})$.

        For $w \in \mathcal L(w^\ast)$ and $w' \neq w^\ast$, we have $A_H(w, w') = 0$, for each $v' \in U'$ we have $A_G(v, v') = 1$.
        Therefore
        \begin{align*}
            B(S', T') &= \sum_{\substack{v' \in U'\\w \in \mathcal L(w^\ast), w' \in W \setminus \{w^\ast\}}} \Big( \underbrace{A_G(v, v')}_{=1} - \underbrace{A_H(w, w')}_{=0}\Big)Q(v, w)Q(v', w') \\
            &= \sum_{\substack{v' \in U'\\w \in \mathcal L(w^\ast), w' \in W \setminus \{w^\ast\}}} Q(v, w)Q(v', w') \\
            &= \underbrace{Q(v, \mathcal L(w^\ast))}_{> \frac{\eta}{2}}  \sum_{v' \in U'} \underbrace{Q(v', W \setminus \{w^\ast\})}_{\geq 1- \delta} \\
            &\geq \mfrac{\eta}{2}(1-\delta)\left( M - \left\lfloor \mfrac{1}{\delta}\right\rfloor \right)
        \end{align*}
        Choosing $\delta \ceq \frac{1}{2}$ now guarantees that $B(S', T') \geq \frac{1}{8}\eta M$ given that $M \geq 4$ which holds if $n \geq 2$ or $k \geq 4$.

        In either case we have shown that $\norm{B}_\cut \geq \frac{1}{8}\eta M$.
        Note that $\maxcut(F) \leq \frac{n^2}{4}$ for any graph $F$.
        Hence the claim follows if $\frac{1}{8}\eta M \geq \frac{n^2}{2}$.
        Since $M \geq kn^2$, this holds given that $kn^2 \geq \frac{4n^2}{\eta}$ which is equivalent to $k \geq \frac{4}{\eta}$.
    \end{claimproof}

    \begin{claim}
        \label{subclaim:cut_base_to_distinct_base_bound-rev}
        Let $v \in \mathcal B_G$ and $w, w' \in \mathcal B_H$ be  distinct with $Q(v, w), Q(v, w') \geq \tau$ for some $\tau> 0$.
        If $k \geq \max\left\{ \frac{2}{\tau}, 4(n-1) \right\}$, then $\norm{B(G, H, Q)}_\cut \geq 2\cdot \maxcut(F)$.
    \end{claim}
    \begin{claimproof}
        Let $U \ceq N_G(v)$, $I \ceq N_H(w) \cap N_H(w')$ and observe that $|U| \geq M$ and $|I| \leq n-1$.
        To see the latter, let $A$ denote the $2m \times n$ matrix constructed by Alon and Naor from the input graph $F$ (compare \cite[Proposition 3.2]{cut_grothendieck}).
        In this matrix, every edge of $F$ contributes two rows, each with exactly one
        entry $+1$ and one entry $-1$. Hence, for every vertex $u$ of $F$, the column
        of $u$ contains exactly $\deg_F(u) \le n-1$ entries $+1$ and $\deg_F(u)$
        entries $-1$. Since $G'$ contains the edges of the positive entries and $H'$
        those of the negative entries, column vertices have degree at most $n-1$ and
        row vertices have degree $1$ in both $G'$ and $H'$. Thus $G'$ and $H'$ have
        maximum degree at most $n-1$. Leaves are adjacent to only one vertex, so they
        never lie in $I$, and hence $|I| \le n-1$.
        For $0 < \delta < 1$ there are at most $\lfloor 1/\delta \cdot |I| \rfloor$ vertices $x \in U$ with $Q(x, I) > \delta$, so for at least $|U| - \lfloor 1/\delta \rfloor \cdot |I|$ vertices, we have that $Q(x, W \setminus I) \geq 1 - \delta$.
        Let $U' \ceq \{ x \in U \mid Q(x, W \setminus I) \geq 1 - \delta \}$.
        For all $y \in W \setminus I$, there is a $w_y \in \{w, w'\}$ such that $yw_y \notin E(H)$. Let $\overline{w_y}$ denote the unique element in $\{w, w'\} \setminus \{w_y\}$.
        Now set $S' \ceq \{(v, w), (v, w')\}$ and $T' \ceq U' \times (W \setminus I)$.
        Then
        \begin{align*}
            B(S', T') &= \sum_{z \in \{w, w'\}} \sum_{v' \in U', z' \in W \setminus I} \Big( \underbrace{A_G(v, v')}_{=1} - A_H(z, z') \Big) Q(v, z)Q(v', z') \\
            &\geq \sum_{v' \in U'} \sum_{z' \in W \setminus I} \underbrace{Q(v, w_{z'})}_{\geq \tau} Q(v', z') \\
            &\geq \tau \sum_{v' \in U'} Q(v', W \setminus I) \\
            &\geq \tau(1-\delta)\left( M - \left\lfloor \mfrac{1}{\delta} \right\rfloor \cdot |I| \right)
        \end{align*}
        where the first inequality holds by choosing for each $z'$, the $z \in \{w, w'\}$ such that $z = w_{z'}$. The summand with $\overline{w_{z'}}$ is non-negative, so omitting it yields a valid lower bound.
        Choosing $\delta \ceq \frac{1}{2}$, we have $B(S', T') \geq \frac{1}{4}\tau M$ given that $M \geq 4(n-1) \geq 4|I|$, since in this case $M - \lfloor1/\delta\rfloor \cdot |I| \geq M/2$.
        This implies that $\norm{B}_\cut \geq \frac{1}{4}\tau M$.
        Now the claim follows, since $M \geq kn^2 \geq \frac{2n^2}{\tau}$ implies that $\norm{B}_\cut \geq \frac{n^2}{2} \geq 2\cdot \maxcut(F)$.
    \end{claimproof}

    Let $\eta, \eps > 0$ be sufficiently small.
    Invoking \cref{subclaim:cut_base_to_distinct_base_bound-rev} with $\tau \ceq \frac{\eps}{n'-1}$, for every $v \in \mathcal B_G$, there are no two distinct vertices $w, w' \in \mathcal B_H$ with $Q(v, w), Q(v, w') \geq \tau$. Since $|\mathcal B_H| = n'$, it follows that for each $v \in \mathcal B_G$ there is a $w \in \mathcal B_H$ with $Q(v, \mathcal L_H \cup \{w\}) \geq 1 - (n'-1)\tau = 1 - \eps$.
    Moreover, by \cref{subclaim:cut_base_to_leaf_bound-rev}, $Q(v, \mathcal L_H) \leq \eta$ which implies that $Q(v, w) \geq 1 - \rho$ where $\rho \ceq \eps + \eta$.

    Recall the enumerations $v_1, \ldots, v_{n'}$ and $w_1, \ldots, w_{n'}$ where $\pi(v_i) = w_i$ for all $i \in [n']$.
    The next claim shows that if $Q$ does not \enquote{respect} this mapping approximately, then the cut norm is already large.
    \begin{claim}
        \label{subclaim:cut_no_permutation_on_base_vertices-rev}
        Let $i, j \in [n']$ be distinct with $Q(v_i, w_j) \geq 1 - \rho$. Then, if $k \geq \frac{1}{2(1-\rho)}$, it holds that $\norm{B(G, H, Q)}_\cut \geq 2 \cdot \maxcut(F)$.
    \end{claim}
    \begin{claimproof}
        We show the claim for the case $i < j$; the other case is identical after reversing the roles of the two graphs.
        Let $\ell_i, \ell_j$ denote the number of leaves attached to vertices $v_i, v_j$ ($w_i, w_j$), respectively. Recall that for each $i \in [n']$ we have $\ell_i = i(n^2 + n)k$.
        Define $R \ceq V(G) \setminus N_G(v_i)$ and choose $S' \ceq \{(v_i, w_j)\}$ and $T' \ceq R \times \mathcal L(w_j)$.
        Then
        \begin{align*}
            B(S', T') &= \sum_{x \in R, y \in \mathcal L(w_j)} \Big( \underbrace{A_G(v_i, x) - A_H(w_j, y)}_{=-1} \Big) Q(v_i, w_j)Q(x, y) \\
            &\leq -(1-\rho) \sum_{x \in R} Q(x, \mathcal L(w_j))
        \end{align*}
        Note that $|R| \geq N - (\ell_i + n)$ since $v_i$ has $\ell_i$ neighbours that are leaves and at most $n-1$ neighbours within $\mathcal B_G$ due to the properties of the matrix of Alon and Naor.
        Moreover, $|V(H) \setminus \mathcal L(w_j)| = N - \ell_j $. Intuitively, there is more mass in $R$ than what can possibly fit into $V(H) \setminus \mathcal L(w_j)$. In particular this implies that the mass transported from $R$ to $\mathcal L(w_j)$ is at least $\ell_j  - (\ell_i + n)$, so
        \[
            \sum_{x \in R} Q(x, \mathcal L(w_j)) \geq \ell_j - (\ell_i + n)
        \]
        Combined with the previous inequality, this gives
        \begin{align*}
            |B(S', T')| &\geq (1-\rho)(\ell_j - \ell_i - n) \\
            &\geq (1-\rho) \big((n^2 + n)k - n \big) \\
            &\geq (1-\rho)n^2k
        \end{align*}
        Thus choosing $k$ sufficiently large such that
        \[
            2 \cdot \maxcut(F) \leq \mfrac{n^2}{2} \leq (1-\rho)n^2k
        \]
        concludes the proof.
        This is equivalent to
        $k \geq \frac{1}{2(1-\rho)}$.
    \end{claimproof}

    To conclude the proof, we now show that if $Q$ is approximately the identity bijection (on the base vertices), then the cut norm of $B(G, H, Q)$ cannot be significantly smaller than $2 \cdot \maxcut(F)$.
    This is the last case we need to consider and thus proves \cref{equation:cut_approx_bound-rev}.
    The basic idea is that we take the maximiser $(S, T)$ under the bijection $\pi$ realising the value $2\cdot\maxcut(F)$ and show that the same pair yields a sufficiently close value under the assumption that $Q$ is \enquote{close} to being the bijection $\pi$.
    \begin{claim}
        \label{subclaim:close_to_identity_cut_norm_lower_bound-rev}
        Let $\delta \in (0, 1)$ and $0 \leq \alpha \leq \frac{\delta}{(2n')^2}$.
        If $Q(v_i, w_i) \geq 1 - \alpha$ for all $i \in [n']$, then
        \[
            \norm{B(G, H, Q)}_\cut \geq 2(1-\delta) \cdot \maxcut(F).
        \]
    \end{claim} 
    \begin{claimproof}
        Write $B \ceq B(G, H, Q)$.
        Let $S, T \subseteq \mathcal B_G$ with
        \[
            R \ceq \left| \left( A_G - A_H^{\pi^{-1}} \right)(S, T) \right| = 2 \cdot \maxcut(F).
        \]
        For $s \in S, t \in T$, let $A(s, t) \ceq A_G(s, t) - A_H(\pi(s), \pi(t))$. Moreover, let $\sigma \in \{-1, +1\}$ be the sign such that $\sigma A(S, T) = R$.
        We choose $S' \ceq S \times W$ and $T' \ceq T \times W$.
        For $s \in S \cup T$, we write $q_s \ceq Q(s, \pi(s))$.

        We decompose $B(S', T')$ into two parts: The first part ($M$) is the contribution of terms corresponding to the identity alignment $\pi$, the second part ($E$) is the error term accounting for the contribution of all other terms.
        Hence, to bound $B(S', T')$, we bound $M$ and $E$ individually.

        Since $|A(s, t)| \leq 1$, we have
        \begin{align*}
            \sigma M &= \sum_{s \in S, t \in T} \sigma A(s, t) q_sq_t \\
            &= \sum_{s \in S, t \in T} \sigma A(s, t) - \sum_{s \in S, t \in T} \sigma A(s, t)(1 - q_sq_t) \\
            &\geq R - \sum_{s \in S, t \in T} (1 - q_sq_t) \\
            &\geq R - 2\alpha|S||T|.
        \end{align*}
        For the last inequality, we used
        \[
            1 - q_sq_t \leq 1 - (1-\alpha)^2 = 2\alpha - \alpha^2 \leq 2 \alpha.
        \]
        Next, we estimate the contribution of terms that do not correspond to mapping according to $\pi$.
        For each pair $(s, t)$, the total mass under $Q$ of all non-identity aligned pairs $(w, w') \neq (\pi(s), \pi(t))$ is at most $1 - q_sq_t \leq 2\alpha$. Since every adjacency difference is bounded by one in absolute value, summing this over all pairs $(s, t)$ yields $|E| \leq 2\alpha |S||T|$.

        Put together, we obtain
        \begin{align*}
            |B(S', T')| &\geq \sigma B(S', T') \\
            &= \sigma M + \sigma E \\
            &\geq R - 2\alpha |S||T| - |E| \\
            &\geq R - 4\alpha |S||T|.
        \end{align*}
        Now since $S, T \subseteq \mathcal B_G$, it holds that $|S||T| \leq n'^2$ and hence
        \[
            \norm{B}_\cut \geq |B(S', T')| \geq R - 4\alpha n'^2 \geq 2(1-\delta)\cdot \maxcut(F). \qedhere
        \]
    \end{claimproof}

    Fix a constant $c$ with $1 \leq c < \frac{17}{16}$ and choose a
    constant $0 < \delta < 1-\frac{16c}{17}$.
    Observe that
    \cref{subclaim:cut_base_to_leaf_bound-rev,subclaim:cut_base_to_distinct_base_bound-rev,subclaim:cut_no_permutation_on_base_vertices-rev,subclaim:close_to_identity_cut_norm_lower_bound-rev}
    exhaustively cover the space of possible matrices $Q$.
    It remains to choose $k$ sufficiently large as to ensure that the
    preconditions of these claims are all satisfied.
    For simplicity, in the following we do not optimise the values chosen.
    Since $n' = 2m+n \leq n^2$, we may choose
    $\alpha \ceq \frac{\delta}{200n^4}
        \leq \frac{\delta}{(2n')^2}$,
    $\eta \ceq \frac{\alpha}{2} = \frac{\delta}{400n^4}$,
    $\eps \ceq \frac{\alpha}{2} = \frac{\delta}{400n^4}$
    and hence $\tau = \frac{\alpha}{2(n'-1)}$ and
    $\rho = \alpha = \frac{\delta}{200n^4}$.
    For convenience, we instead choose
    $\tau \ceq \frac{\alpha}{4n^2}
        \leq \frac{\alpha}{2(n'-1)}$,
    so that $\tau = \frac{\delta}{800n^6}$.

    Then, it suffices to choose $k$ such that
    \begin{align*}
        k &\geq \max\left\{
            \mfrac{4}{\eta},
            \mfrac{2}{\tau},
            4(n-1),
            \mfrac{1}{2(1-\rho)}
        \right\}
    \end{align*}
    to ensure that the preconditions of
    \cref{subclaim:cut_base_to_leaf_bound-rev,subclaim:cut_base_to_distinct_base_bound-rev,subclaim:cut_no_permutation_on_base_vertices-rev,subclaim:close_to_identity_cut_norm_lower_bound-rev}
    are satisfied.
    Since $\delta < 1$ and $n \geq 1$, we have
    $\rho = \alpha \leq \frac{1}{200} \leq \frac{1}{2}$ and hence
    $\frac{1}{2(1-\rho)} \leq 1$.
    Plugging in the values, we get
    \[
        k \geq \max\left\{
            \mfrac{1600n^4}{\delta},
            \mfrac{1600n^6}{\delta},
            4(n-1),
            1
        \right\},
    \]
    and we choose
    $k \ceq \left\lceil \frac{1600n^6}{\delta} \right\rceil$.
    This means that the resulting graphs are of order at most
    \begin{align*}
        \sum_{i=1}^{n'} (1+\ell_i)
        &\leq \sum_{i=1}^{n'}
            \left(1+i(n^2+n)
                \left\lceil\mfrac{1600n^6}{\delta}\right\rceil
            \right)
        \\
        &\leq n'
            + \frac{n'(n'+1)}{2} \cdot 2n^2
                \cdot \left\lceil\mfrac{1600n^6}{\delta}\right\rceil
        \\
        &\leq n^2 + 2n^6
            \left(\mfrac{1600n^6}{\delta}+1\right)
        = O\left(\mfrac{n^{12}}{\delta}\right),
    \end{align*}
    which is polynomial in the order of the input graph $F$,
    since $\delta$ is fixed.

    By the preceding claims and the upper bound established earlier,
    \[
        2(1-\delta)\maxcut(F)
        \leq \dOT{\cut}(G,H)
        \leq 2\maxcut(F).
    \]
    Hence, a polynomial-time $c$-approximation algorithm for
    $\dOT{\cut}$ would approximate the value $\maxcut(F)$
    with multiplicative error at most
    \[
        \mfrac{c}{1-\delta} < \mfrac{17}{16},
    \]
    contradicting the inapproximability result for \textsc{MaxCut}, assuming that $\P \neq \NP$.
    Since $c < \frac{17}{16}$ was arbitrary, this concludes the proof.
\end{proof}

\begin{remark}[Hardness for Other Operator $p \to q$-norms]
    \cref{lemma:lower_bound_operator_p_blowup,claim:lower_bounding_A_GHQ} carry over to $\norm{\cdot}_{p \to q}$ with $b(p)$ replaced by $r \ceq \max\{1/q,\, 1 - 1/p\}$.
    For $p \leq q$, the value of YES-instances under the canonical bijection remains $1$, so the proof of \cref{theorem:ot_op_p_general_hardness} yields inapproximability within $2^r - \eps$ whenever $(p, q) \neq (1, \infty)$.
    For $p > q$, the value of YES-instances grows like $m^{1/q - 1/p}$, so the single-column bound of \cref{claim:lower_bounding_A_GHQ} no longer yields a gap. 
    In the proof of \cref{theorem:ot_cut_hardness}, leaves force every \enquote{good} $Q$ to be close to the canonical bijection. Similarly to our proof above, one can show that this gives $(1 - \frac{1}{\poly(n)})\norm{D}_{p \to q} \leq \dOT{p \to q}(G, H) \leq \norm{D}_{p \to q}$ for any symmetric $D \in \{-1, 0, 1\}^{n' \times n'}$ with zero diagonal.
    Hence, hardness reduces to approximating $\norm{D}_{p \to q}$.
    For $p = \infty > q > 1$, \NP-hardness then follows from \textsc{MaxCut}, as the oriented incidence matrix $M$ of $F$ satisfies $\norm{M}_{\infty \to q} = 2\cdot\maxcut(F)^{1/q}$. For $1 < q < p < \infty$ the complexity remains open.
\end{remark}